\documentclass[preprint,12pt]{elsarticle}
\usepackage{subfigure}
\usepackage{lineno,hyperref}
\usepackage{indentfirst}
\usepackage{bm}
\usepackage{threeparttable}
\usepackage{multirow}
\usepackage{mathrsfs}
\usepackage{amsmath}
\usepackage{array}
\usepackage{amssymb}
\usepackage{graphicx}
\usepackage{url}
\usepackage{float}
\usepackage{hyperref}
\usepackage{epstopdf}
\usepackage{booktabs}
\usepackage{color}
\usepackage{lscape}
\usepackage{fancyhdr}
\usepackage[top=2.5cm, bottom=2.5cm, left=2cm, right=2cm]{geometry}
\modulolinenumbers[5]

\journal{Elsevier}

\biboptions{numbers,sort&compress}

\begin{document}

\begin{frontmatter}

\title{Adaptive criterion and modification of wave-particle decomposition in UGKWP method for high-speed flow simulation}

\author{Junzhe Cao$^{a,}$$^b$}
\ead{jcaobb@connect.ust.hk}

\author{Yufeng Wei$^a$}
\ead{yweibe@connect.ust.hk}

\author{Wenpei Long$^a$}
\ead{wlongab@connect.ust.hk}

\author{Chengwen Zhong$^b$}
\ead{zhongcw@nwpu.edu.cn}

\author[]{Kun Xu$^{a,}$$^{c,}$$^{d,}$$^*$\corref{mycorrespondingauthor}}
\ead{makxu@ust.hk}

\address{$^a$Department of Mathematics, Hong Kong University of Science and Technology, Hong Kong, China\\
$^b$School of Aeronautics, Northwestern Polytechnical University, Xi'an, Shaanxi 710072, China\\
$^c$Department of Mechanical and Aerospace Engineering, Hong Kong University of Science and Technology, Hong Kong, China\\
$^d$HKUST Shenzhen Research Institute, Shenzhen, 518057, China}

\begin{abstract}
Benefitting from the direct modeling of physical laws in a discretized space and the automatic decomposition of hydrodynamic waves and particles, the unified gas-kinetic wave-particle (UGKWP) method offers notable advantages in various multiscale physics, such as hypersonic flow, plasma transport and radiation transport. Aiming at achieving a more suitable and efficient wave-particle decomposition in high-speed flow simulation and enhancing the performance in the drastic scale variation region, in this work, the scale adaptive criterion is studied and the flux evolution of UGKWP method is modified. Specifically, besides the perspective of time which is naturally considered in the time-dependent distribution function of UGKWP method, two more criteria from views of space and gradient are utilized to identify the local scale, and to reduce the computational consumption of particles on describing the near-equilibrium microscopic gas distribution function. Moreover, corresponding to the coefficients in the time integration flux of unified gas-kinetic scheme (UGKS), the evolution of hydrodynamic wave is modified to be more consistent with particles, which is essential when the scale changes intensely in different cells. A variety of test cases are conducted to validate the performance of the adaptive UGKWP method, including hypersonic flows around a cylinder at multiple inflow Knudsen numbers, hypersonic flow over a slender cavity, side-jet impingement on hypersonic flow and three-dimensional hypersonic flows over a $70^{\circ}$ blunted cone with a cylindrical sting.
\end{abstract}


\end{frontmatter}

\section{Introduction}\label{Sec:introduction}
The multiscale non-equilibrium flows are receiving increasing attention~\cite{dsmcre,dm}, with the development of hypersonic vehicles and micro-electromechanical system~\cite{wpwp,tiangong,wy,gsisale}. To illustrate, continuum and rarefied flow regimes can occur simultaneously in different areas of near-space vehicles, mainly because of the strong compression caused by the bow shock at the head of the blunt body, the thickened Knudsen layer along the wall, and intense expansion around the wake~\cite{jiangaia}. Since the nonlinear constitutive relation in the non-equilibrium flow can hardly be recovered by Navier--Stokes (N-S) equations at the hydrodynamic scale, a variety of numerical methods based on the more fundamental Boltzmann equation have been developed~\cite{dvmre,imexre,gkuare,ugksre,dugksre,gsisre,dsmcre,apdsmcre,uspbgkre,trmcre,wpre}. Following the direct modeling methodology~\cite{dm}, an effective multiscale unified gas-kinetic scheme (UGKS) has been proposed and developed for multiscale hypersonic flow simulations~\cite{ugksre,ugks1,ugks2,ugks3,ugks4}, further extended to other physics such as micro flow~\cite{ugks5}, plasma transport~\cite{ugks6} and radiative transport~\cite{ugks7}. For the flux of UGKS, the free transport of molecules is coupled with the collision by the integral solution along the characteristic line. As a result, the discrete scales of UGKS in the physical space and time are no longer limited to be less than the molecular mean free path and mean collision time, but recovering the hydrodynamic level of N-S solver in the near-continuum flow regime. This advantage has been analyzed in the unified preserving (UP) way~\cite{up}.

On the framework of UGKS, the unified gas-kinetic wave-particle (UGKWP) method is proposed by employing the stochastic particle to discretize the microscopic gas distribution function instead~\cite{wpre,ugkwp2}. For the hypersonic flow, since both the peak value of temperature and the inflow Mach number ($\rm{Ma_{\infty}}$) are high, the distribution function is too broad globally and too sharp in some local region. As a result, the deterministic discrete velocity space in the UGKS is prohibitively expensive, while the Monte Carlo-type stochastic particle approach is rather efficient. However, its efficiency goes down in the low-speed flow simulation, suffering from the intrinsic statistical noise. On one hand, in the UGKWP method, the non-equilibrium phenomenon is described by particles in the rarefied flow regime. On the other hand, in the continuum flow regime, the UGKWP method can recover the gas-kinetic scheme (GKS)~\cite{gks} without extra calculation and memory consumption. There are few particles in the domain, and the gas evolution is mainly constructed by deterministic hydrodynamic waves. In the intermediate transition regime, particles coexist with hydrodynamic waves in a single cell, whose evolutions are accurately controlled by the integral solution of kinetic model equation. In recent years, the UGKWP method has been studied~\cite{ugkwp3,ugkwp4,suwp,dugkwp} and developed to other complex physics including thermal non-equilibrium flow~\cite{ugkwp6}, gas-solid particle two-phase flow~\cite{ugkwp7}, radiation transport~\cite{ugkwp8}, partially ionized plasma~\cite{ugkwp9} and turbulent flow~\cite{ugkwp10}. Especially with the flexibility of particles, the UGKWP method has advantages in the direct modeling methodology~\cite{ugkwp11}.

The wave-particle formulation provides an appropriate framework for multiscale modeling and numerical method. However, in the original UGKWP method, only the view of time is considered for adjusting the weights between hydrodynamic wave and particle, which brings about a great deal of extra computation consumption of particles. For example, although the free inflow is always at the equilibrium state which can be easily evolved by hydrodynamic wave, a lot of particles are used there by using the criterion $e^{-\Delta t/\tau}$ in the original UGKWP method. As Ref.~\cite{ugkwp4}, nearly a half number of particles should be saved due to this reason. On the other hand, the time step $\Delta t$ is a global variable, and it is restricted by the minimum mesh size. When refined meshes are used to describe the boundary layer, $e^{-\Delta t/\tau}$ will be a rather larger value, and more regions will be identified to be rarefied. Ref.~\cite{ugkwp4} studies how to introduce other criteria into the UGKWP method, and the widely used $\rm{Kn_{GLL}}$~\cite{KnGLL1,KnGLL2,KnGLL3} is implemented for significant improvement in efficiency. In this study, the scale of physical space is additionally considered, where the local mesh scale is taken as the observation scale. Further efficiency improvement is achieved by detailing the scale assessment criteria. Besides, modifications are introduced in regions of drastic scale variation, which makes the evolution of hydrodynamic wave more consistent with particles, enhancing both the stability and accuracy.

The remainder of this paper is organized as follows: the basis of gas kinetic theory is introduced in Sec.~\ref{sec:gkt}. Sec.~\ref{sec:wp} is the construction of the original UGKWP method. Adaptive criteria and modifications are introduced in Sec.~\ref{sec:awp}. Numerical test cases are conducted in Sec.~\ref{sec:cases}, and conclusions are in Sec.~\ref{sec:conclusion}.

\section{Gas kinetic theory}\label{sec:gkt}
In the gas-kinetic theory, the physical system is described by the microscopic gas distribution function $f\left(\rho,\boldsymbol{x},\boldsymbol{u},\boldsymbol{\xi},t\right)$, where $\boldsymbol{x}$ denotes the position, $t$ denotes the time, $\rho$ is the macroscopic density, $\boldsymbol{u}$ denotes the molecular velocity and internal energy is represented by the equivalent velocity $\boldsymbol{\xi}$. As to the equilibrium state, the distribution function is totally determined by macroscopic variables, known as the Maxwellian distribution function:
\begin{equation}\label{eq:maxwellian}
g = \rho{\left( {\frac{1}{{2\pi RT}}} \right)^{\frac{{D + 3}}{2}}}\exp \left( { - \frac{{\left|\boldsymbol{c}\right|^2 + \left|\boldsymbol{\xi}\right|^2}}{2RT}} \right),
\end{equation}
where $R$ is the specific gas constant, $T$ is the macroscopic temperature, $D$ is the degree of freedom about the rotational and vibrational motions of molecules, $\boldsymbol{c}=\boldsymbol{u}-\boldsymbol{U}$ is the peculiar velocity (the macroscopic velocity $\boldsymbol{U}$ is known as the mean velocity of all particles). The macroscopic conservative variables $\boldsymbol{W}=\left(\rho, \rho \boldsymbol{U}, \rho E\right)^T$(density, momentum and energy) can be directly obtained from the distribution function $f$ by the following relation:
\begin{equation}\label{eq:gkt_moment0}
\begin{aligned}
&\boldsymbol{W} = \int_{\mathbb{R}^D}\int_{\mathbb{R}^3} \boldsymbol{\Psi}f {\rm d}\boldsymbol{u} {\rm d}\boldsymbol{\xi},\\
&\boldsymbol{\Psi}=\left({ 1,\boldsymbol{u}, {1\over 2}\left|\boldsymbol{u}\right|^2+{1\over 2}\left|\boldsymbol{\xi}\right|^2 }\right).
\end{aligned}
\nonumber
\end{equation}
The integral is called the ``moment'' in the gas-kinetic theory. The stress ${\boldsymbol{P}}$ and heat flux $\boldsymbol{Q}$ can also be obtained by the following moments:
\begin{equation}\label{eq:gkt_moment1}
\begin{aligned}
& {\boldsymbol{P}} = \int_{\mathbb{R}^D}\int_{\mathbb{R}^3} \boldsymbol{cc}(f-g) {\rm d}\boldsymbol{u} {\rm d}\boldsymbol{\xi},\\
& {\boldsymbol{Q}} = \int_{\mathbb{R}^D}\int_{\mathbb{R}^3} \boldsymbol{c}\left[{ {1\over 2}\left({ \left|\boldsymbol{c}\right|^2 + \left|\boldsymbol{\xi}\right|^2 }\right)f }\right] {\rm d}\boldsymbol{u} {\rm d}\boldsymbol{\xi}.
\nonumber
\end{aligned}
\end{equation}
In this study, the Shakhov model~\cite{shakhov} is used as the kinetic model equation, which recovers viscosity $\mu$ and Prandtl number $\rm{Pr}$ in the continuum flow regime, as follows:
\begin{equation}\label{eq:shakhov}
\begin{aligned}
& \frac{\partial f}{\partial t}+\boldsymbol{u}\cdot\frac{\partial f}{\partial \boldsymbol{x}}=\frac{g^S-f}{\tau},\\
& g^S=g\left[ {1+(1-{\rm{Pr}})\frac{\boldsymbol{c}\cdot\boldsymbol{Q}}{5pRT}\left( {\frac{\left|\boldsymbol{c}\right|^2+\left|\boldsymbol{\xi}\right|^2}{RT}-5} \right)} \right],
\end{aligned}
\end{equation}
where $p=\rho RT$ is the macroscopic pressure and $\tau=\mu/p$ is the relaxation time. When ${\rm{Pr}}=1$, it goes back to Bhatnagar--Gross--Krook (BGK) model~\cite{bgk}.

Additionally, the molecular mean free path $\lambda$ is defined by the variable soft sphere (VSS) model,
\begin{equation}\label{eq:lambda}
\begin{aligned}
\lambda &= \frac{1}{\beta}\frac{\mu}{p}\sqrt{\frac{RT}{2\pi}},\\
\beta  &= \frac{{5(\alpha  + 1)(\alpha  + 2)}}{{4\alpha (5 - 2\omega )(7 - 2\omega )}},
\end{aligned}
\end{equation}
where $\alpha$ donates the molecular scattering factor and $\omega$ donates the viscosity index. In this study, the variable hard sphere (VHS) model is used to calculate viscosity, $\mu\sim T^{\omega}$, and $\alpha$ is set to be $1$. The inflow Knudsen number ($\rm{Kn_{\infty}}$) is defined as $\lambda_{\infty}/L$, where $L$ is the reference length.

\section{Unified gas-kinetic wave-particle method}\label{sec:wp}
In the general finite volume method (FVM) framework, conservations of both macroscopic variables and microscopic gas distribution functions are considered,
\begin{equation}\label{eq:conservation}
\begin{aligned}
\boldsymbol{W}^{n+1}_i &= \boldsymbol{W}^{n}_i - \frac{1}{\Omega_i}\sum\limits_{j\in \mathcal{M}\left(i\right)}\boldsymbol{F}_jS_j,\\
f^{n+1}_i &= f^{n}_i - \frac{1}{\Omega_i}\sum\limits_{j\in \mathcal{M}\left(i\right)}\int^{\Delta t}_0\boldsymbol{u}\cdot\boldsymbol{n}_jf_jS_j{\rm d}t + \int^{\Delta t}_0\mathcal{J}\left(f,f\right){\rm d}t,
\nonumber
\end{aligned}
\end{equation}
where ``$i$'' is the label of cell, $\mathcal{M}\left(i\right)$ is a gather of interfaces surrounding cell ``$i$'', $\Omega_i$ is the area or volume of cell ``$i$'', $S_j$ is the length or area of interface ``$j$'', $\boldsymbol{n}_j$ is the outer unit normal vector. In the Shakhov model equation as Eq.~\eqref{eq:shakhov}, the collision term is modeled as $\mathcal{J}\left(f,f\right)=\frac{g^S-f}{\tau}$. Macroscopic flux $\boldsymbol{F}_j$ is calculated as,
\begin{equation}\label{eq:macrof}
\boldsymbol{F}_j = \int_{\mathbb{R}^D} \int_{\mathbb{R}^3}\int^{\Delta t}_0\boldsymbol{u}\cdot\boldsymbol{n}_jf_j\boldsymbol{\Psi} {\rm d}t{\rm d}\boldsymbol{u}{\rm d}\boldsymbol{\xi},
\end{equation}
where $\Delta t$ is the time step. Setting the origin of coordinates at the center of an interface, the integral solution along the characteristic line can be derived as,
\begin{equation}\label{eq:integral}
f\left( {\boldsymbol{0},t} \right) = \frac{1}{\tau}\int^t_0 g^S\left[{ -\boldsymbol{u}\left({ t-\tilde{t} }\right),\tilde{t} }\right]e^{\frac{\tilde{t}-t}{\tau}} d\tilde{t} + e^{-t/\tau}f\left( {-\boldsymbol{u}t,0} \right).
\end{equation}
In this equation, both collision and free transport are considered about. The first term denotes the accumulation of equilibrium state. The second term denotes the transport of non-equilibrium initial state. Expanding $g^S$ and $f$ as,
\begin{equation}\label{eq:taylor}
\begin{aligned}
g^S\left({ \boldsymbol{x},t }\right)&=g^S_0+\frac{\partial g^S}{\partial \boldsymbol{x}}\cdot\boldsymbol{x}+\frac{\partial g^S}{\partial t}t,\\
f\left({ \boldsymbol{x},0 }\right)&=f_0+\frac{\partial f}{\partial \boldsymbol{x}}\cdot\boldsymbol{x},
\nonumber
\end{aligned}
\end{equation}
it can be derived from Eq.~\eqref{eq:integral} that,
\begin{equation}\label{eq:integrala1}
f\left( {\boldsymbol{0},t} \right) = \varepsilon_ag^S_0 + \varepsilon_b\frac{\partial g^S}{\partial \boldsymbol{x}}\cdot\boldsymbol{u}+\varepsilon_c\frac{\partial g^S}{\partial t}+\varepsilon_df_0 + \varepsilon_e\frac{\partial f}{\partial \boldsymbol{x}}\cdot\boldsymbol{u},
\end{equation}
where,
\begin{equation}\label{eq:integrala2}
\begin{aligned}
\varepsilon_a &= 1-e^{-t/\tau},\\
\varepsilon_b &= te^{-t/\tau}-\tau\left(1-e^{-t/\tau}\right),\\
\varepsilon_c &= t-\tau\left(1-e^{-t/\tau}\right),\\
\varepsilon_d &= e^{-t/\tau},\\
\varepsilon_e &= -te^{-t/\tau}.
\nonumber
\end{aligned}
\end{equation}
Then the integrated flux can be derived by substituting Eq.~\eqref{eq:integrala1} into Eq.~\eqref{eq:macrof},
\begin{equation}\label{eq:integralb1}
\boldsymbol{F} = \boldsymbol{F}^{eq}+\boldsymbol{F}^{fr},
\end{equation}
where,
\begin{equation}\label{eq:integralb2}
\begin{aligned}
\boldsymbol{F}^{eq} =& \int_{\mathbb{R}^D} \int_{\mathbb{R}^3}{\boldsymbol{\Psi}\left(\delta_ag^S_0+\delta_b\frac{\partial g^S}{\partial \boldsymbol{x}}\cdot\boldsymbol{u}+ \delta_c\frac{\partial g^S}{\partial t} \right)\left(\boldsymbol{u}\cdot\boldsymbol{n}\right) {\rm d}\boldsymbol{u}} {\rm d}\boldsymbol{\xi},\\
\boldsymbol{F}^{fr} =& \int_{\mathbb{R}^D} \int_{\mathbb{R}^3}{ \boldsymbol{\Psi}\left(\delta_df_0+\delta_e\frac{\partial f}{\partial \boldsymbol{x}}\cdot\boldsymbol{u} \right)} \left(\boldsymbol{u}\cdot\boldsymbol{n}\right) {\rm d}\boldsymbol{u} {\rm d}\boldsymbol{\xi},
\end{aligned}
\end{equation}
and,
\begin{equation}\label{eq:integralb3}
\begin{aligned}
\delta_a &= \Delta t-\tau\left( {1-e^{-\Delta t/\tau}} \right),\\
\delta_b &= 2\tau^2\left( {1-e^{-\Delta t/\tau}} \right)-\tau\Delta t-\tau\Delta te^{-\Delta t/\tau},\\
\delta_c &= \frac{\Delta t^2}{2}-\tau\Delta t+\tau^2\left( {1-e^{-\Delta t/\tau}} \right),\\
\delta_d &= \tau\left( {1-e^{-\Delta t/\tau}} \right),\\
\delta_e &= \tau\Delta te^{-\Delta t/\tau}-\tau^2\left( {1-e^{-\Delta t/\tau}} \right).
\end{aligned}
\end{equation}
The $\boldsymbol{F}^{eq}$ term in Eq.~\eqref{eq:integralb1} plays a major role in the continuum flow regime, where the gas evolution is constructed by deterministic hydrodynamic waves. The $\boldsymbol{F}^{fr}$ term gradually dominates the gas evolution when the flow gets rarefied. Particles are employed for this part of simulation.

\subsection{Wave evolution}\label{sec:macro}
As the first term in Eq.~\eqref{eq:integralb1}, $\boldsymbol{F}^{eq}$ can be calculated as the GKS~\cite{gks}, where the first step is to calculate $g^S_0$ and its derivatives $\frac{\partial g^S}{\partial \boldsymbol{x}}$, $\frac{\partial g^S}{\partial t}$. It is proved in Ref.~\cite{xxc} that $\frac{\partial g^S}{\partial \boldsymbol{x}}$ and $\frac{\partial g^S}{\partial t}$ can be simplified by $\frac{\partial g}{\partial \boldsymbol{x}}$ and $\frac{\partial g}{\partial t}$. And for simplicity, instead of $g^S_0$, $g_0$ along with the Prandtl fix method in the GKS~\cite{may} is used. Calculated by the sufficient collision as follows, $g_0$ denotes the equilibrium state at the interface,
\begin{equation}\label{eq:mac1}
\int_{\mathbb{R}^D} \int_{\mathbb{R}^3} \boldsymbol{\Psi} g_0 {\rm d}\boldsymbol{u} {\rm d}\boldsymbol{\xi} = \boldsymbol{W}_0 = \int_{\mathbb{R}^D} \int_{\boldsymbol{u}\cdot\boldsymbol{n}>0} \boldsymbol{\Psi} g_L {\rm d}\boldsymbol{u} {\rm d}\boldsymbol{\xi} + \int_{\mathbb{R}^D} \int_{\boldsymbol{u}\cdot\boldsymbol{n}<0} \boldsymbol{\Psi} g_R {\rm d}\boldsymbol{u} {\rm d}\boldsymbol{\xi},
\nonumber
\end{equation}
where $g_L$, $g_R$, and $g_0$ are calculated from macroscopic variables $\boldsymbol{W}_L$, $\boldsymbol{W}_R$ and $\boldsymbol{W}_0$ through Eq.~\eqref{eq:maxwellian}. $\boldsymbol{W}_L$ and $\boldsymbol{W}_R$ are left-hand and right-hand limits at the interface after reconstruction. Meanwhile, spatial derivation $\frac{\partial g}{\partial \boldsymbol{x}}$ and temporal derivation $\frac{\partial g}{\partial t}$ are derived as:
\begin{equation}\label{eq:mac2}
\begin{aligned}
\frac{\partial g}{\partial \boldsymbol{x}} &= \boldsymbol{a}g,\\
\frac{\partial g}{\partial t} &= Ag,
\nonumber
\end{aligned}
\end{equation}
where,
\begin{equation}
\begin{aligned}
\boldsymbol{a}&=\frac{1}{g}\frac{\partial g}{\partial \boldsymbol{x}}=\frac{\partial \left[ {\rm{ln}}(g) \right]}{\partial \boldsymbol{x}},\\
A&=\frac{1}{g}\frac{\partial g}{\partial t}=\frac{\partial \left[ {\rm{ln}}(g) \right]}{\partial t}.
\nonumber
\end{aligned}
\end{equation}
For the spatial derivative, $\boldsymbol{a}$ is written in the form of:
\begin{equation}
\boldsymbol{a}=\boldsymbol{a}_0+\boldsymbol{a}_1u_1+\boldsymbol{a}_2u_2+\boldsymbol{a}_3u_3+\boldsymbol{a}_4\frac{|\boldsymbol{u}|^2+|\boldsymbol{\xi}|^2}{2},
\nonumber
\end{equation}
and,
\begin{equation}
\begin{aligned}
\boldsymbol{a}_0 &= \frac{ 2\frac{\partial\rho}{\partial\boldsymbol{x}}\Lambda + \rho\left[{ -4\Lambda^2\boldsymbol{U}\cdot\frac{\partial\boldsymbol{U}}{\partial\boldsymbol{x}} + \left({3+D-2|\boldsymbol{U}|^2\Lambda}\right)\frac{\partial\Lambda}{\partial\boldsymbol{x}} }\right] }{2\rho\Lambda},\\
\boldsymbol{a}_1 &= 2\left( {\frac{\partial U_1}{\partial\boldsymbol{x}}\Lambda + U_1\frac{\partial\Lambda}{\partial\boldsymbol{x}}} \right),\\
\boldsymbol{a}_2 &= 2\left( {\frac{\partial U_2}{\partial\boldsymbol{x}}\Lambda + U_2\frac{\partial\Lambda}{\partial\boldsymbol{x}}} \right),\\
\boldsymbol{a}_3 &= 2\left( {\frac{\partial U_3}{\partial\boldsymbol{x}}\Lambda + U_3\frac{\partial\Lambda}{\partial\boldsymbol{x}}} \right),\\
\boldsymbol{a}_4 &= -2\frac{\partial\Lambda}{\partial\boldsymbol{x}},
\nonumber
\end{aligned}
\end{equation}
where $\Lambda=\frac{1}{2RT}$. When using the Prandtl fix method, $\Lambda$ derivation is modified as follows,
\begin{equation}\label{eq:mac3}
\begin{aligned}
&\frac{\partial\Lambda}{\partial\boldsymbol{x}}\Rightarrow\frac{\partial\Lambda}{\partial\boldsymbol{x}}/{\rm{Pr}}.
\nonumber
\end{aligned}
\end{equation}

On the other hand, the time derivative is calculated obeying the conservation law on the right side of BGK-type model:
\begin{equation}
\int_{\mathbb{R}^D} \int_{\mathbb{R}^3} gA \boldsymbol{\Psi} {\rm d}\boldsymbol{u} {\rm d}\boldsymbol{\xi} = -\int_{\mathbb{R}^D} \int_{\mathbb{R}^3} g \boldsymbol{a}\cdot\boldsymbol{u} \boldsymbol{\Psi} {\rm d}\boldsymbol{u} {\rm d}\boldsymbol{\xi} = \boldsymbol{b}.
\nonumber
\end{equation}
Spreading $A$ into $A=A_0+A_1u_1+A_2u_2+A_3u_3+A_4\frac{|\boldsymbol{u}|^2+|\boldsymbol{\xi}|^2}{2}$, it is rewritten as:
\begin{equation}
A_{\beta} M_{\alpha\beta} = b_{\alpha},
\nonumber
\end{equation}
where subscripts $\alpha$ and $\beta$ here denote components, and Einstein summation convention is used. For example, as to the two-dimensional case,
\begin{equation}
\boldsymbol{M}=
\left(
\begin{array}{cccc}
1    &U_1                       &U_2                      &B_1 \\
U_1  &U_1^2+\frac{1}{2\Lambda}  &U_1U_2                   &B_2 \\
U_2  &U_1U_2                    &U_2^2+\frac{1}{2\Lambda} &B_3 \\
B_1  &B_2                       &B_3                      &B_4
\end{array} \right),
\nonumber
\end{equation}
where
\begin{equation}
\begin{aligned}
&B_1 = \frac{1}{2}\left( {U_1^2+U_2^2+\frac{D+3}{2\Lambda}} \right),\\
&B_2 = \frac{1}{2}\left( {U_1^3+U_1U_2^2+\frac{D+5}{2\Lambda}U} \right),\\
&B_3 = \frac{1}{2}\left( {U_2^3+U_1^2V+\frac{D+5}{2\Lambda}V} \right),\\
&B_4 = \frac{1}{4}\left[ {\left({U_1^2+U_2^2}\right)^2 + \frac{D+5}{\Lambda}\left({U_1^2+U_2^2}\right) + \frac{D^2+8D+15}{4\Lambda^2}} \right].
\nonumber
\end{aligned}
\end{equation}
And the result is:
\begin{equation}
\begin{aligned}
&A_4 = \frac{8\Lambda^2}{D+3}\left[ {b_4-U_1b_2-U_2b_3-(B_1+U_1^2+U_2^2)b_1} \right],\\
&A_3 = 2\Lambda(b_3-U_2b_1)-U_2A_4,\\
&A_2 = 2\Lambda(b_2-U_1b_1)-U_1A_4,\\
&A_1 = b_1-U_1A_2-U_2A_3-B_1A_4.
\nonumber
\end{aligned}
\end{equation}

Additionally, details of integrations are provided for calculating $\boldsymbol{F}^{eq}$ in Eq.~\eqref{eq:integralb2}, as follows,
\begin{equation}
\begin{aligned}
&\int_{\mathbb{R}} {\left( {\frac{1}{{2\pi RT}}} \right)^{\frac{{1}}{2}}}\exp \left( { - \frac{|u-U|^2}{{2RT}}} \right) du = 1,\\
&\int_{\mathbb{R}} u {\left( {\frac{1}{{2\pi RT}}} \right)^{\frac{{1}}{2}}}\exp \left( { - \frac{|u-U|^2}{{2RT}}} \right) du = U,\\
&\int_{u>0} {\left( {\frac{1}{{2\pi RT}}} \right)^{\frac{{1}}{2}}}\exp \left( { - \frac{|u-U|^2}{{2RT}}} \right) du = \frac{1}{2}\rm{erfc}\left( {-\sqrt{\Lambda}U} \right),\\
&\int_{u>0} u {\left( {\frac{1}{{2\pi RT}}} \right)^{\frac{{1}}{2}}}\exp \left( { - \frac{|u-U|^2}{{2RT}}} \right) du = \frac{U}{2}\rm{erfc}\left( {-\sqrt{\Lambda}U} \right) + \frac{1}{2}\frac{e^{-\Lambda U^2}}{\sqrt{\pi \Lambda}},\\
&\int_{u<0} {\left( {\frac{1}{{2\pi RT}}} \right)^{\frac{{1}}{2}}}\exp \left( { - \frac{|u-U|^2}{{2RT}}} \right) du = \frac{1}{2}\rm{erfc}\left( {\sqrt{\Lambda}U} \right),\\
&\int_{u<0} u {\left( {\frac{1}{{2\pi RT}}} \right)^{\frac{{1}}{2}}}\exp \left( { - \frac{|u-U|^2}{{2RT}}} \right) du = \frac{U}{2}\rm{erfc}\left( {\sqrt{\Lambda}U} \right) - \frac{1}{2}\frac{e^{-\Lambda U^2}}{\sqrt{\pi \Lambda}},
\nonumber
\end{aligned}
\end{equation}
and,
\begin{equation}
\begin{aligned}
&\int_{\mathbb{R}/u>0/u<0} u^{n+2} {\left( {\frac{1}{{2\pi RT}}} \right)^{\frac{{1}}{2}}}\exp \left( { - \frac{|u-U|^2}{{2RT}}} \right) du \\
= &U\int_{\mathbb{R}/u>0/u<0} u^{n+1} {\left( {\frac{1}{{2\pi RT}}} \right)^{\frac{{1}}{2}}}\exp \left( { - \frac{|u-U|^2}{{2RT}}} \right) du \\ + &\frac{n+1}{2\Lambda}\int_{\mathbb{R}/u>0/u<0} u^{n} {\left( {\frac{1}{{2\pi RT}}} \right)^{\frac{{1}}{2}}}\exp \left( { - \frac{|u-U|^2}{{2RT}}} \right) du,
\nonumber
\end{aligned}
\end{equation}
where $\rm{erfc}()=1-\rm{erf}()$ is the complementary error function.

\subsection{Particle evolution}\label{sec:micro}
In this method, numerical particles are employed to describe the free-transport part of distribution function, whose parameters are: particle mass $m$, location $\boldsymbol{x}$, velocity $\boldsymbol{u}$ and specific internal thermal energy $e=0.5\left|\boldsymbol{\xi}\right|^2$. As Eq.~\eqref{eq:integral}, the distribution function after $\Delta t$ is a combination of the initial state $f_0$ and equilibrium state $g^S$. Firstly, for a particle from the initial state, the probability for keeping free-transport is $e^{-\frac{\Delta t}{\tau}}$, otherwise it collides with other particles and is relaxed into $g^S$. As a result, the cumulative distribution of a free-transport particle is $e^{-\frac{\Delta t}{\tau}}$. By taking a random number $\epsilon$ uniformly distributing in $\left(0,1\right)$, the free-transport time $t_{f,k}$ of a particle labeled by ``$k$'' can be calculated by,
\begin{equation}\label{eq:mic1}
t_{f,k} = {\rm{min}}\left(-\tau_k{\rm{ln}}\left(\epsilon\right),\Delta t\right).
\end{equation}
Then the location of the particle can be renewed by,
\begin{equation}\label{eq:mic2}
\boldsymbol{x}_k \Rightarrow \boldsymbol{x}_k + \boldsymbol{u}_kt_{f,k},
\end{equation}
and the contribution of all particles to the macroscopic conserved variables is,
\begin{equation}\label{eq:mic3}
\boldsymbol{W}_i^{fr,p} = \boldsymbol{W}_i^{p,n+1}-\boldsymbol{W}_i^{p,n},
\end{equation}
and,
\begin{equation}\label{eq:mic4}
\boldsymbol{W}_i^{p,n} = \frac{1}{\Omega_i}\sum\limits_{k\in \mathcal{N}^{n}\left(i\right)}m_{p,k}\boldsymbol{\Psi}_k,
\nonumber
\end{equation}
where $\mathcal{N}\left(i\right)$ is a gather of particles within cell ``$i$''. Besides, after the free transport, if $t_{f,k}<\Delta t$, the particle is relaxed into $g^S$, called collisional particles. Otherwise, if $t_{f,k}=\Delta t$, the particle will be kept, called collisionless particles.

Secondly, according to Eq.~\eqref{eq:integral}, $e^{-\frac{\Delta t}{\tau_i}}$ proportion of particles should be sampled from the hydrodynamic wave part, $\boldsymbol{W}_i^{h}=\boldsymbol{W}_i-\boldsymbol{W}_i^{p}$ with $t_{f,k}=\Delta t$.  The total mass density of the sampled collisionless particle is,
\begin{equation}\label{eq:mic5}
\rho_i^{hp} = e^{-\frac{\Delta t}{\tau_i}}\rho_i^{h}.
\end{equation}
Unless $\rho_i^{h}=0$ when no particle will be sampled, the number of sampled particles is calculated by,
\begin{equation}\label{eq:mic6}
N^{hp}_i = \lceil{\frac{\rho^{hp}_i}{\rho_i}N^{hp,{\rm{ref}}}}\rceil,
\end{equation}
where $N^{hp,{\rm{ref}}}$ is a given reference number, which is set to be $100$ in all cases of this paper. The corresponding mass of a particle is,
\begin{equation}\label{eq:mic7}
m_k = \frac{\rho_i^{hp}\Omega_i}{N^{hp}_i}.
\nonumber
\end{equation}
The particle position is set according to a uniform distribution probability within cell ``$i$''. The velocity is got by acception-rejection sampling, as follows. Firstly, the velocity according with the Maxwellian distribution function is calculated by,
\begin{equation}\label{eq:maxw}
\begin{aligned}
u_{k,1} &= U_{i,1} + \sqrt{2RT_i}{\rm{cos}}(2\pi \epsilon_{a1})\sqrt{-{\rm{ln}}(\epsilon_{a2})},\\
u_{k,2} &= U_{i,2} + \sqrt{2RT_i}{\rm{cos}}(2\pi \epsilon_{b1})\sqrt{-{\rm{ln}}(\epsilon_{b2})},\\
u_{k,3} &= U_{i,3} + \sqrt{2RT_i}{\rm{cos}}(2\pi \epsilon_{c1})\sqrt{-{\rm{ln}}(\epsilon_{c2})},
\nonumber
\end{aligned}
\end{equation}
where $\epsilon$ is a random number distributing uniformly in $\left(0,1\right)$. Aiming at the internal thermal energy, two reduced functions of the Shakhov model are derived as,
\begin{equation}\label{eq:thermal}
\begin{aligned}
\int_{\mathbb{R}^D} g^S {\rm d}\boldsymbol{\xi} &= \left[1+\left(1-{\rm{Pr}}\right)\frac{\boldsymbol{c}\cdot\boldsymbol{Q}}{5pRT}\left(\frac{\left|\boldsymbol{c}\right|^2}{RT}-5\right)\right]\rho\left(\frac{1}{2\pi RT}\right)^{\frac{3}{2}}e^{-\frac{\left|\boldsymbol{c}\right|^2}{2RT}},\\
\int_{\mathbb{R}^D} \left|\boldsymbol{\xi}\right|^2 g^S {\rm d}\boldsymbol{\xi} &= DRT\left[1+\left(1-{\rm{Pr}}\right)\frac{\boldsymbol{c}\cdot\boldsymbol{Q}}{5pRT}\left(\frac{\left|\boldsymbol{c}\right|^2}{RT}-5\right)\right]\rho\left(\frac{1}{2\pi RT}\right)^{\frac{3}{2}}e^{-\frac{\left|\boldsymbol{c}\right|^2}{2RT}}.
\nonumber
\end{aligned}
\end{equation}
As a result, the criterion of the acceptance-rejection method is suggested to be:
\begin{equation}\label{eq:ajshak}
\frac{1+(1-{\rm{Pr}})\frac{\boldsymbol{c}_k\cdot\boldsymbol{Q}_i}{5p_iRT_i}\left( {\frac{|\boldsymbol{c}_k|^2}{RT_i}-5} \right)}{1+(1-{\rm{Pr}})\frac{20|\boldsymbol{Q}_i|}{p_i\sqrt{RT_i}}},
\nonumber
\end{equation}
and its specific internal thermal energy is,
\begin{equation}\label{eq:aje}
e_k=0.5DRT_i.
\end{equation}

Additionally, since the analytical macroscopic flux of newly sampled particles corresponds to the second-order DVM, the free transport fluxes contributed from the collisional particles of $\left(\boldsymbol{W}^h-\boldsymbol{W}^{hp}\right)$ can be calculated as,
\begin{equation}\label{eq:ffrwave}
\begin{aligned}
&\boldsymbol{F}^{fr,wave}=\boldsymbol{F}^{fr}_{\rm{UGKS}}\left(\boldsymbol{W}^h\right)-\boldsymbol{F}^{fr}_{\rm{DVM}}\left(\boldsymbol{W}^{hp}\right)\\
=&\int_{\mathbb{R}^D}\int_{\mathbb{R}^3}\boldsymbol{\Psi}\left(\delta_dg_0^h+\delta_e\frac{\partial g^h}{\partial \boldsymbol{x}}\cdot\boldsymbol{u}\right) \left(\boldsymbol{u}\cdot\boldsymbol{n}\right) {\rm d}\boldsymbol{u} {\rm d}\boldsymbol{\xi}\\
-&e^{\frac{\Delta t}{\tau}}\int_{0}^{\Delta t}\int_{\mathbb{R}^D}\int_{\mathbb{R}^3}\boldsymbol{\Psi}\left(g_0^h+t\frac{\partial g^h}{\partial \boldsymbol{x}}\cdot\boldsymbol{u}\right) \left(\boldsymbol{u}\cdot\boldsymbol{n}\right) {\rm d}\boldsymbol{u} {\rm d}\boldsymbol{\xi}{\rm d}t\\
=&\int_{\mathbb{R}^D}\int_{\mathbb{R}^3}\boldsymbol{\Psi}\left[\left(\delta_d-\Delta te^{\frac{\Delta t}{\tau}}\right)g_0^h+\left(\delta_e+\frac{\Delta t^2}{2}e^{\frac{\Delta t}{\tau}}\right)\frac{\partial g^h}{\partial \boldsymbol{x}}\cdot\boldsymbol{u}\right] \left(\boldsymbol{u}\cdot\boldsymbol{n}\right) {\rm d}\boldsymbol{u} {\rm d}\boldsymbol{\xi}.
\end{aligned}
\end{equation}
So finally, the macroscopic update equation for the UGKWP method is,
\begin{equation}\label{eq:renew}
\boldsymbol{W}^{n+1}_i = \boldsymbol{W}^{n}_i - \frac{1}{\Omega_i}\sum\limits_{j\in \mathcal{M}\left(i\right)}\boldsymbol{F}_j^{eq}S_j - \frac{1}{\Omega_i}\sum\limits_{j\in \mathcal{M}\left(i\right)}\boldsymbol{F}_j^{fr,wave}S_j+\boldsymbol{W}_i^{fr,p}.
\end{equation}

\subsection{Algorithm of UGKWP}\label{sec:sum1}
\begin{figure}[H]
	\centering
	\subfigure[]{\label{fig1a}
			\includegraphics[width=0.22 \textwidth]{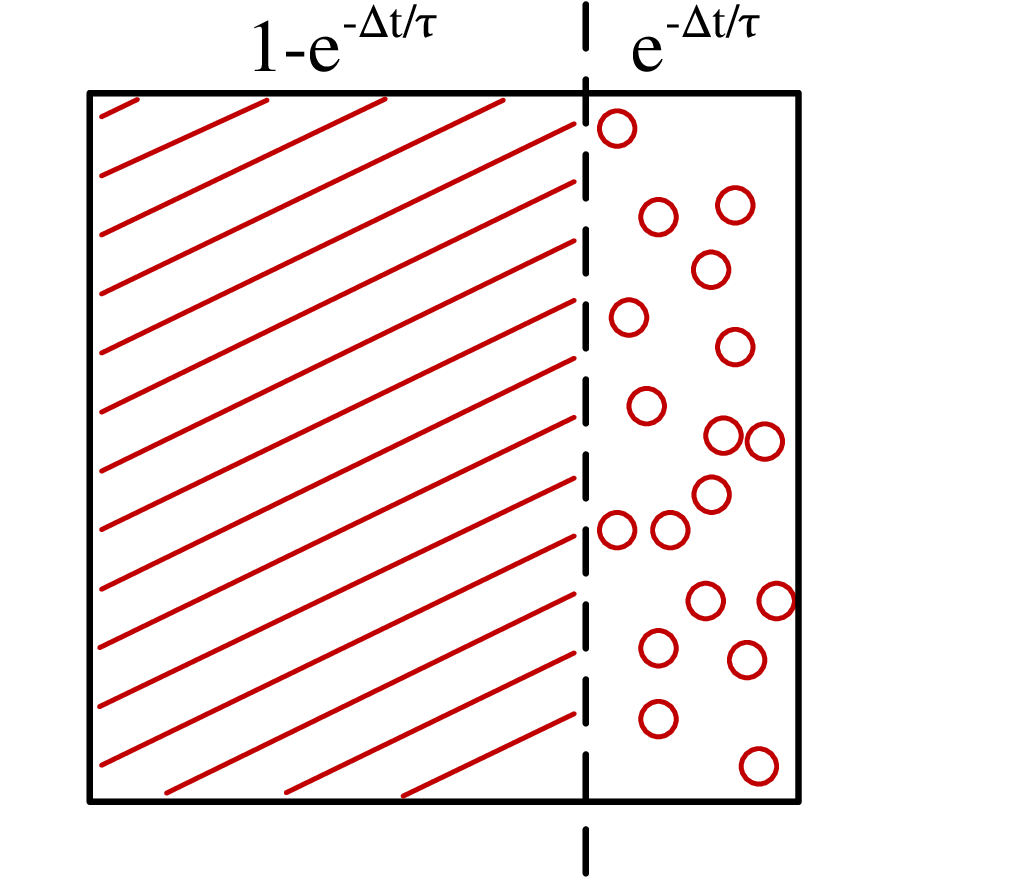}
		}
    \subfigure[]{\label{fig1b}
    		\includegraphics[width=0.22 \textwidth]{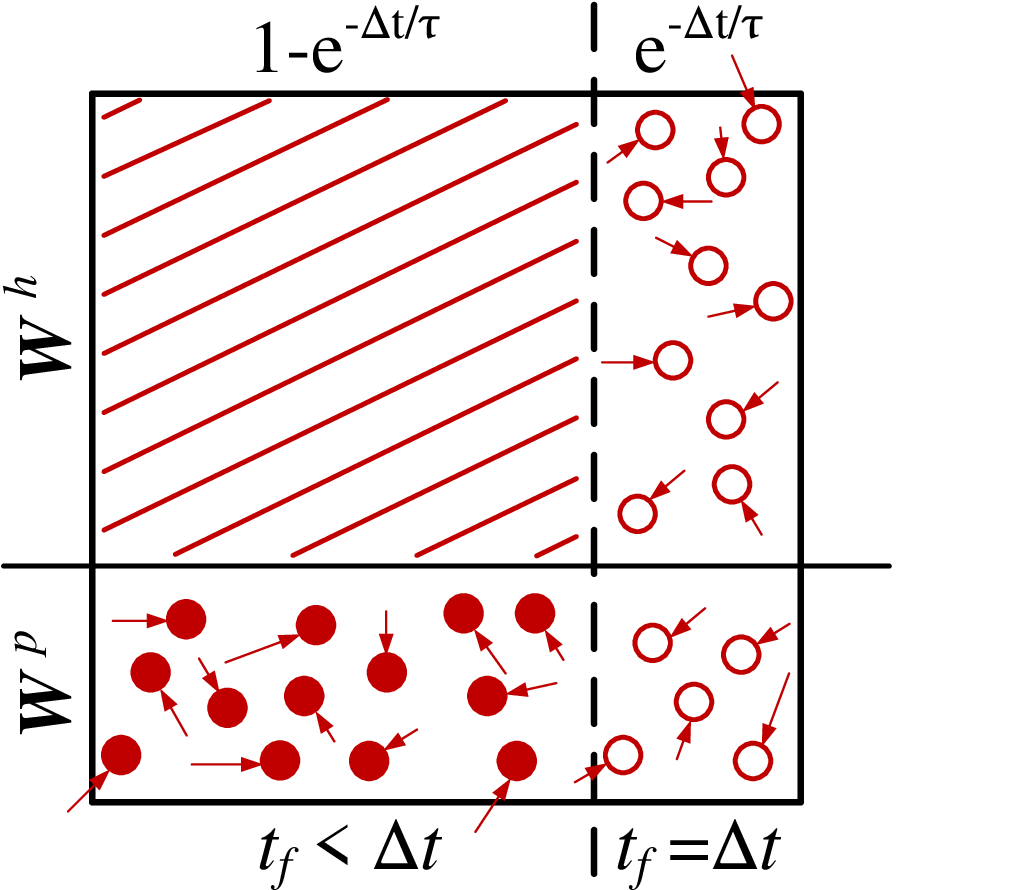}
    	}
    \subfigure[]{\label{fig1c}
    		\includegraphics[width=0.22 \textwidth]{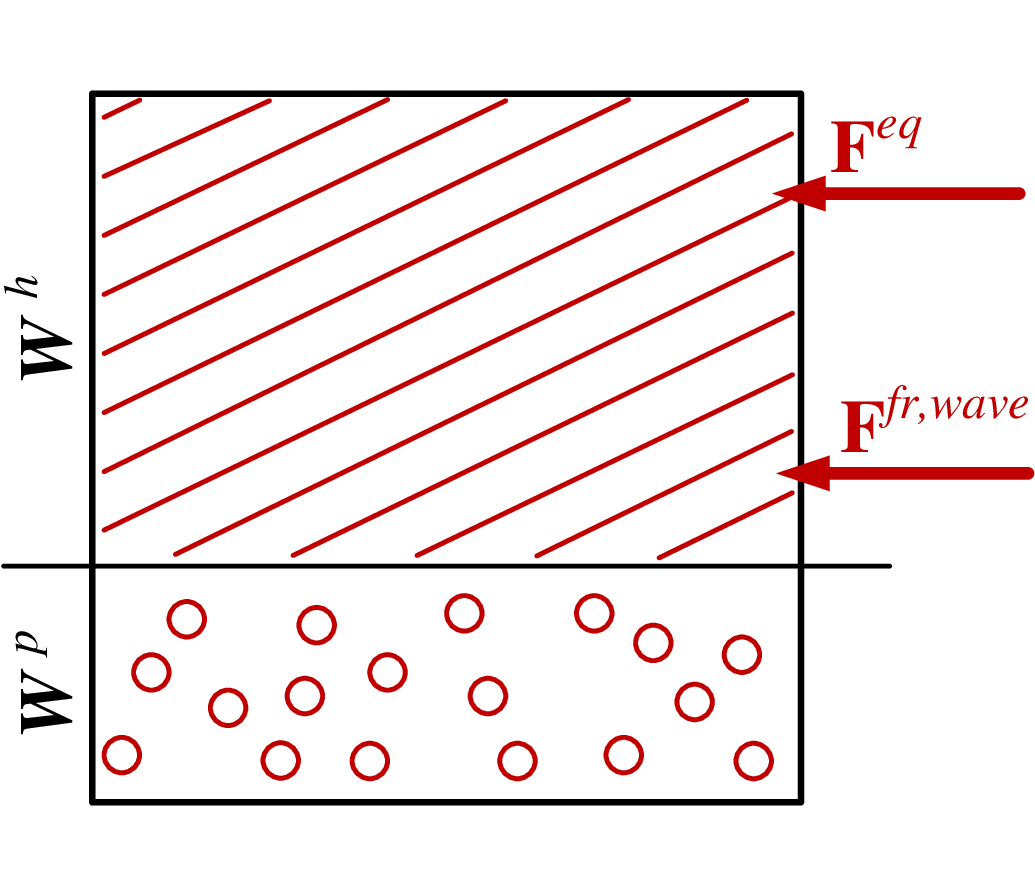}
    	}
    \subfigure[]{\label{fig1d}
    		\includegraphics[width=0.22 \textwidth]{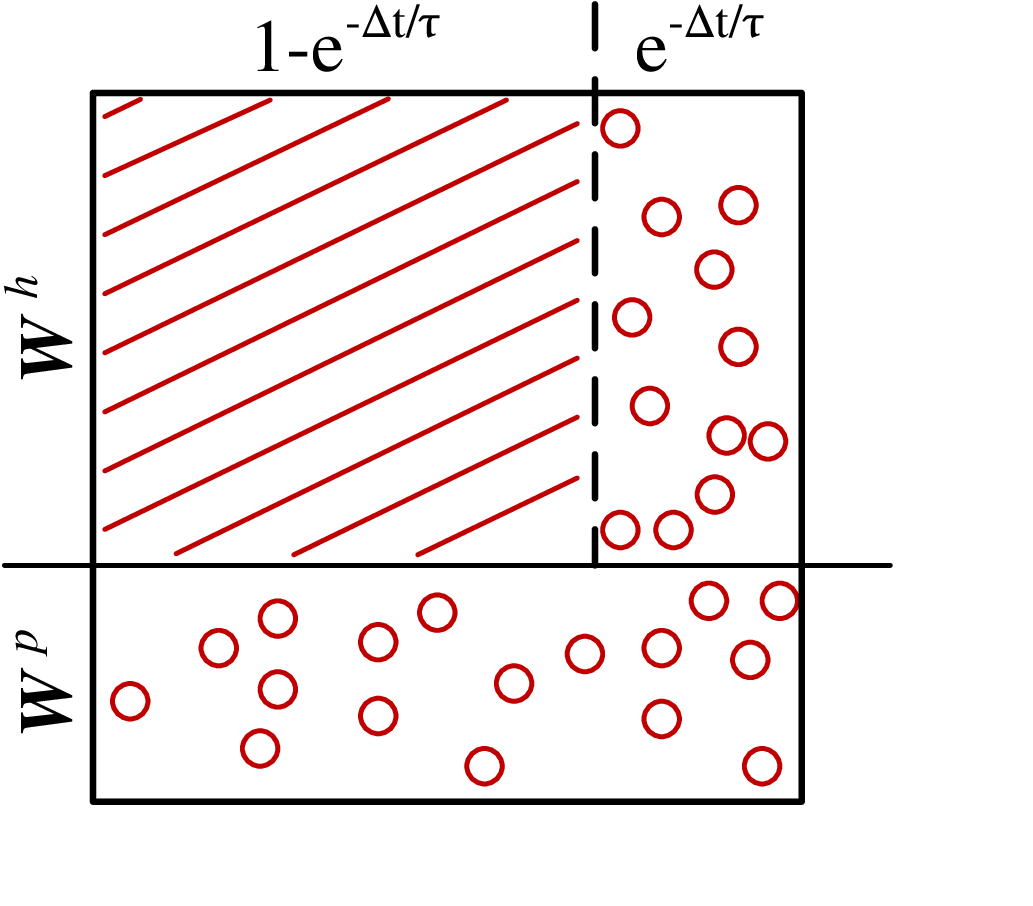}
    	}
	\caption{\label{fig1} Diagram to illustrate the algorithm of UGKWP method: (a) Initial state at $n=0$ step, (b) step $2$, (c) step $3$, (d) initial state at other steps.}
\end{figure}

Regarding as Fig.~\ref{fig1}, the following part is a summary of the algorithm of UGKWP method.
\begin{description}
    \item[Step (1)] Initial state. As the result of the previous time step $n-1$ in Fig.~\ref{fig1d}, numerical particles $\boldsymbol{W}^p$ coexist with hydrodynamic wave $\boldsymbol{W}^h$. Then collisionless particles $\boldsymbol{W}^{hp}$ are sampled as Eq.~\eqref{eq:mic5}, with detailed parameters from Eq.~\eqref{eq:mic6} to Eq.~\eqref{eq:aje}. For the first step, $\boldsymbol{W}^p=\boldsymbol{0}$, as shown in Fig.~\ref{fig1a}.
    \item[Step (2)] Free transport. Firstly, divide particles $\boldsymbol{W}^p$ into two parts as Eq.~\eqref{eq:mic1}. As Fig.~\ref{fig1b}, collisionless particles are denoted by hollow circles, and collisional particles are denoted by solid circles. Then transport all particles as Eq.~\eqref{eq:mic2}, and sum their contribution to the macroscopic conserved variables as Eq.~\eqref{eq:mic3}. Meanwhile, the free transport fluxes $\boldsymbol{F}_j^{fr,wave}$ contributed from the collisional particles of $\left(\boldsymbol{W}^h-\boldsymbol{W}^{hp}\right)$ is also calculated as Eq.~\eqref{eq:ffrwave}. Finally, delete the collisional particles which are denoted by solid circles.
    \item[Step (3)] Collision. Compute the $\boldsymbol{F}^{eq}_j$ term in Eq.~\eqref{eq:integralb1} by equations in Sec.~\ref{sec:macro}. Then renew $\boldsymbol{W}^{n+1}$ as Eq.~\eqref{eq:renew}.
    \item[Step (4)] If the simulation continues, go to Step $\left(1\right)$, where collisionless particles $\boldsymbol{W}^{hp}$ will be sampled.
\end{description}

\section{Adaptive criteria and modifications}\label{sec:awp}
\begin{figure}[H]
	\centering
	\includegraphics[width=0.7 \textwidth]{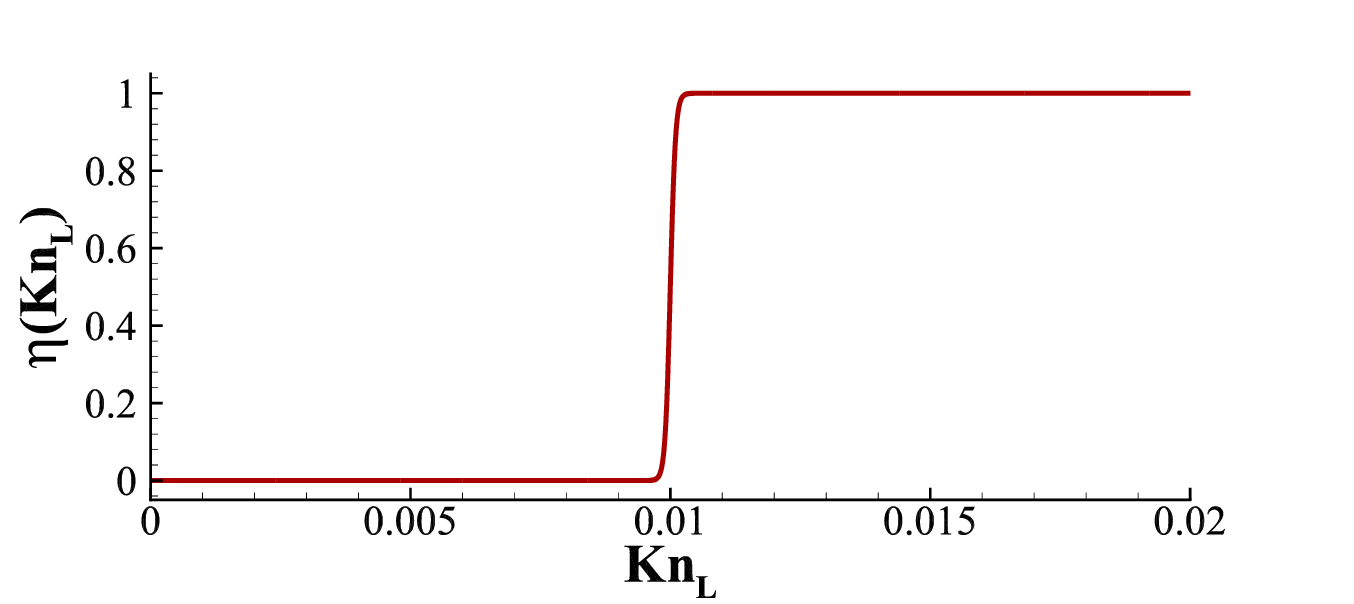}
	\caption{\label{fig2} Diagram to illustrate the function $\eta\left({\rm{Kn_L}}\right)$.}
\end{figure}

For promoting the efficiency, the adaptive UGKWP (AUGKWP) method is developed in Ref.~\cite{ugkwp4}. In the original UGKWP method, the local scale is identified naturally from the view of time. Specifically, in the integration solution Eq.~\eqref{eq:integral}, the weights of hydrodynamic wave and non-equilibrium particle are controlled by the ratio between global time step $\Delta t$ and local relaxation time $\tau$. While only considering about the time is not sufficient to identify the near-continuum region. Apart from $\Delta t \gg \tau$, there are other regions where the gas is at the equilibrium state. For example, even at large $\rm{Kn_{\infty}}$, the free inflow is at the equilibrium state. As a result, a parameter $\eta\left({\rm{Kn_L}}\right)\in\left(0,1\right)$ is used in the AUGKWP method to measure the local equilibrium degree, as follows,
\begin{equation}\label{eq:yita}
\eta\left({\rm{Kn_L}}\right)=\frac{1}{2}\left[{\rm{tanh}}\left(\frac{{\rm{Kn_L}}/{\rm{Kn_{ref}}}-1}{{\rm{Kn_{ref}}}}\right)+1\right],
\end{equation}
where $\rm{Kn_L}$ is the local $\rm{Kn}$ number, whose detailed expression is to be introduced in Sec.~\ref{sec:kngll}. As Fig.~\ref{fig2}, when the flow becomes rarefied, $\eta$ tends to be $1$, and when the flow is in the local continuum regime, $\eta$ tends to be $0$. A critical value set as ${\rm{Kn_{ref}}}=0.01$ is suggested in Ref.~\cite{ugkwp4} and Ref.~\cite{ugks3}. The method is simply modified by exchanging Eq.~\eqref{eq:mic5} and Eq.~\eqref{eq:ffrwave} into,
\begin{equation}\label{eq:mic5b}
\rho_i^{hp} = \eta_i e^{-\frac{\Delta t}{\tau_i}}\rho_i^{h},
\end{equation}
and,
\begin{equation}\label{eq:ffrwaveb}
\boldsymbol{F}^{fr,wave}=\int_{\mathbb{R}^D}\int_{\mathbb{R}^3}\boldsymbol{\Psi}\left[\left(\delta_d-\eta\Delta te^{\frac{\Delta t}{\tau}}\right)g_0^h+\left(\delta_e+\eta\frac{\Delta t^2}{2}e^{\frac{\Delta t}{\tau}}\right)\frac{\partial g^h}{\partial \boldsymbol{x}}\cdot\boldsymbol{u}\right] \left(\boldsymbol{u}\cdot\boldsymbol{n}\right) {\rm d}\boldsymbol{u} {\rm d}\boldsymbol{\xi}.
\end{equation}
Multiplied by $\eta_i$, particles are restrained from being sampled in Eq.~\eqref{eq:mic5b}, so that their evolution follows the hydrodynamic wave way, as $g^h$ in Eq.~\eqref{eq:ffrwaveb}. Thus the computational consumption of these particles can be reduced. In the $\eta=0$ region, the method returns to the GKS. In the remainder of this section, it is discussed how to get a reasonable $\rm{Kn_L}$, and modifications are introduced to deal with the intensely changing region, mainly affected by $\eta$.

\subsection{$\rm{Kn_{\rm{GLL}}}$, $\rm{Kn_{\lambda}}$ and exponential moving average}\label{sec:kngll}
In this work, $\rm{Kn_L}$ is calculated by following equations,
\begin{equation}\label{eq:knl}
\begin{aligned}
{\rm{Kn_L}} &= {\rm{min}}\left({\rm{Kn_{GLL}}},{\rm{Kn_{\lambda}}}\right), \\
{\rm{Kn_{GLL}}} &= \frac{\lambda}{\rho/\left|\nabla\rho\right|}, \\
{\rm{Kn_{\lambda}}} &= \frac{\lambda}{L_{\lambda,{\rm{ref}}}}.
\end{aligned}
\end{equation}
As in Ref.~\cite{ugkwp4,KnGLL1,KnGLL2,KnGLL3}, ${\rm{Kn_{GLL}}}$ is widely used to measure the non-equilibrium degree, which mainly has advantages in the equilibrium state of the free inflow. The gradient of density is used to find the intensely changing region, and the mean free path is also taken into account. However, ${\rm{Kn_{GLL}}}$ is not enough to identify near-continuum regions at most. For example, $\rho/\left|\nabla\rho\right|$ is too large when density rises up rapidly around the stagnation point and post-shock. Taking the stagnation point as an example, density there is always the maximum in the domain, which means it is always the most near-continuum region. But even in a very low $\rm{Kn_{\infty}}$ case, $\rm{Kn_{GLL}}$ is large there. As a result, $\rm{Kn_{\lambda}}$ is introduced for improvement. Reference length $L_{\lambda,{\rm{ref}},i}$ in cell ``$i$'' is calculated in terms of its mesh size, as follows,
\begin{equation}
\begin{aligned}
L_{\lambda,{\rm{ref}},i}=\left\{ \begin{array}{l}
C\sqrt{\Omega_i},\quad {\rm{for \ 2-D \ case}},\\
C\sqrt[3]{\Omega_i},\quad {\rm{for \ 3-D \ case}}.
\end{array}\right.
\end{aligned}
\end{equation}
Together with $L_{\lambda,{\rm{ref}}}$, local mean free path $\lambda$ is taken to measure the local rarefied degree. The free parameter is set to be $C=20$. $\rm{Kn_{\lambda}}$ can clearly identify the regions around the stagnation point or post-shock whether at the equilibrium state. At the stagnation point, rather refined meshes are needed to predict the heat flux accurately, especially at low $\rm{Kn_{\infty}}$. If this region is misjudged to be rarefied, a huge number of particles will be used, and the noise there makes it a hard task to develop corresponding implicit schemes.

As mentioned in Ref.~\cite{principle}, ``A multiscale scheme should recover the kinetic Boltzmann equation when the mesh size is on the order of the mean free path and recover the hydrodynamic equations when the mesh size is much larger than the mean free path. Numerical multiscale modeling is more complicated than the theoretical fluid dynamics.'' In the numerical multiscale modeling, the mesh size is taken as an observation size to identify the local scale. Besides the numerical scheme, mesh generation should also be considered to be an early procedure to determine the local scale of different regions. For example, even though shock wave is a typical non-equilibrium phenomenon, in the low $\rm{Kn_{\infty}}$ case, the under-resolved mesh is used and the scheme should recover the hydrodynamic equations there. More discussion about numerical multiscale modeling can be referred to Ref.~\cite{principle}, and in this study, $\rm{Kn_{\lambda}}$ is used to consider about the mesh size.

Furthermore, to get rid of the adverse effects of particle noise, the exponential moving average (EMA) method is utilized like Ref.~\cite{ema}. Firstly,
\begin{equation}
\rho^{{\rm{EMA}},n+1}=\left(1-\frac{1}{N_{{\rm{EMA}}}}\right)\rho^{{\rm{EMA}},n}+\frac{1}{N_{{\rm{EMA}}}}\rho^{n+1},
\nonumber
\end{equation}
is used to calculate $\rm{Kn_{GLL}}$ in Eq.~\eqref{eq:knl}, along with its gradient. The parameter $N_{{\rm{EMA}}}$ is set to be $100$ in this work. The EMA method can be regarded as the relaxation of variables, which reduces the noise from stochastic particles by a great extent. Different from the time average method, the EMA variables and their gradients evolve together with the flow, which brings about little extra computation consumption. Without the reduction of noise by EMA method, $\rm{Kn_{GLL}}$ will be much larger than the averaged value, which even introduces particles in the uniform flow at low $\rm{Kn_{\infty}}$. The other position needing EMA method is the calculation of $\eta$ in Eq.~\eqref{eq:yita}, as follows. Otherwise, some cells will face the problem that $\eta\rightarrow 1$ in step $n$, but $\eta\rightarrow 0$ in step $n+1$, and it happens repeatedly, which makes the simulation to blow up very easily.
\begin{equation}\label{eq:ema}
\eta^{{\rm{EMA}},n+1}=\left(1-\frac{1}{N_{\rm{EMA}}}\right)\eta^{{\rm{EMA}},n}+\frac{1}{N_{{\rm{EMA}}}}\eta^{n+1},
\nonumber
\end{equation}

\subsection{Modification of $\boldsymbol{F}^{eq}$ and $\boldsymbol{F}^{fr,wave}$ terms in the flow field regions of drastic scale variation}\label{sec:fw}
\begin{figure}[H]
	\centering
	\includegraphics[width=0.65 \textwidth]{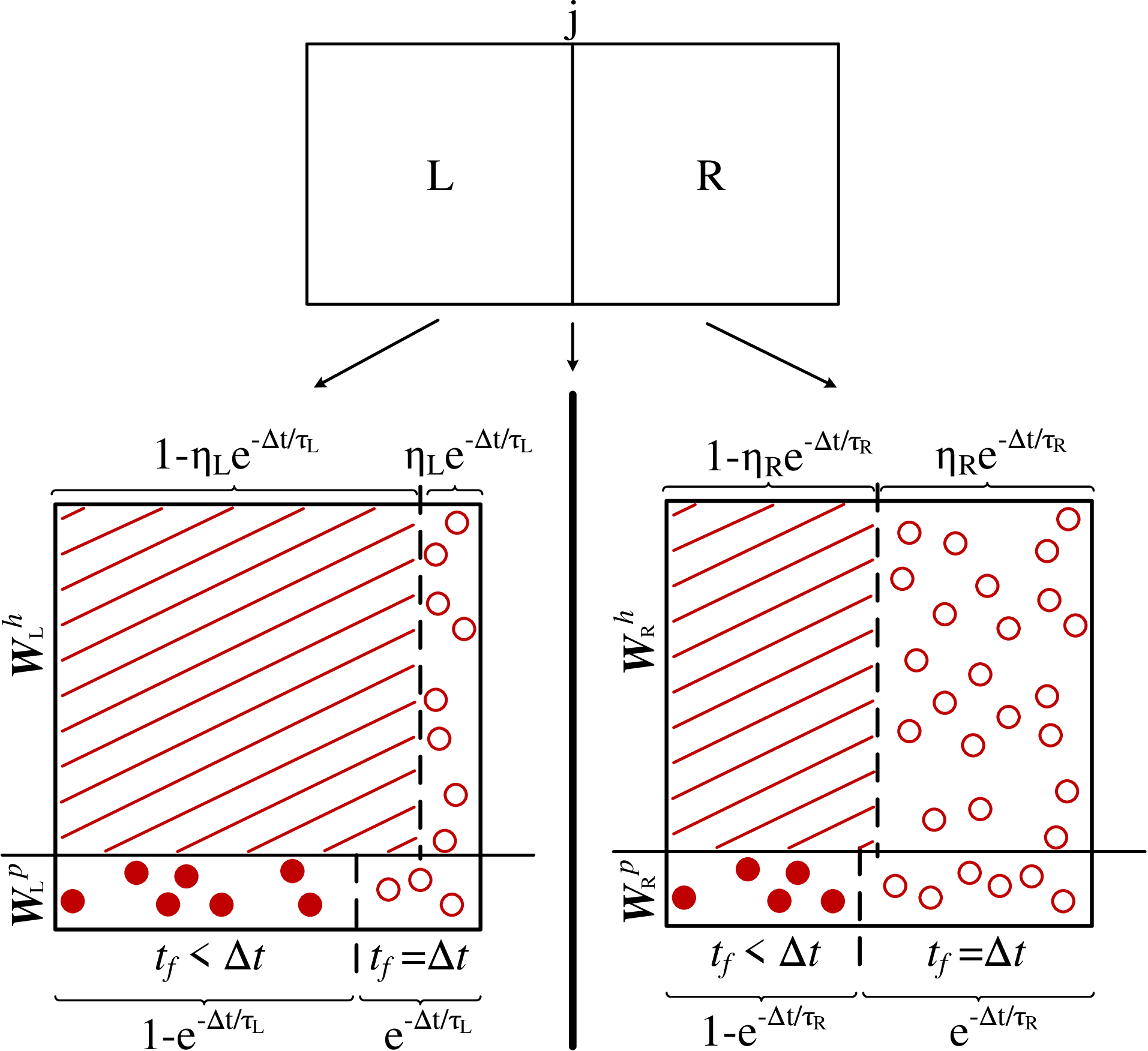}
	\caption{\label{fig3} Diagram to illustrate the intense change of $\eta$ and $\tau$ in two adjacent cells.}
\end{figure}

In the AUGKWP method, $\eta$ may change intensely in different cells, and there may be significant increase of $\tau$ around the stagnation point and shock as well. As Fig.~\ref{fig3}, according to Fig.~\ref{fig1b}, two adjacent cells are taken for an example, labeled by ``$L$'' and ``$R$''. The interface between them is labeled by ``$j$''. Cell ``$L$'' is closer to the equilibrium state, with $\tau_L < \tau_R$ and $\eta_L < \eta_R$.

When calculating $t_f$ as Eq.~\eqref{eq:mic1} and sampling particles as Eq.~\eqref{eq:mic5b}, the location of used $\tau$ and $\eta$ is within the cell, $\tau_{L/R}$ and $\eta_{L/R}$. Corresponding to the integrated flux in Eq.~\eqref{eq:integralb1}, the $2$nd-accuracy analytical expression of particles contribution to the macroscopic flux is derived as,
\begin{equation}
\begin{aligned}
\boldsymbol{F}^{p}_j=&\int_{\mathbb{R}^D}\int_{\boldsymbol{u}\cdot\boldsymbol{n}_j>0}\boldsymbol{\Psi}\left[\eta_L\Delta te^{\frac{\Delta t}{\tau_L}}g_L^h-\eta_L\frac{\Delta t^2}{2}e^{\frac{\Delta t}{\tau_L}}\frac{\partial g^h}{\partial \boldsymbol{x}}|_L\cdot\boldsymbol{u}+\delta_{d,L}f_L^p+\delta_{e,L}\frac{\partial f^p}{\partial \boldsymbol{x}}|_L\cdot\boldsymbol{u}\right] \left(\boldsymbol{u}\cdot\boldsymbol{n}_j\right) {\rm d}\boldsymbol{u} {\rm d}\boldsymbol{\xi} \\
+&\int_{\mathbb{R}^D}\int_{\boldsymbol{u}\cdot\boldsymbol{n}_j<0}\boldsymbol{\Psi}\left[\eta_R\Delta te^{\frac{\Delta t}{\tau_R}}g_R^h-\eta_R\frac{\Delta t^2}{2}e^{\frac{\Delta t}{\tau_R}}\frac{\partial g^h}{\partial \boldsymbol{x}}|_R\cdot\boldsymbol{u}+\delta_{d,R}f_R^p+\delta_{e,R}\frac{\partial f^p}{\partial \boldsymbol{x}}|_R\cdot\boldsymbol{u}\right] \left(\boldsymbol{u}\cdot\boldsymbol{n}_j\right) {\rm d}\boldsymbol{u} {\rm d}\boldsymbol{\xi},
\end{aligned}
\end{equation}
and,
\begin{equation}
\boldsymbol{W}_i^{fr,p} = -\frac{1}{\Omega_i}\sum\limits_{j\in \mathcal{M}\left(i\right)}\boldsymbol{F}_j^{p}S_j.
\end{equation}
Nevertheless, when calculating $\boldsymbol{F}^{eq}$ and $\boldsymbol{F}^{fr,wave}$ as Eq.~\eqref{eq:integralb2} and Eq.~\eqref{eq:ffrwaveb}, $\tau_j$ and $\eta_j$ at the interface are used. For example, in the case as Fig.~\ref{fig3}, $\tau_L < \tau_j < \tau_R$ and $\eta_L < \eta_j < \eta_R$. For simplicity, only considering about the first order term, we have,
\begin{equation}
\begin{aligned}
&\eta_L\Delta te^{\frac{\Delta t}{\tau_L}} < \eta_j\Delta te^{\frac{\Delta t}{\tau_j}}, \\
&\delta_{d,L} < \delta_{d,j},
\end{aligned}
\end{equation}
which means compared with the case that all coefficients are taken at interface ``$j$'' like the UGKS, more particles go out from cell ``$L$'' into cell ``$R$''. Similarly,
\begin{equation}
\begin{aligned}
&\eta_R\Delta te^{\frac{\Delta t}{\tau_R}} > \eta_j\Delta te^{\frac{\Delta t}{\tau_j}}, \\
&\delta_{d,R} > \delta_{d,j},
\end{aligned}
\end{equation}
which means less particles go out of cell ``$R$'' into cell ``$L$''. When $\eta$ and $\tau$ change intensely, mass will be wrongly and endlessly stored in cell ``$L$''. In order to make the particle coefficients consistent with hydrodynamic wave coefficients, Eq.~\eqref{eq:integralb2} and Eq.~\eqref{eq:ffrwaveb} are modified by splitting into two sides as well, as follows,
\begin{equation}\label{eq:integralb2b}
\begin{aligned}
\boldsymbol{F}^{eq}_j =& \int_{\mathbb{R}^D} \int_{\boldsymbol{u}\cdot\boldsymbol{n}_j>0}{\boldsymbol{\Psi}\left(\delta_{a,L}g_0+\delta_{b,L}\frac{\partial g}{\partial \boldsymbol{x}}\cdot\boldsymbol{u}+ \delta_{c,L}\frac{\partial g}{\partial t} \right)\left(\boldsymbol{u}\cdot\boldsymbol{n}_j\right) {\rm d}\boldsymbol{u}} {\rm d}\boldsymbol{\xi}\\
+&\int_{\mathbb{R}^D} \int_{\boldsymbol{u}\cdot\boldsymbol{n}_j<0}{\boldsymbol{\Psi}\left(\delta_{a,R}g_0+\delta_{b,R}\frac{\partial g}{\partial \boldsymbol{x}}\cdot\boldsymbol{u}+ \delta_{c,R}\frac{\partial g}{\partial t} \right)\left(\boldsymbol{u}\cdot\boldsymbol{n}_j\right) {\rm d}\boldsymbol{u}} {\rm d}\boldsymbol{\xi},
\end{aligned}
\end{equation}
and,
\begin{equation}\label{eq:ffrwavebb}
\begin{aligned}
\boldsymbol{F}^{fr,wave}_j=&\int_{\mathbb{\mathbb{R}}^D}\int_{\boldsymbol{u}\cdot\boldsymbol{n}_j>0}\boldsymbol{\Psi}\left[\left(\delta_{d,L}-\eta_L\Delta te^{\frac{\Delta t}{\tau_L}}\right)g_0^h+\left(\delta_{e,L}+\eta_L\frac{\Delta t^2}{2}e^{\frac{\Delta t}{\tau_L}}\right)\frac{\partial g^h}{\partial \boldsymbol{x}}\cdot\boldsymbol{u}\right] \left(\boldsymbol{u}\cdot\boldsymbol{n}_j\right) {\rm d}\boldsymbol{u} {\rm d}\boldsymbol{\xi} \\
+&\left[1-\eta\left({\rm{Kn}}_{\lambda,L}\right)\right]\hat{\delta}_{f,L}\int_{\mathbb{\mathbb{R}}^D}\int_{\boldsymbol{u}\cdot\boldsymbol{n}_j>0}\boldsymbol{\Psi}\left[\frac{\partial g^h}{\partial \boldsymbol{x}}\cdot\boldsymbol{u}+\frac{\partial g^h}{\partial t}\right] \left(\boldsymbol{u}\cdot\boldsymbol{n}_j\right) {\rm d}\boldsymbol{u} {\rm d}\boldsymbol{\xi} \\
+&\int_{\mathbb{\mathbb{R}}^D}\int_{\boldsymbol{u}\cdot\boldsymbol{n}_j<0}\boldsymbol{\Psi}\left[\left(\delta_{d,R}-\eta_R\Delta te^{\frac{\Delta t}{\tau_R}}\right)g_0^h+\left(\delta_{e,R}+\eta_R\frac{\Delta t^2}{2}e^{\frac{\Delta t}{\tau_R}}\right)\frac{\partial g^h}{\partial \boldsymbol{x}}\cdot\boldsymbol{u}\right] \left(\boldsymbol{u}\cdot\boldsymbol{n}_j\right) {\rm d}\boldsymbol{u} {\rm d}\boldsymbol{\xi}\\
+&\left[1-\eta\left({\rm{Kn}}_{\lambda,R}\right)\right]\hat{\delta}_{f,R}\int_{\mathbb{\mathbb{R}}^D}\int_{\boldsymbol{u}\cdot\boldsymbol{n}_j<0}\boldsymbol{\Psi}\left[\frac{\partial g^h}{\partial \boldsymbol{x}}\cdot\boldsymbol{u}+\frac{\partial g^h}{\partial t}\right] \left(\boldsymbol{u}\cdot\boldsymbol{n}_j\right) {\rm d}\boldsymbol{u} {\rm d}\boldsymbol{\xi},
\end{aligned}
\end{equation}
where,
\begin{equation}
\hat{\delta}_f = -\tau^2\left( {1-e^{-\Delta t/\tau}} \right).
\nonumber
\end{equation}
The corresponding terms after $\hat{\delta}_f$ are the Chapman-Enskog expansion terms of initial state of the integral solution. As the GKS, in order to recover the viscosity of N-S equation, these terms are introduced in the continuum flow regime, and $\left[1-\eta\left({\rm{Kn}}_{\lambda}\right)\right]$ is an adjusting parameter.

\subsection{Algorithm of AUGKWP}\label{sec:sum2}
\begin{figure}[H]
	\centering
	\subfigure[]{\label{fig4a}
			\includegraphics[width=0.22 \textwidth]{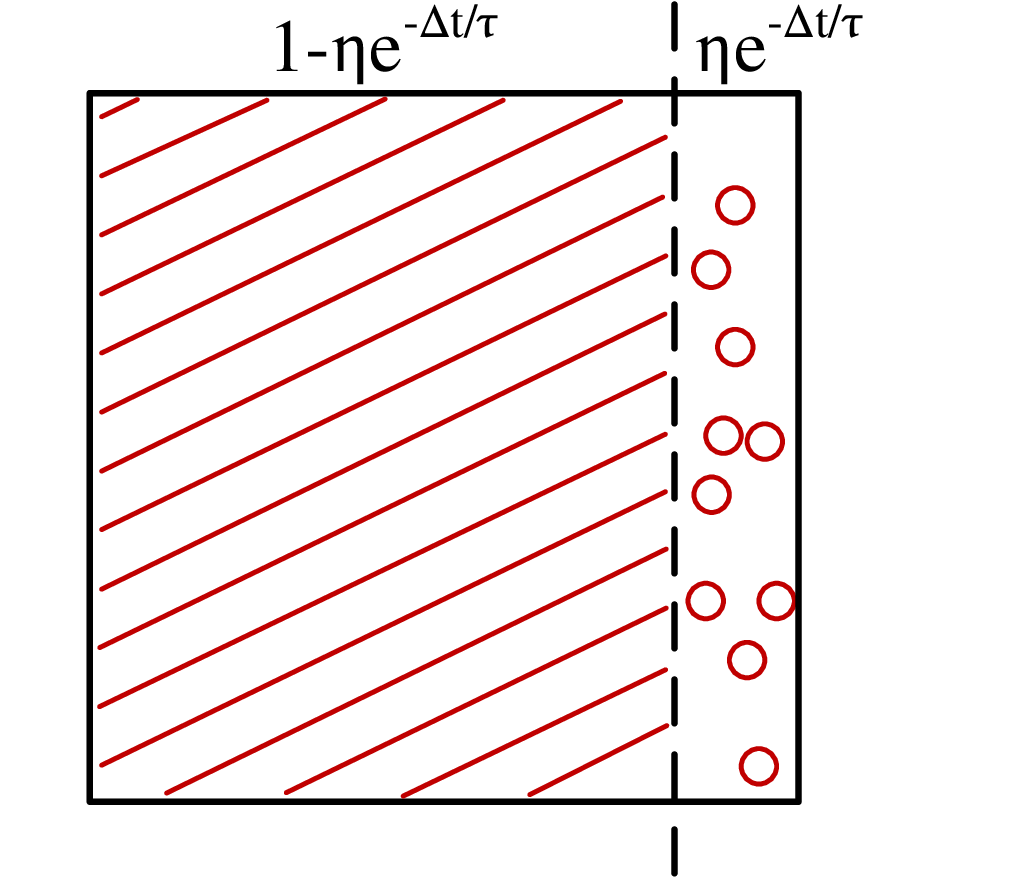}
		}
    \subfigure[]{\label{fig4b}
    		\includegraphics[width=0.22 \textwidth]{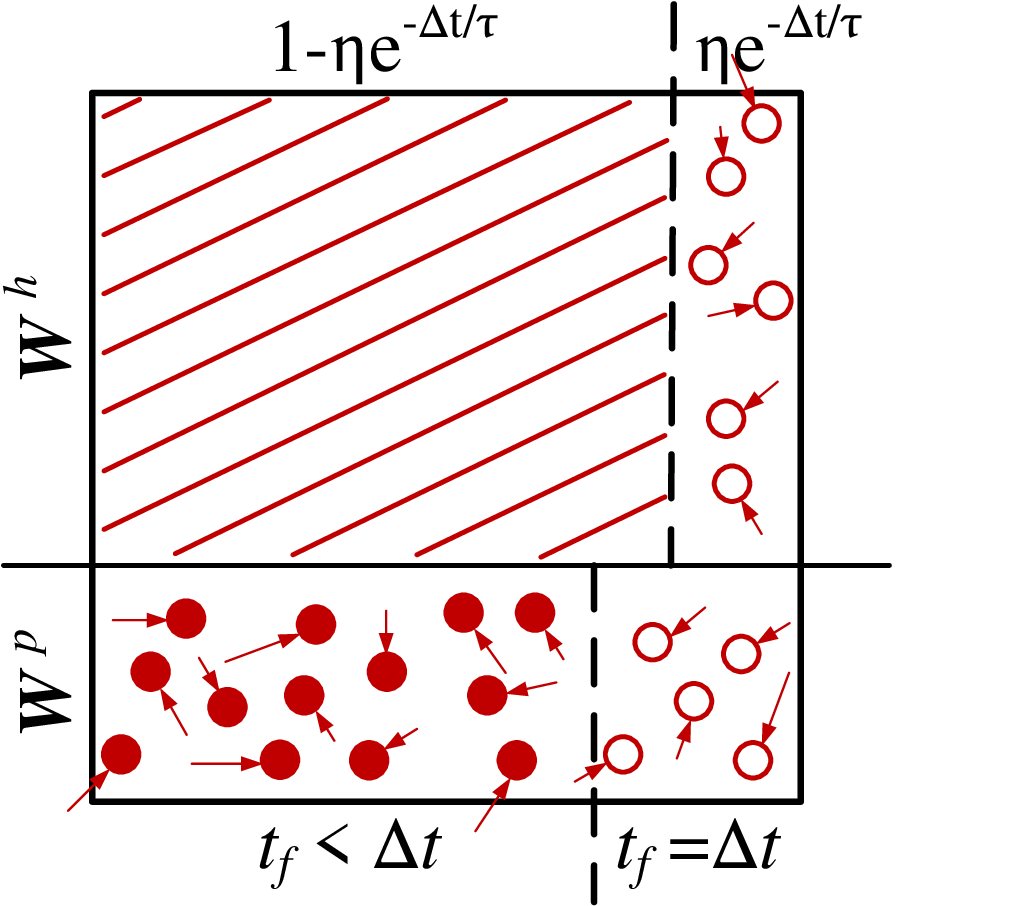}
    	}
    \subfigure[]{\label{fig4c}
    		\includegraphics[width=0.22 \textwidth]{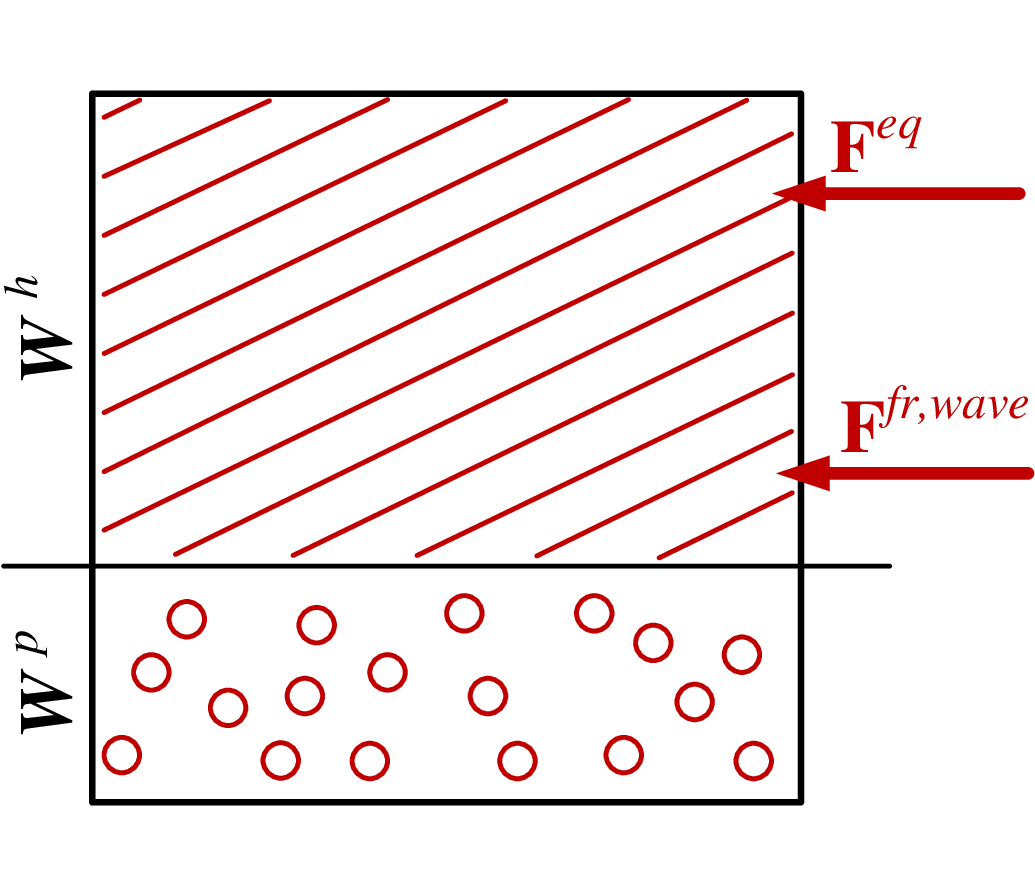}
    	}
    \subfigure[]{\label{fig4d}
    		\includegraphics[width=0.22 \textwidth]{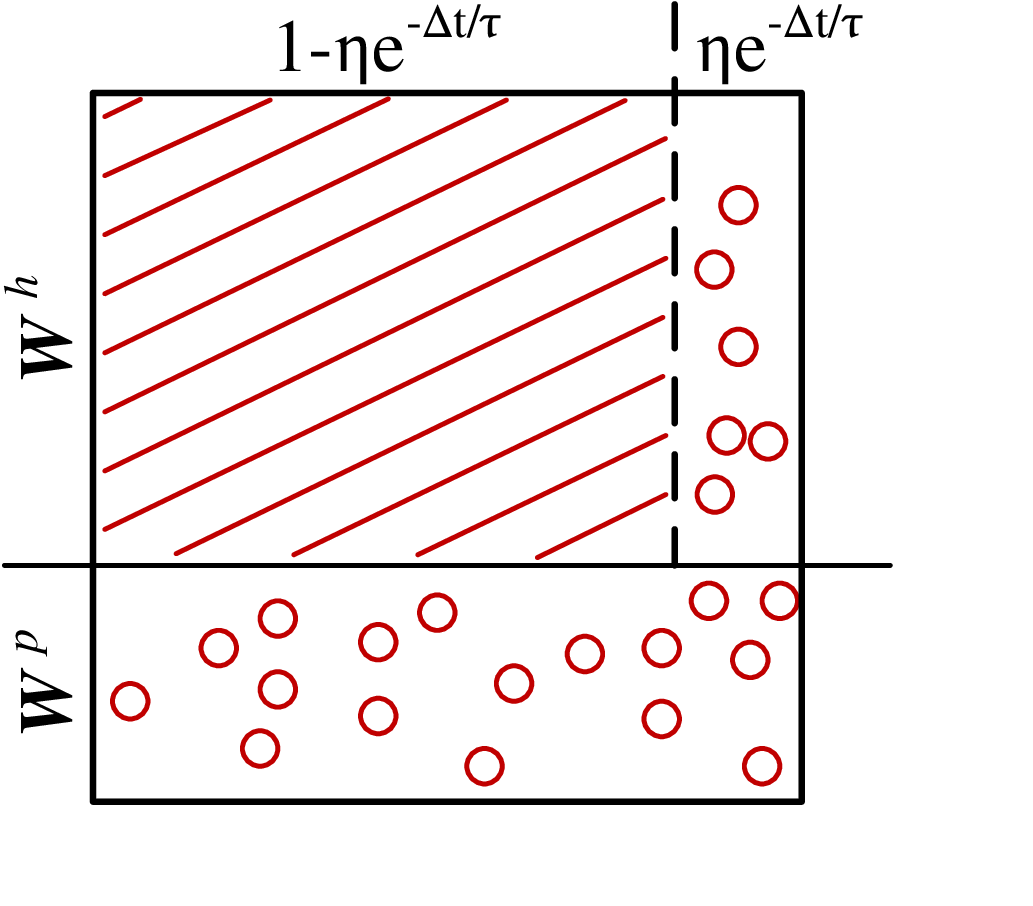}
    	}
	\caption{\label{fig4} Diagram to illustrate the algorithm of AUGKWP method: (a) Initial state at $n=0$ step, (b) step $2$, (c) step $3$, (d) initial state at other steps.}
\end{figure}

Regarding as Fig.~\ref{fig4}, the following part is a summary of the algorithm of AUGKWP method.
\begin{description}
    \item[Step (1)] Initial state. As the result of the previous time step $n-1$ in Fig.~\ref{fig4d}, numerical particles $\boldsymbol{W}^p$ coexist with hydrodynamic wave $\boldsymbol{W}^h$. Then collisionless particles $\boldsymbol{W}^{hp}$ are sampled as Eq.~\eqref{eq:mic5b}, with detailed parameters from Eq.~\eqref{eq:mic6} to Eq.~\eqref{eq:aje}. For the first step, $\boldsymbol{W}^p=\boldsymbol{0}$, as shown in Fig.~\ref{fig4a}.
    \item[Step (2)] Free transport. Firstly, divide particles $\boldsymbol{W}^p$ into two parts as Eq.~\eqref{eq:mic1}. As Fig.~\ref{fig4b}, collisionless particles are denoted by hollow circles, and collisional particles are denoted by solid circles. Then transport all particles as Eq.~\eqref{eq:mic2}, and cumulate their contribution to the macroscopic conserved variables as Eq.~\eqref{eq:mic3}. Meanwhile, the free transport fluxes $\boldsymbol{F}^{fr,wave}_j$ contributed from the collisional particles of $\left(\boldsymbol{W}^h-\boldsymbol{W}^{hp}\right)$ is calculated as Eq.~\eqref{eq:ffrwavebb}. Finally, delete the collisional particles which are denoted by solid circles.
    \item[Step (3)] Collision. Compute the $\boldsymbol{F}^{eq}_j$ term as Eq.~\eqref{eq:integralb2b} by equations in Sec.~\ref{sec:macro}. Then renew $\boldsymbol{W}^{n+1}$ as Eq.~\eqref{eq:renew}.
    \item[Step (4)] If the simulation continues, go to Step $\left(1\right)$, where collisionless particles $\boldsymbol{W}^{hp}$ will be sampled.
\end{description}

\section{Test cases}\label{sec:cases}
A series of cases are conducted to test the performance of AUGKWP method. Firstly, the hypersonic flow around a cylinder at a wide range of $\rm{Kn}_{\infty}$ is simulated for a detailed validation in different flow regimes. The following two cases involve that the non-equilibrium phenomenon is induced by a localized minor perturbation in a globally near-continuum flow regime, such as the hypersonic flow over a slender cavity and the side-jet impingement on hypersonic flow. Finally, the hypersonic flow over a $70^{\circ}$ blunted cone with a cylindrical sting is simulated as a three dimensional case. The flow is in the continuum regime at the windward side, but is rather rarefied in the wake area because of the strong expansion. The full accommodation is applied for the isothermal walls as the boundary condition. The Courant-Friedrichs-Lewy (CFL) number is set to be $0.8$ in all cases. Besides, the spatial reconstruction of macroscopic variables is carried out in the monotone upwind scheme for conservation laws (MUSCL)~\cite{muscl} framework, by the weighted least-square method~\cite{wls} with the Venkatarishnan limiter~\cite{venkata}. More details can be referred to Ref.~\cite{jiri}.

\subsection{Hypersonic flow around a cylinder at a wide range of $\rm{Kn}_{\infty}$}\label{sec:cylinder}
The hypersonic flow around a cylinder is a typical multiscale case from continuum to rarefied regime. The high-$\rm{Ma}$ shock strongly compresses the flow at the windward wall, which makes the local scale tends towards more to continuum, while the flow at the wake area is more rarefied due to the expansion. Since the gradients of temperature and velocity are large near the isothermal wall, accurate prediction of heat flux and shear stress brings about high requirement of numerical method and needs precise mesh there. In this work, for a detailed validation of AUGKWP method, a wide range of $\rm{Kn}_{\infty}$ is taken as the working condition, such as $0.1$, $0.01$ and $0.001$. The reference length is set to be the radius of the cylinder, $1.0$.

In this case, the inflow ${\rm{Ma_{\infty}}}$ is set to be $5$, and the temperature at the wall is set to be the same as the inflow value. The UGKS is taken for comparison and in order to get rid off the influence of different treatment of thermal internal energy and ${\rm{Pr}}$ number, the gas is set to be monatomic and the BGK model is taken as the kinetic model. The dynamic viscosity coefficient is calculated as $\mu\sim T^{\omega}$, where $\omega$ is set to be $0.81$. The mesh used by UGKS are the same as that of AUGKWP method. The height of first layer of mesh adjacent the wall is set to be $0.005$ at ${\rm{Kn}}_{\infty}=0.1$, $0.001$ at ${\rm{Kn}}_{\infty}=0.01$ and $0.0002$ at ${\rm{Kn}}_{\infty}=0.001$. For the circumference, $140$ nodes are set in every ${\rm{Kn}}_{\infty}$.

\begin{figure}[H]
	\centering
	\subfigure[]{
			\includegraphics[width=0.22 \textwidth]{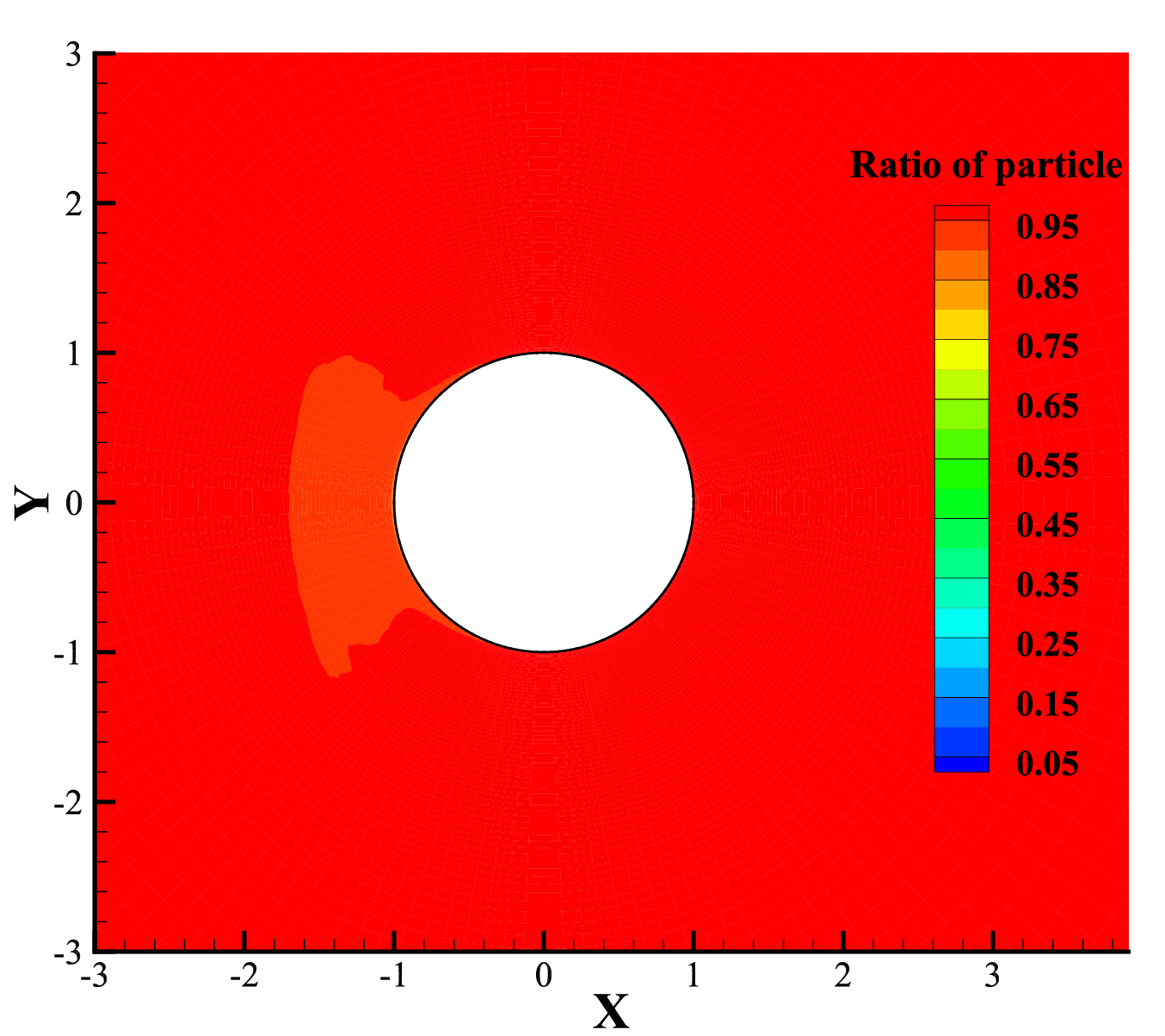}
		}
    \subfigure[]{\label{cylinder_kn0.01_rop_b}
    		\includegraphics[width=0.22 \textwidth]{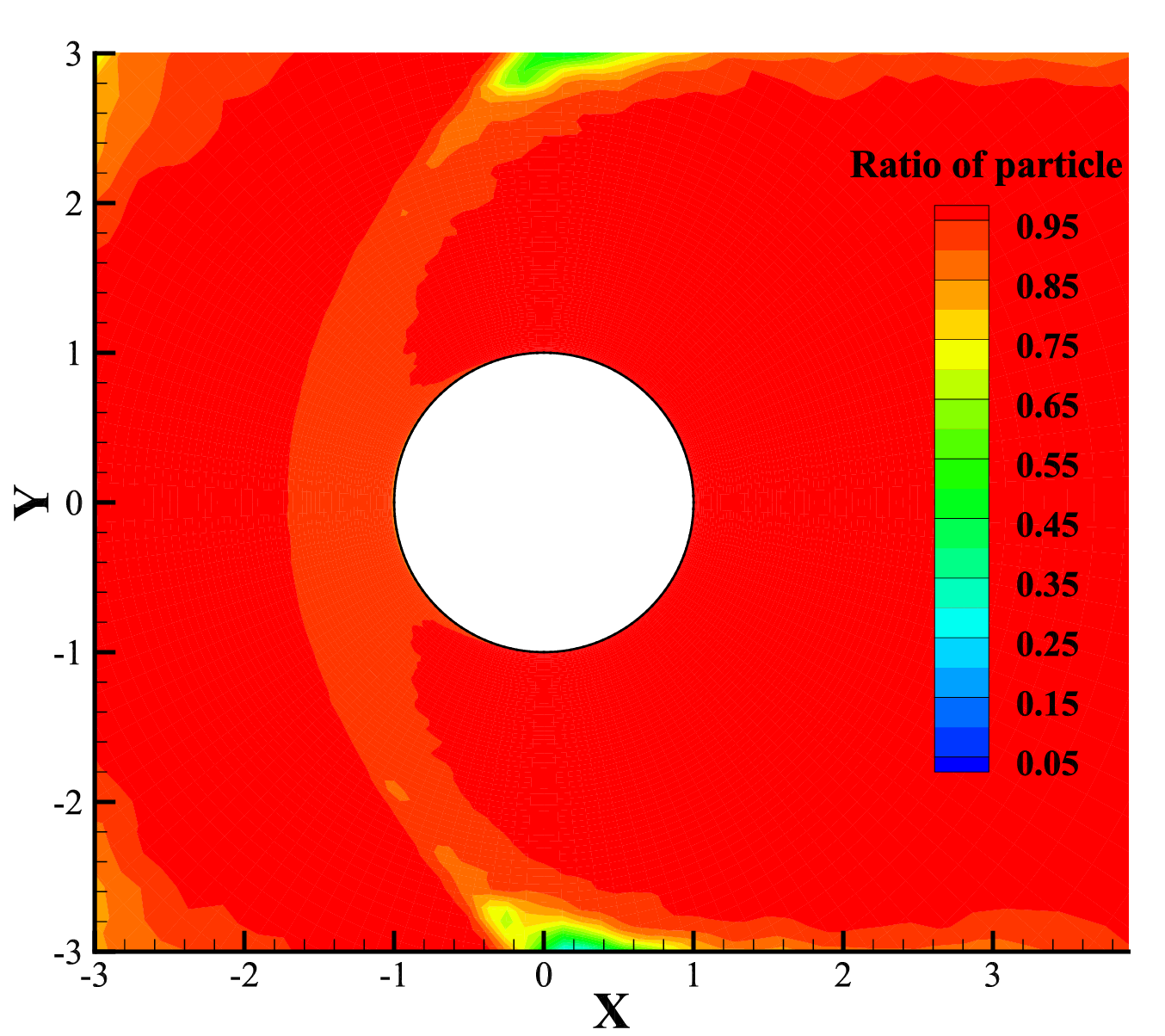}
    	}
    \subfigure[]{
    		\includegraphics[width=0.22 \textwidth]{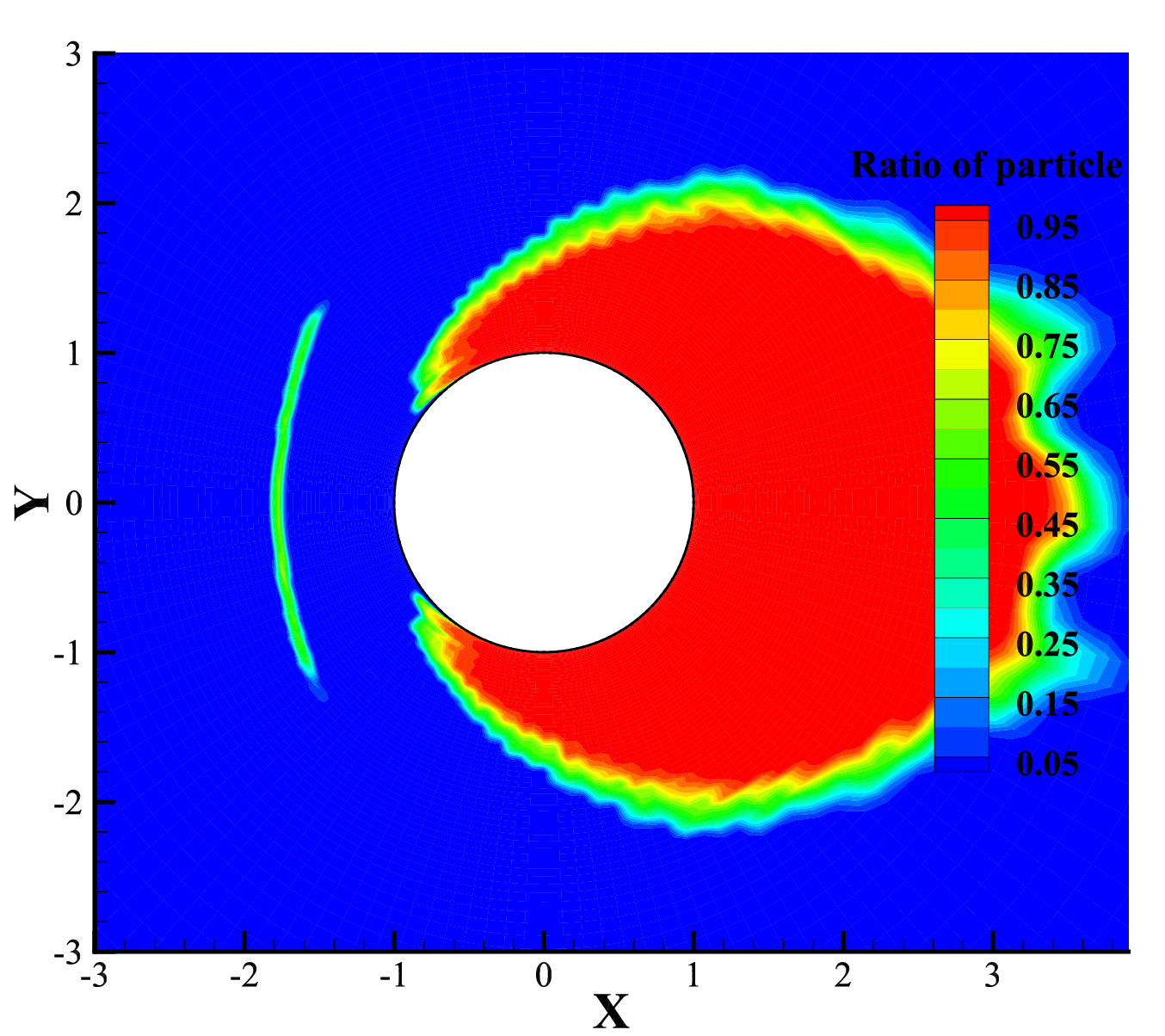}
    	}
    \subfigure[]{
    		\includegraphics[width=0.22 \textwidth]{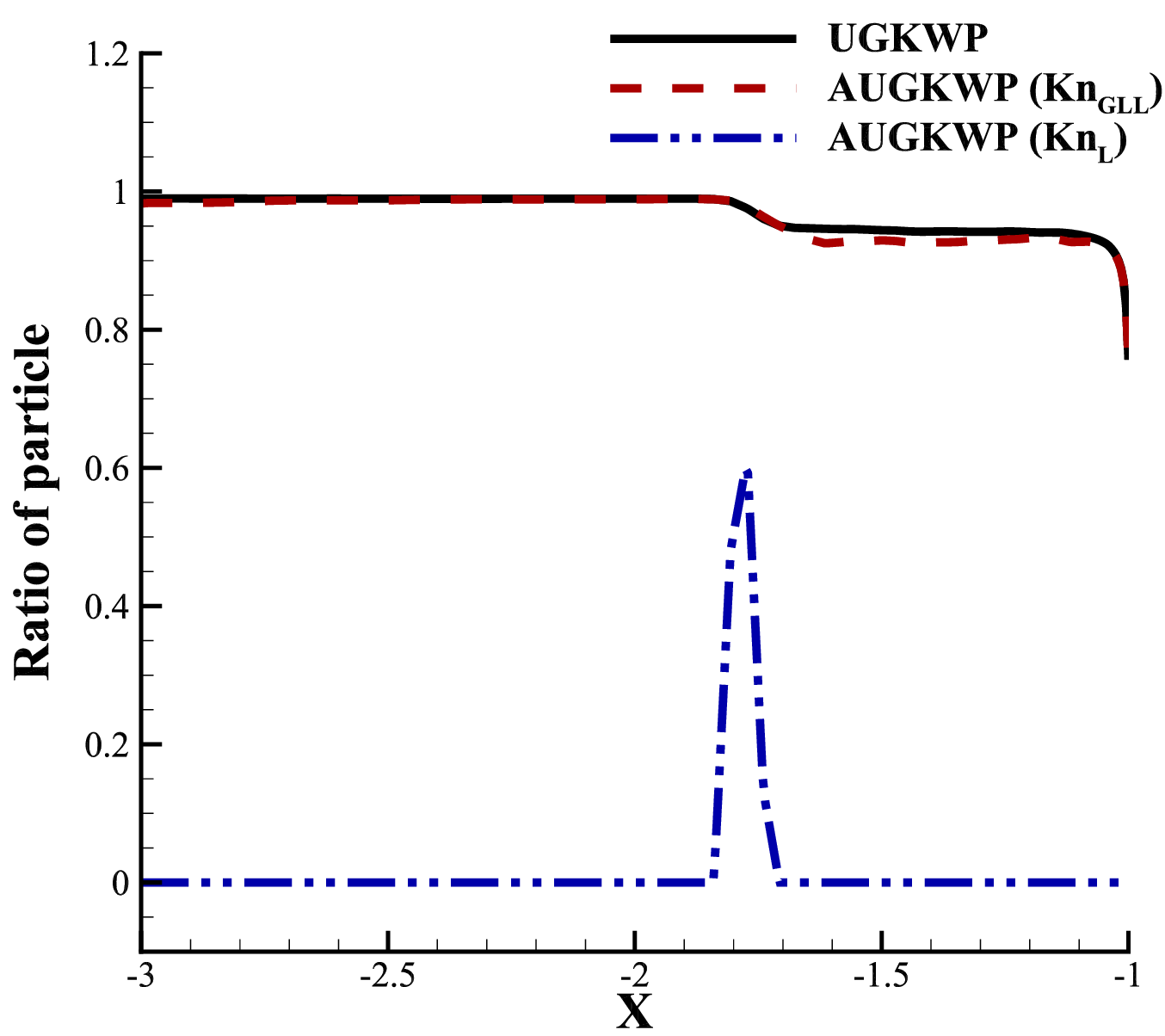}
    	}
	\caption{\label{cylinder_kn0.01_rop} Comparison of ratio of particles at ${\rm{Kn_{\infty}}}=0.01$: (a) Original UGKWP method, (b) AUGKWP method with ${\rm{Kn_{GLL}}}$, (c) AUGKWP method with $\rm{Kn_{L}}$, (d) details along the stagnation line of three approaches.}
\end{figure}

\begin{figure}[H]
	\centering
	\subfigure[]{
			\includegraphics[width=0.3 \textwidth]{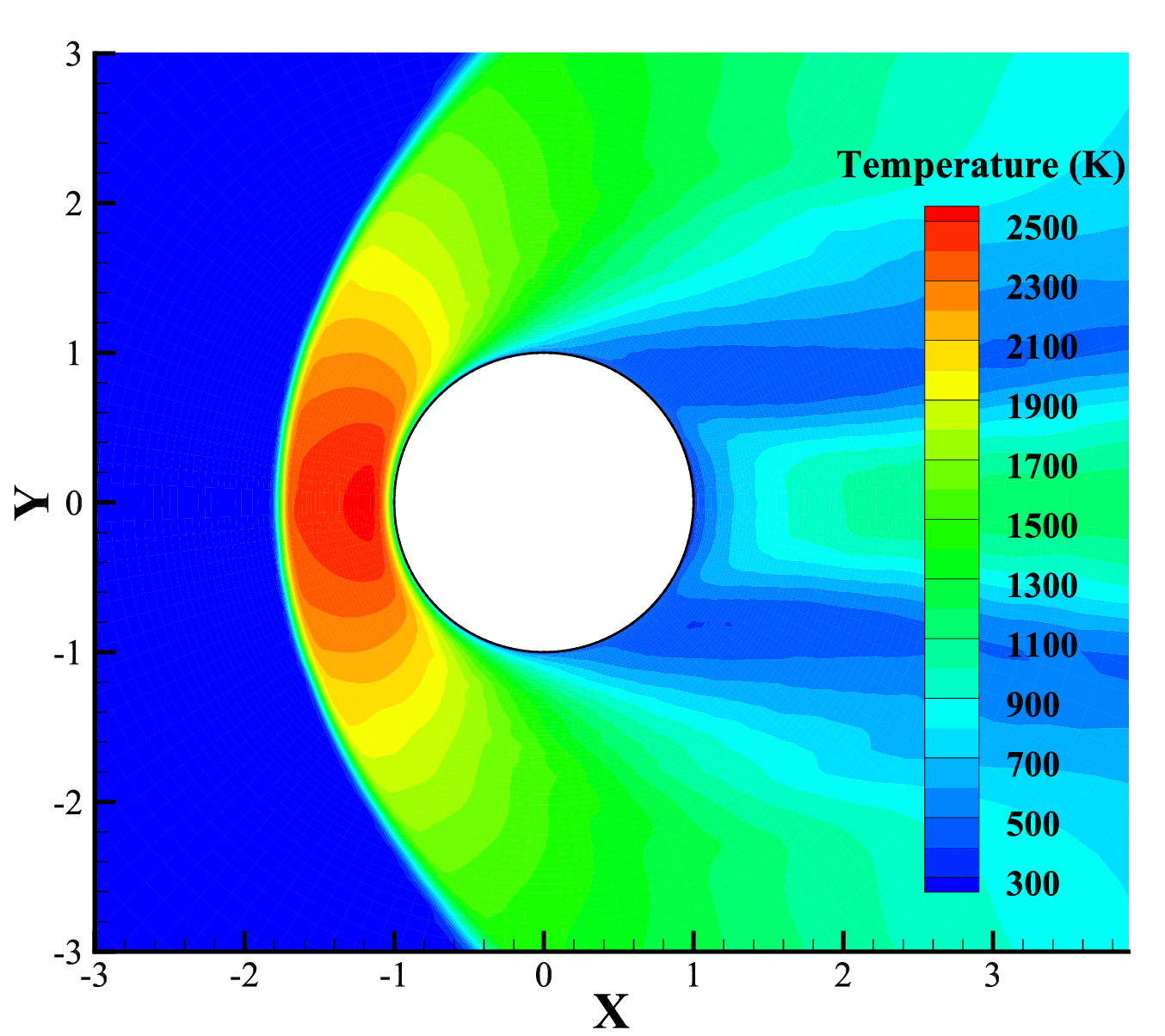}
		}
    \subfigure[]{
    		\includegraphics[width=0.3 \textwidth]{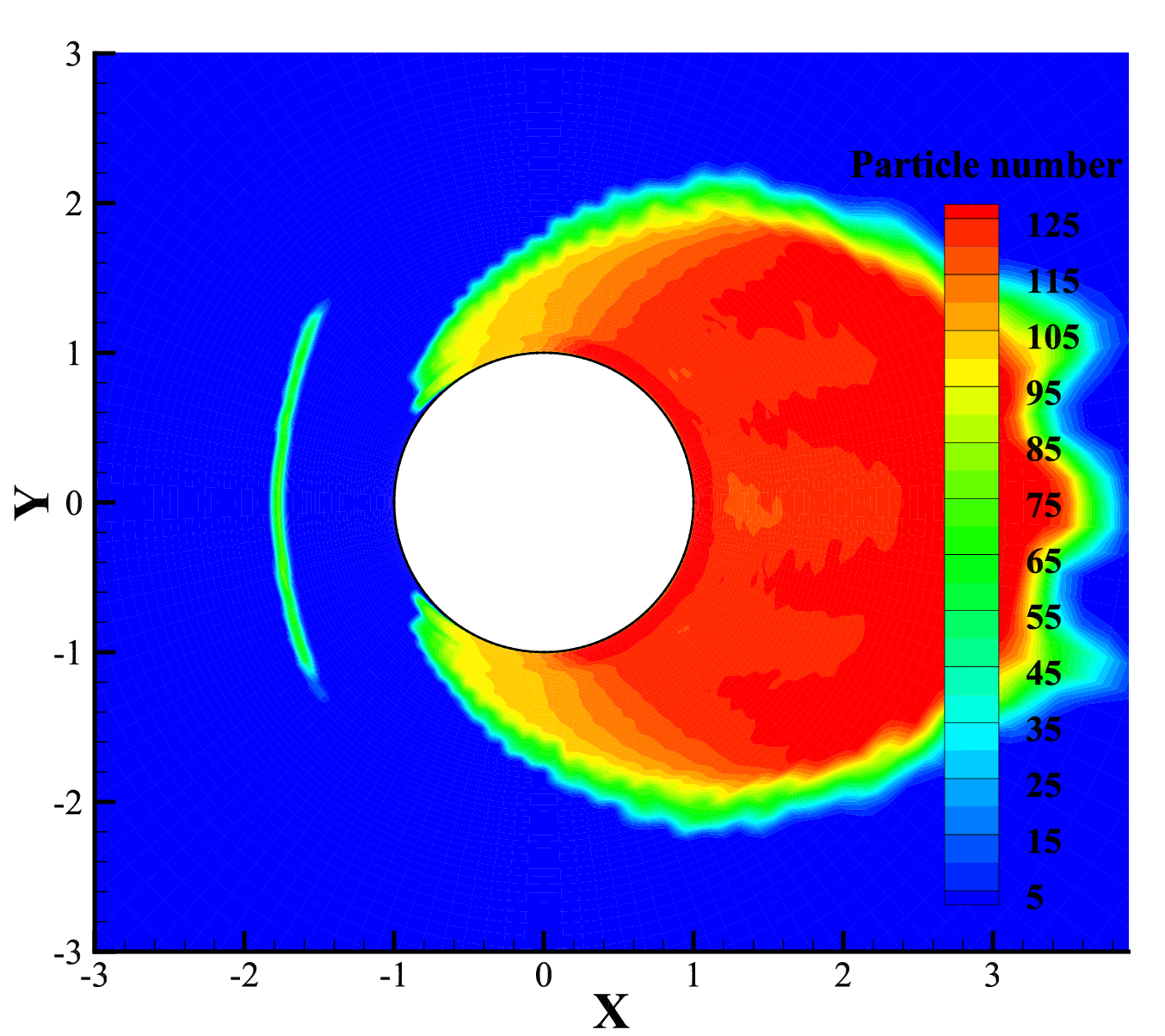}
    	}
    \subfigure[]{
    		\includegraphics[width=0.3 \textwidth]{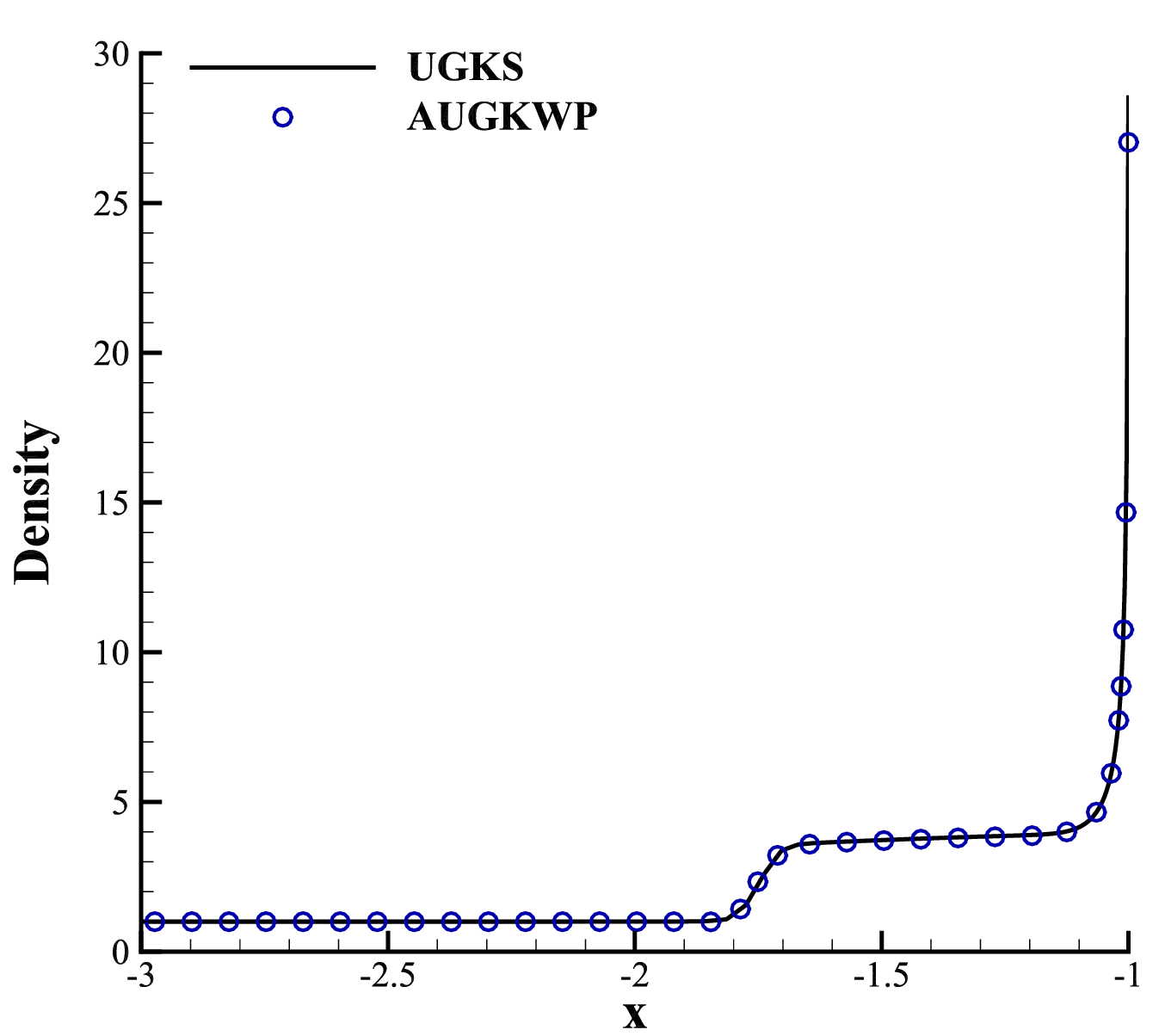}
    	}
    \subfigure[]{\label{cylinder_kn0.01_d}
			\includegraphics[width=0.3 \textwidth]{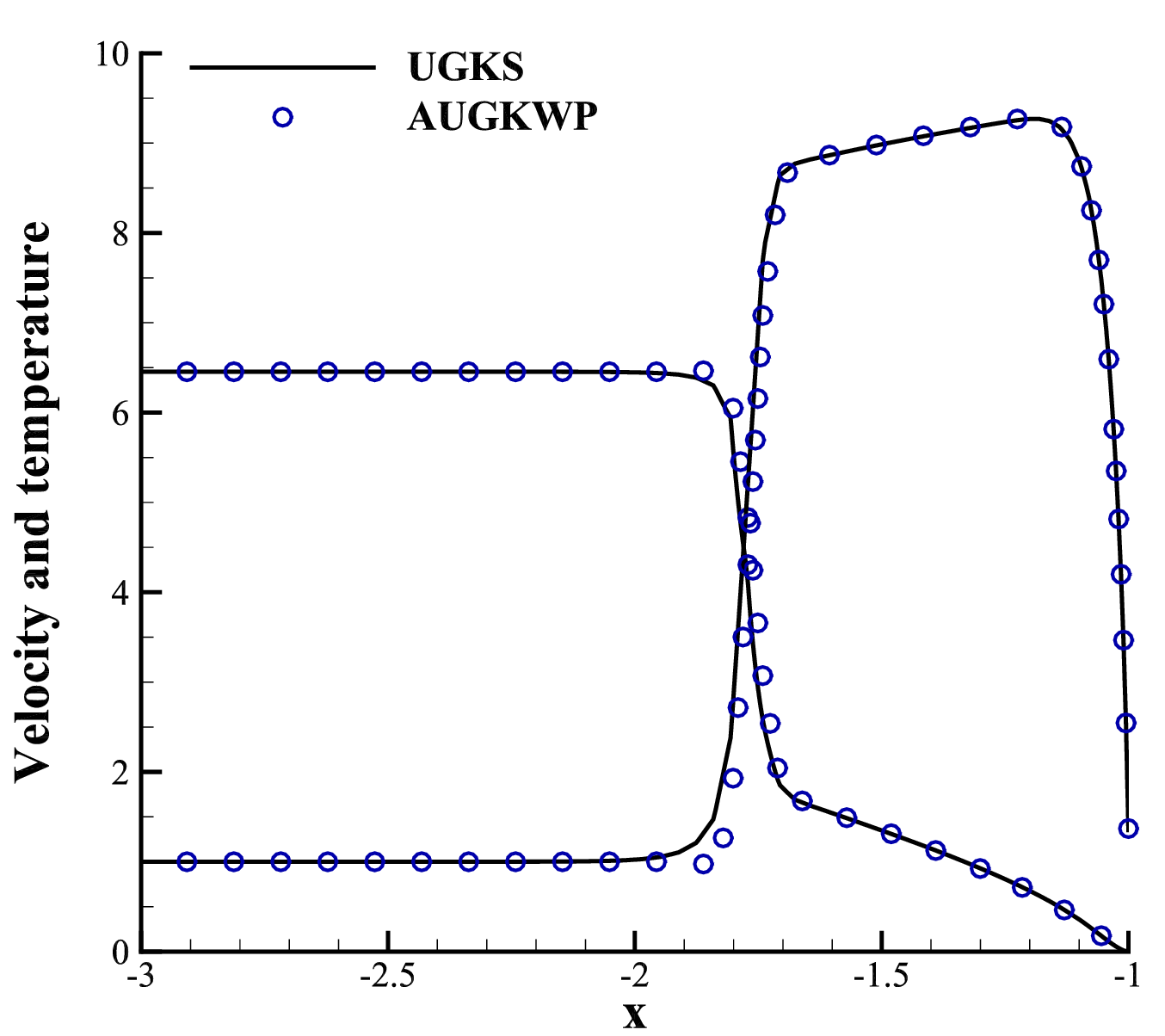}
		}
    \subfigure[]{
    		\includegraphics[width=0.3 \textwidth]{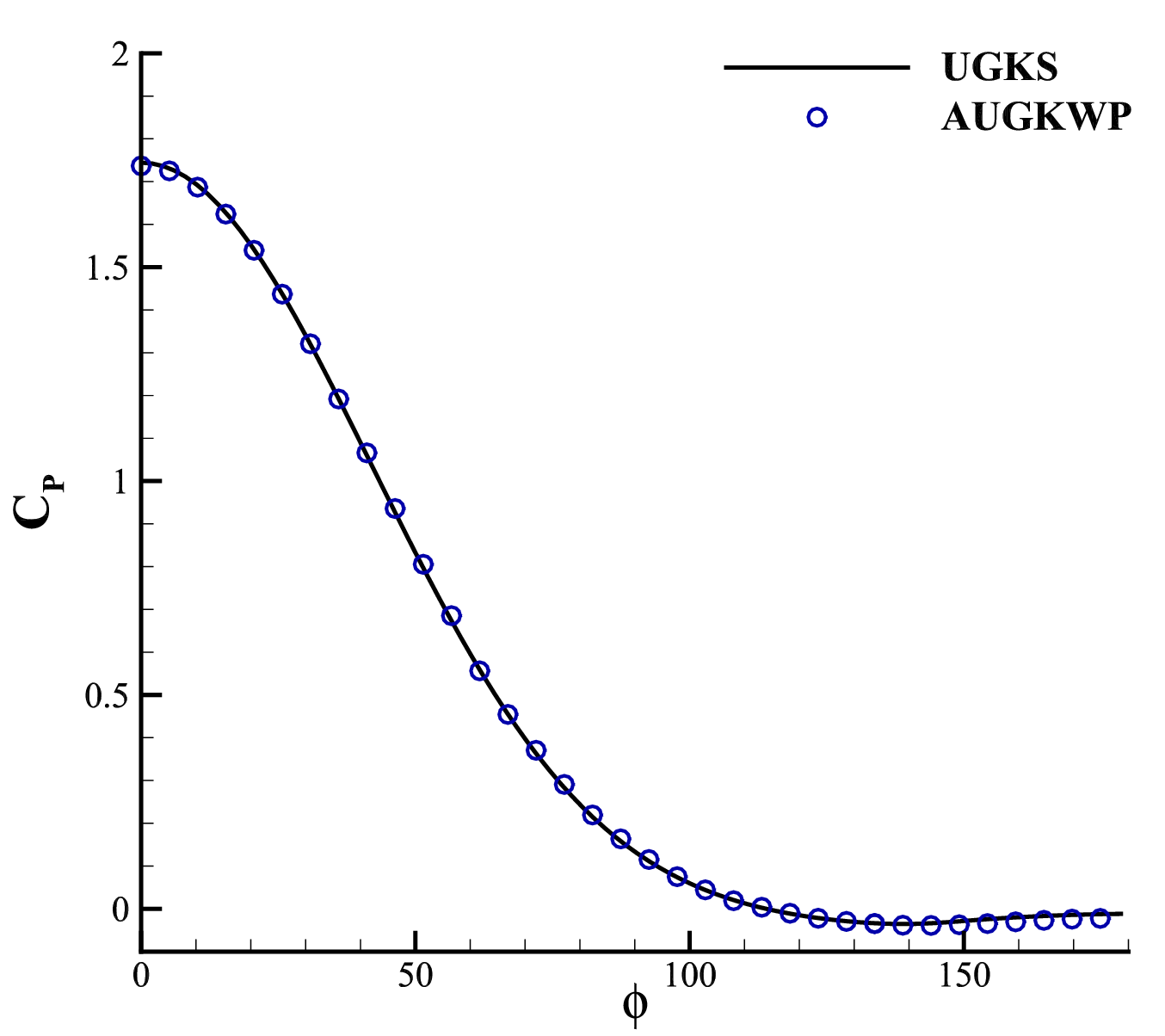}
    	}
    \subfigure[]{
    		\includegraphics[width=0.3 \textwidth]{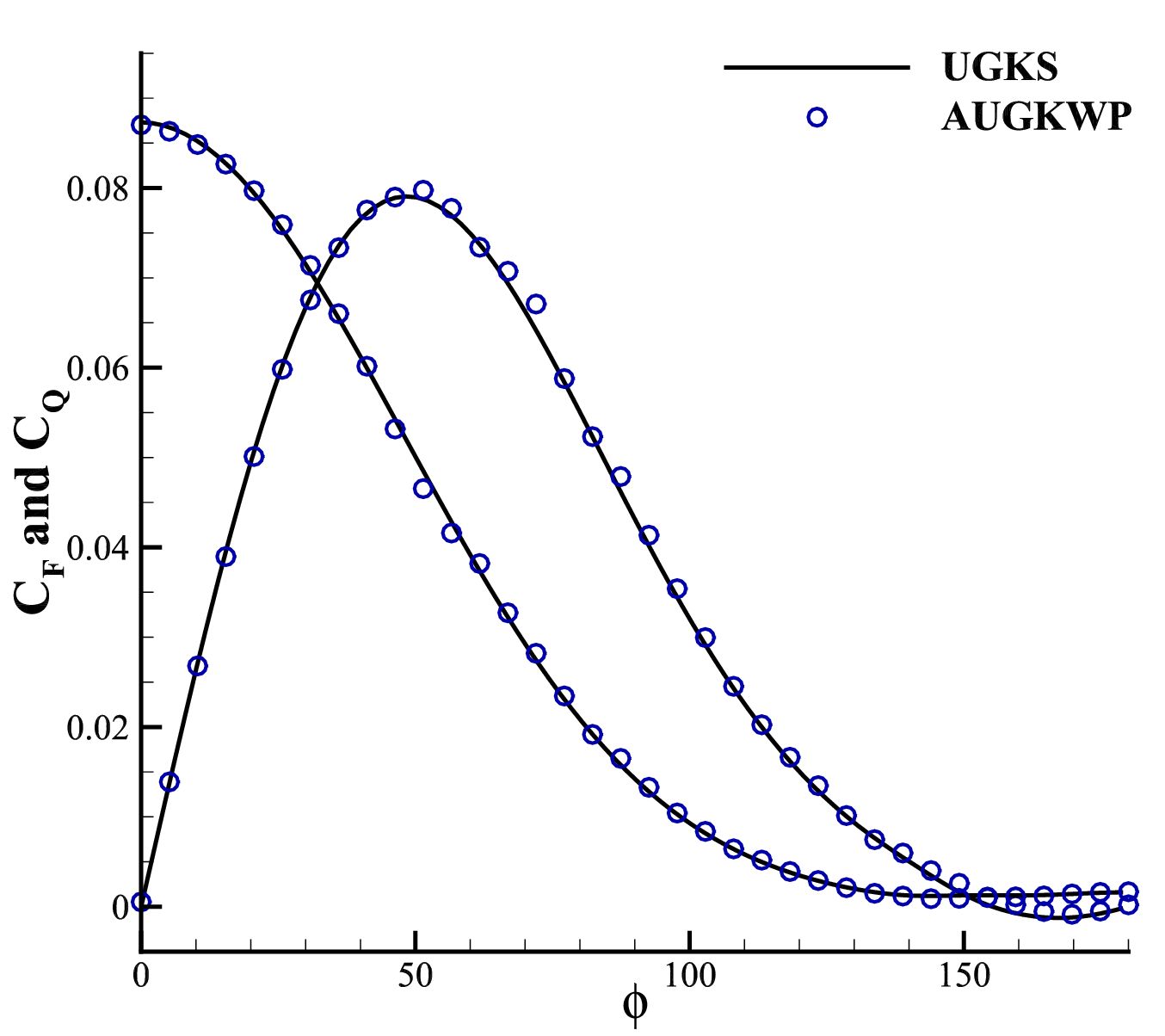}
    	}
	\caption{\label{cylinder_kn0.01} Results of AUGKWP method at ${\rm{Kn_{\infty}}}=0.01$: (a) Temperature contour, (b) particle number contour, (c) density along the stagnation line, (d) velocity and temperature along the stagnation line, (e) pressure coefficient at the wall, (f) shear stress and heat flux coefficients at the wall.}
\end{figure}

As shown in Fig.~\ref{cylinder_kn0.01_rop}, at ${\rm{Kn}}_{\infty}=0.01$, the AUGKWP method saves a large number of particles at the farfield, post-shock and windward wall. There are two regions where particles are employed. One is the wake area, because of the expansion. The other is the pre-shock region, because the mesh there is smaller than the farfield, which decreases $L_{\lambda,{\rm{ref}}}$ in Eq.~\eqref{eq:knl}. The result of only using ${\rm{Kn}}_{GLL}$ in the AUGKWP method is also shown as Fig.~\ref{cylinder_kn0.01_rop_b}. It is noticed that much more particles are saved by using ${\rm{Kn_L}}$ than by ${\rm{Kn_{GLL}}}$ in this ${\rm{Kn_{\infty}}}$. Detailed contours and quantitative comparisons at the stagnation line and wall are shown as Fig.~\ref{cylinder_kn0.01}. For variables at the stagnation line, density and temperature are nondimensionalized by inflow value, and velocity is nondimensionalized by inflow sound speed. For aerodynamic coefficients at the wall, $C_P$ is calculated by $C_P=\frac{p-p_{\infty}}{0.5\rho_{\infty}\left|U\right|_{\infty}^2}$, $C_F$ is nondimensionalized by $0.5\rho_{\infty}\left|U\right|_{\infty}^2$ and $C_H$ is nondimensionalized by $0.5\rho_{\infty}\left|U\right|_{\infty}^3$. Results of AUGKWP method are in line well with those of UGKS, except that a deviation occurs in Fig.~\ref{cylinder_kn0.01_d}. Specifically at the pre-shock region, the temperature predicted by AUGKWP method does not rise as high as the UGKS. It is because the discontinuity is under-resolved by the mesh there. Besides, particles there are released suddenly and then absorbed by hydrodynamic wave again. Such intense change brings about deviation as well.

\begin{figure}[H]
	\centering
	\subfigure[]{\label{cylinder_kn0.001_rop_a}
			\includegraphics[width=0.22 \textwidth]{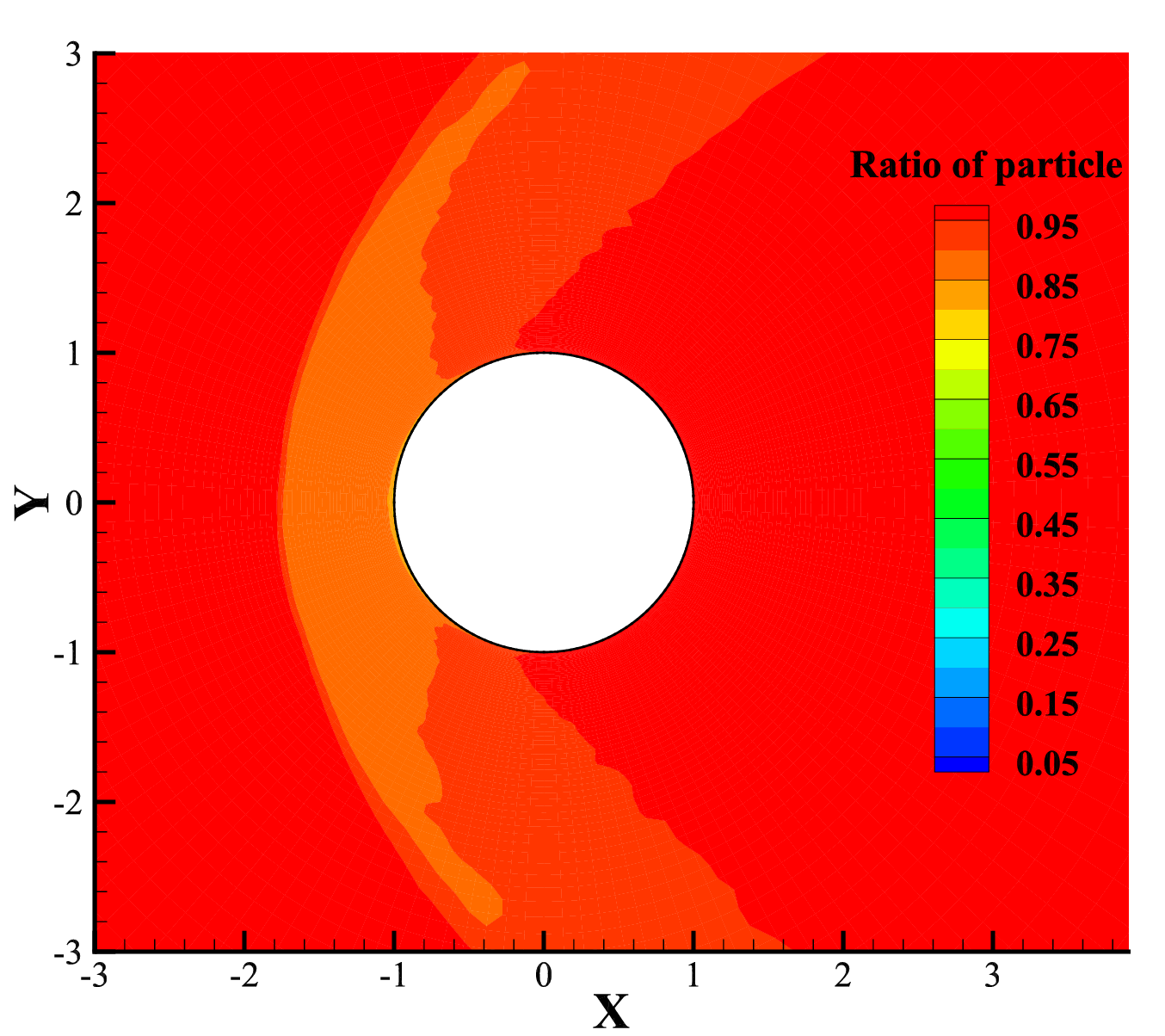}
		}
    \subfigure[]{\label{cylinder_kn0.001_rop_b}
    		\includegraphics[width=0.22 \textwidth]{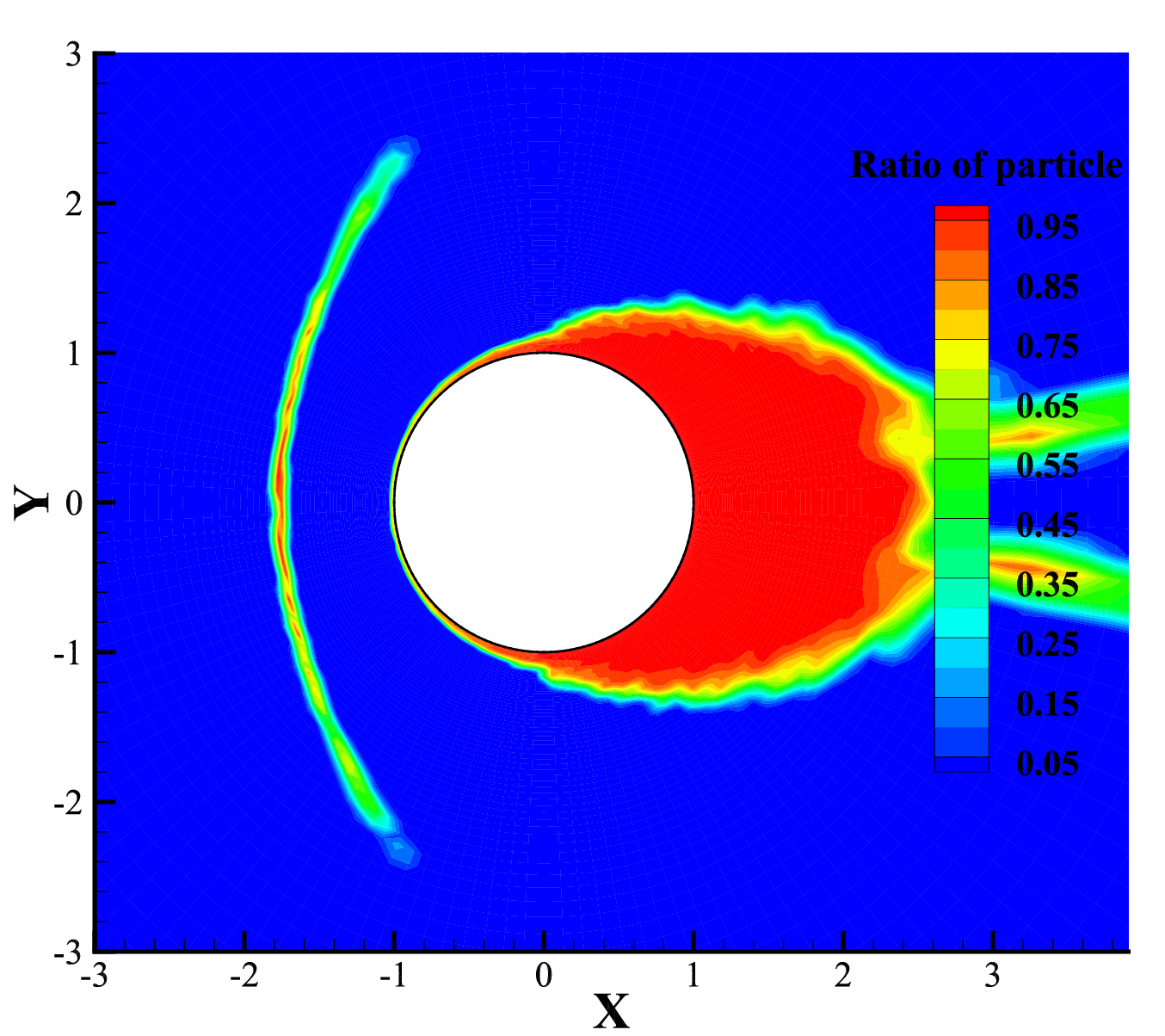}
    	}
    \subfigure[]{\label{cylinder_kn0.001_rop_c}
    		\includegraphics[width=0.22 \textwidth]{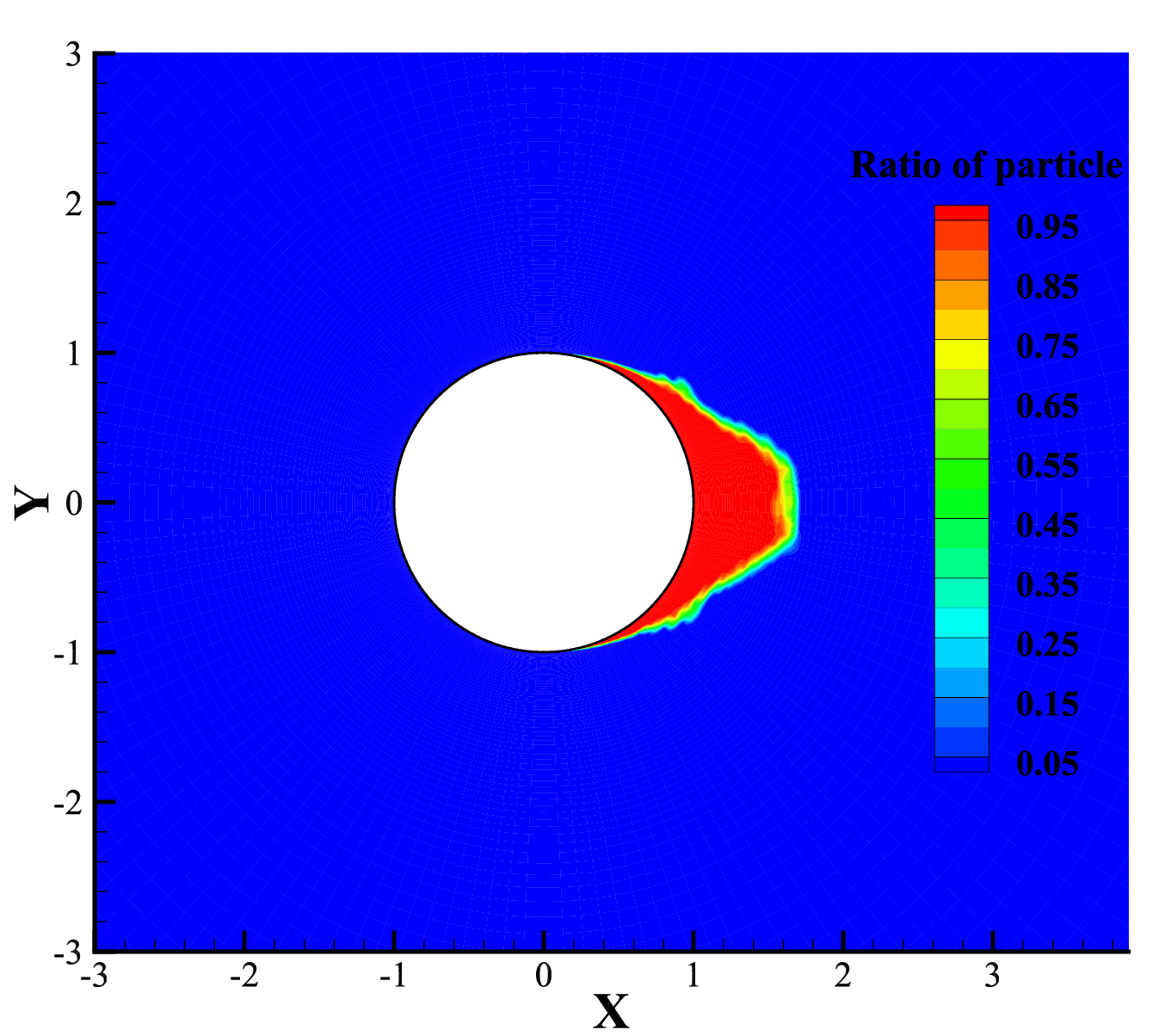}
    	}
    \subfigure[]{
    		\includegraphics[width=0.22 \textwidth]{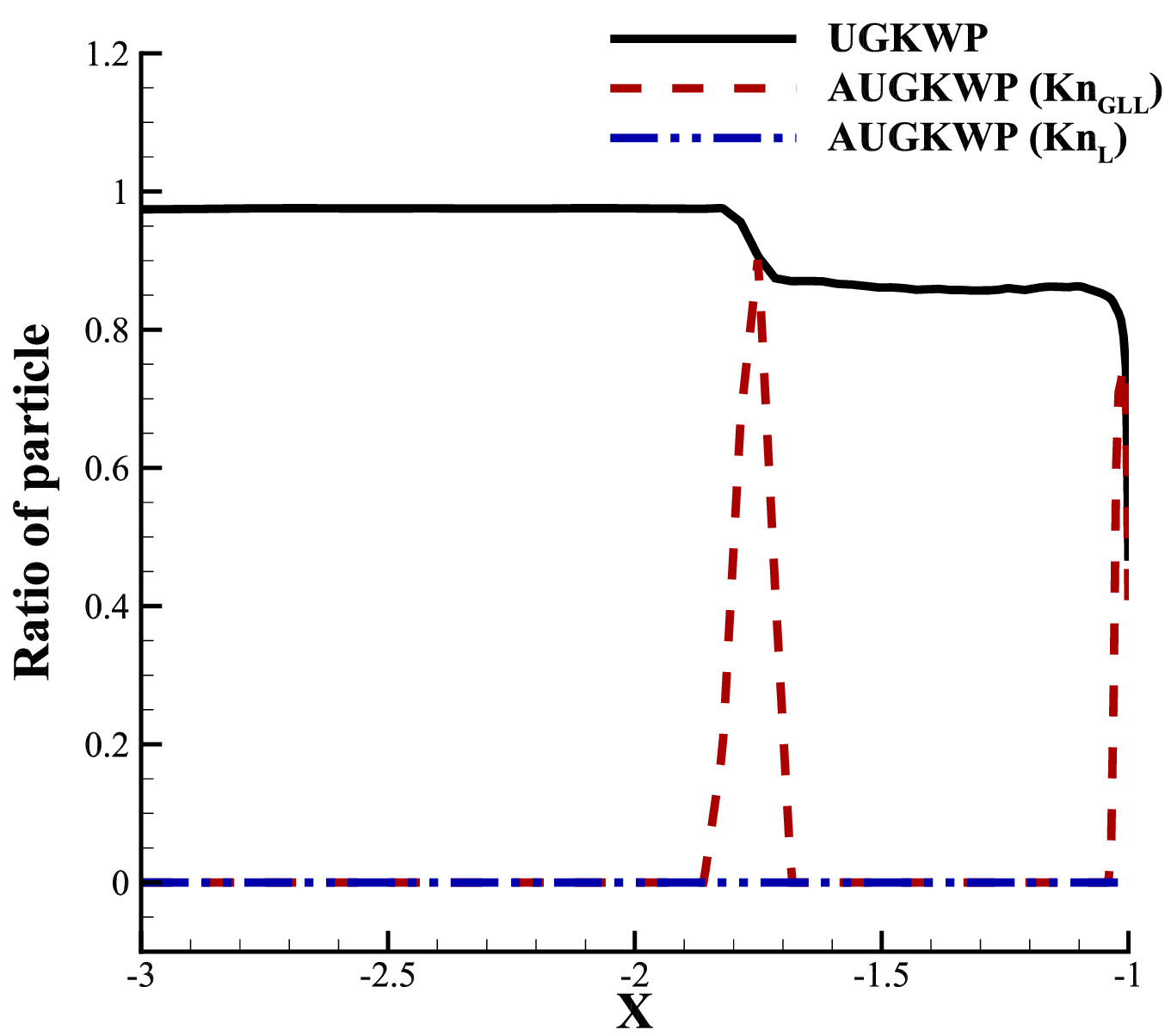}
    	}
	\caption{\label{cylinder_kn0.001_rop} Comparison of ratio of particles at ${\rm{Kn_{\infty}}}=0.001$: (a) Original UGKWP method, (b) AUGKWP method with ${\rm{Kn_{GLL}}}$, (c) AUGKWP method with ${\rm{Kn_{L}}}$, (d) details along the stagnation line of three approaches.}
\end{figure}

\begin{figure}[H]
	\centering
	\subfigure[]{
			\includegraphics[width=0.3 \textwidth]{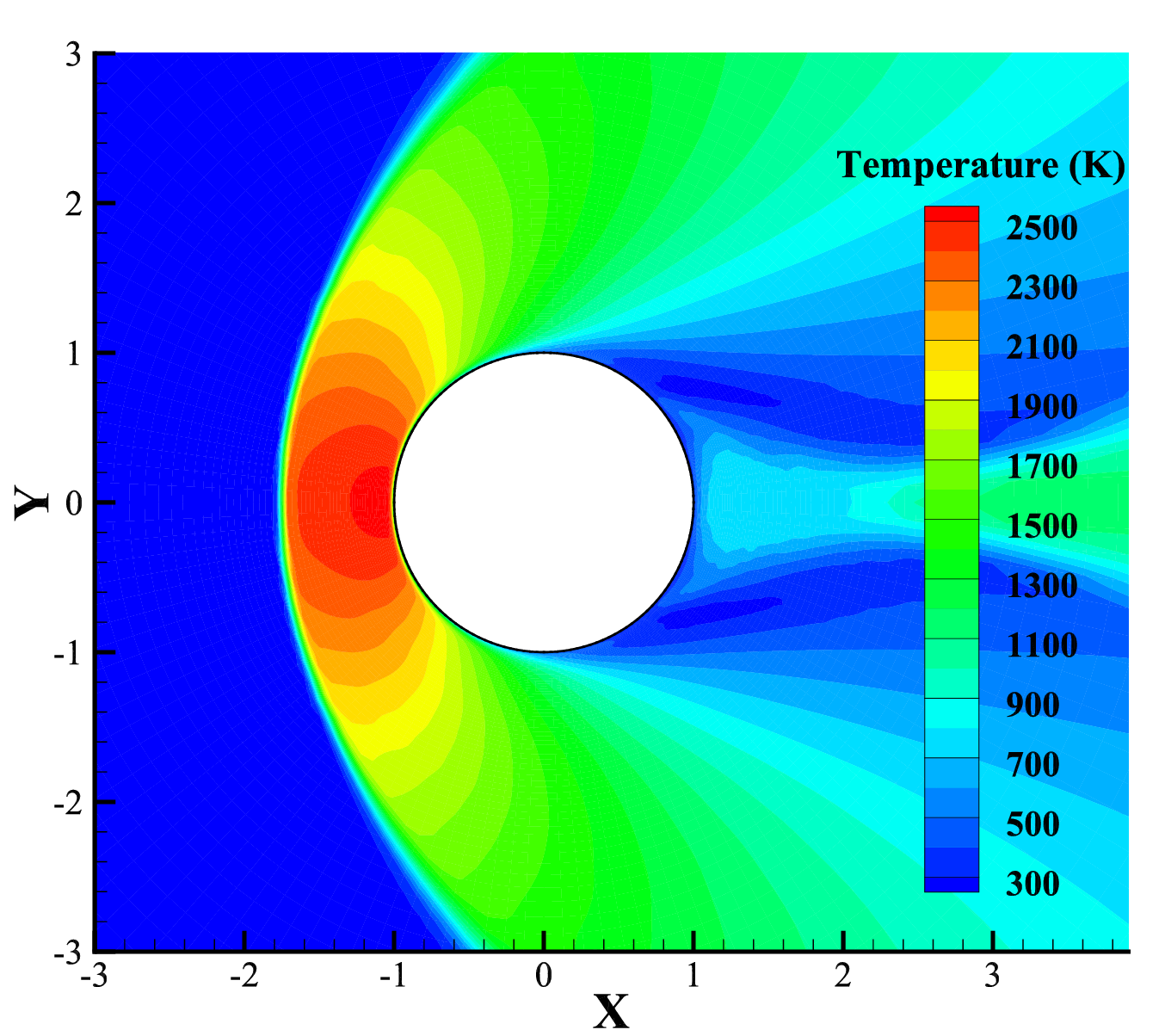}
		}
    \subfigure[]{
    		\includegraphics[width=0.3 \textwidth]{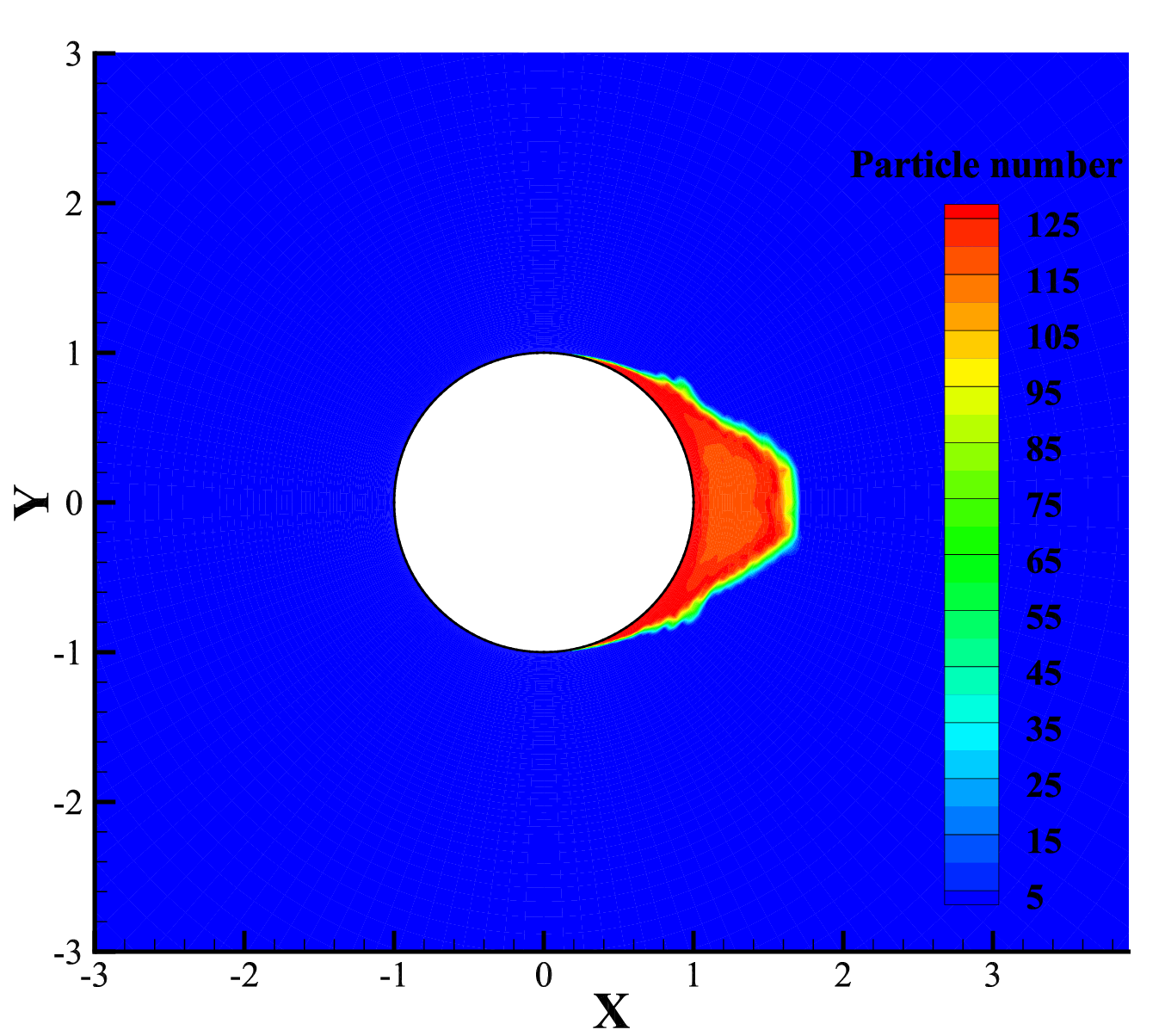}
    	}
    \subfigure[]{
    		\includegraphics[width=0.3 \textwidth]{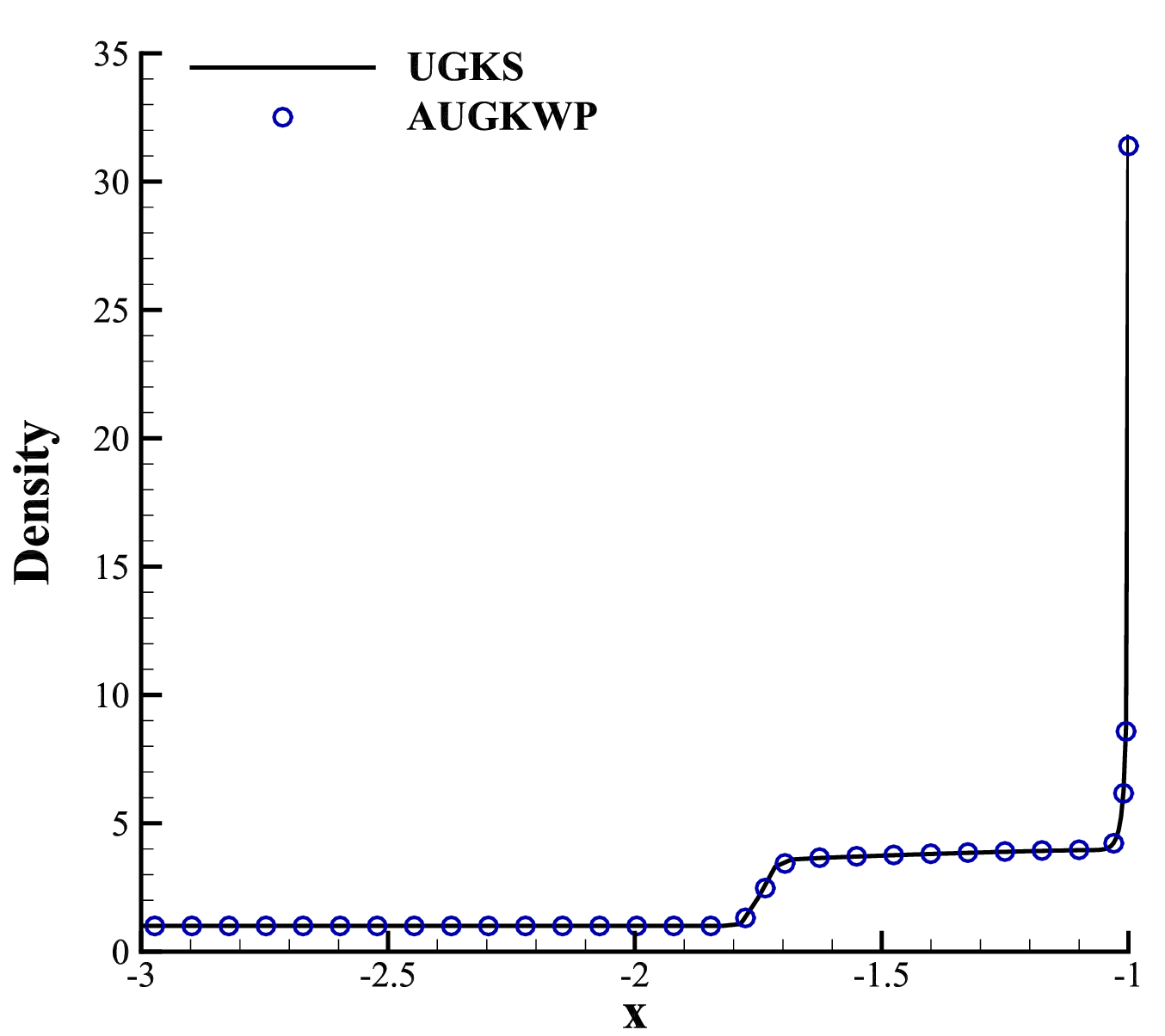}
    	}
    \subfigure[]{
			\includegraphics[width=0.3 \textwidth]{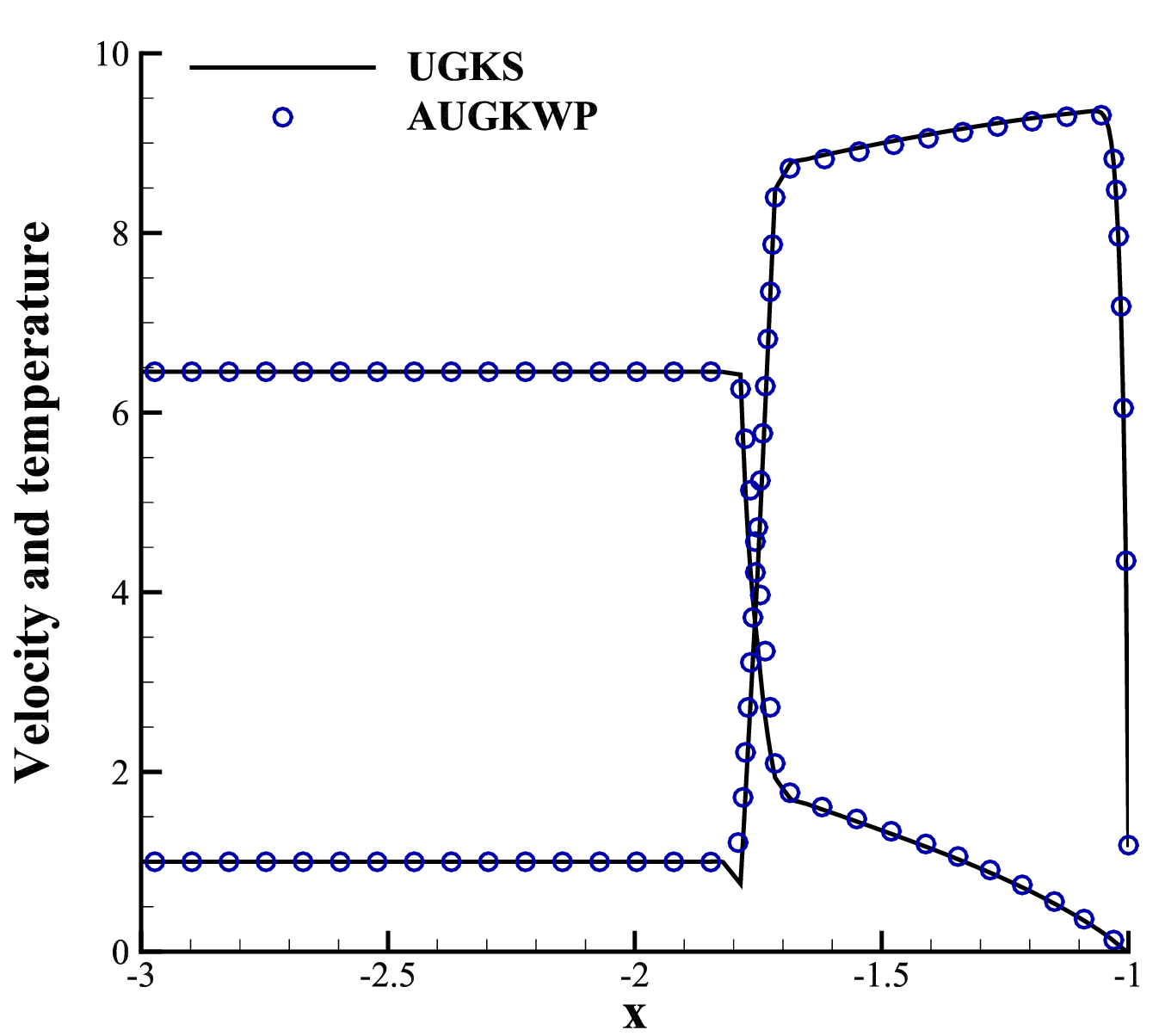}
		}
    \subfigure[]{
    		\includegraphics[width=0.3 \textwidth]{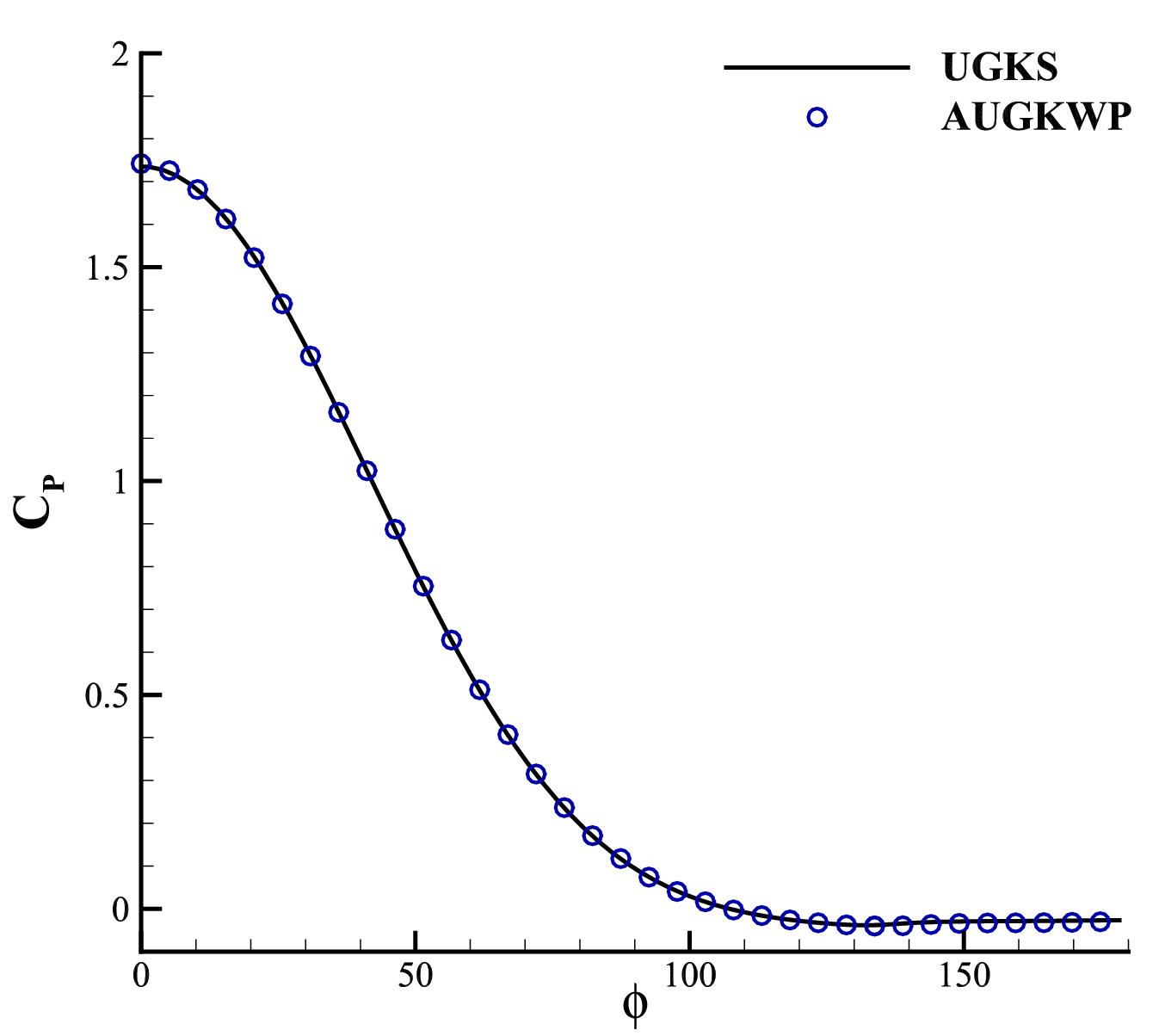}
    	}
    \subfigure[]{
    		\includegraphics[width=0.3 \textwidth]{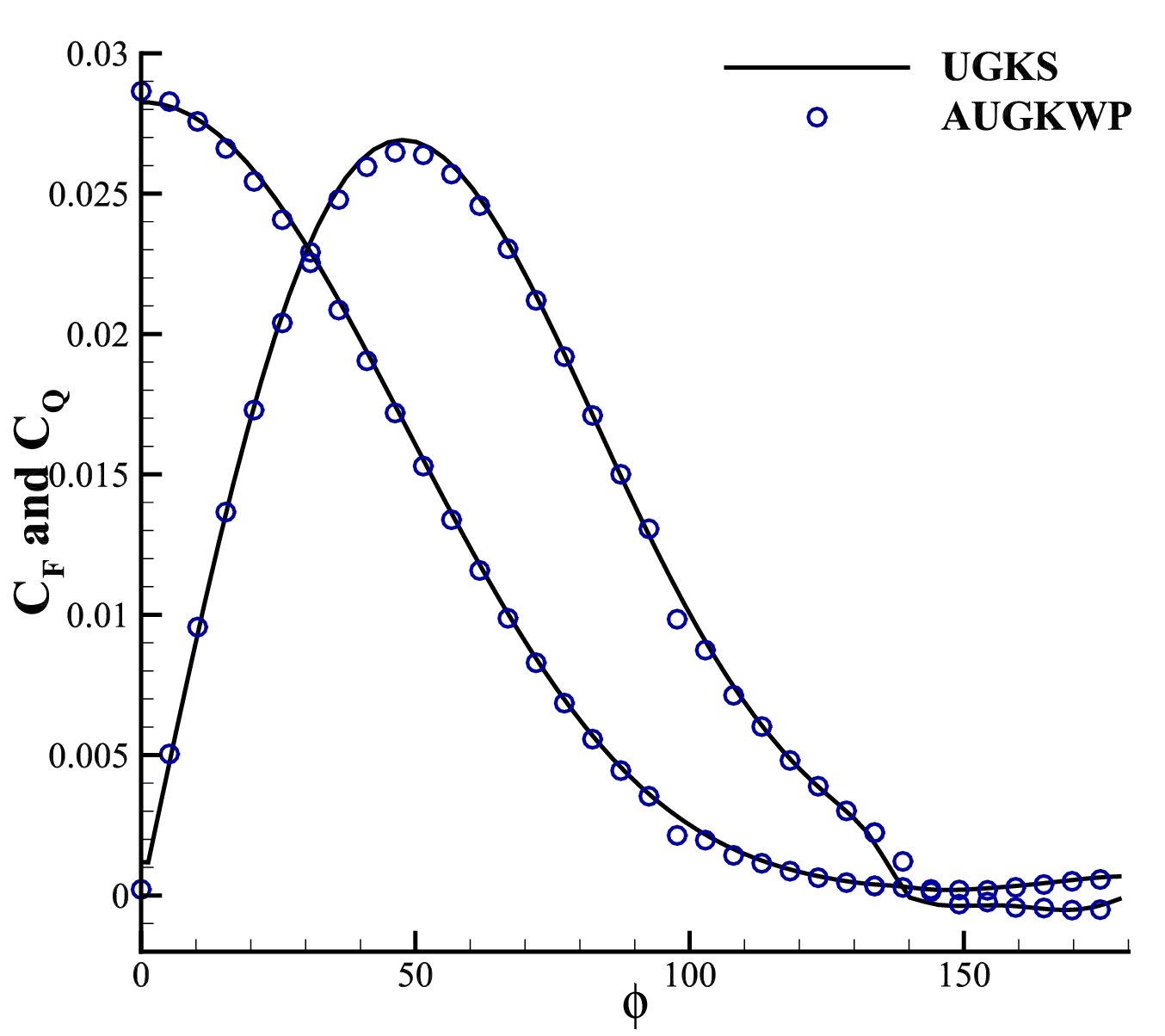}
    	}
	\caption{\label{cylinder_kn0.001} Results of AUGKWP method at ${\rm{Kn_{\infty}}}=0.001$: (a) Temperature contour, (b) particle number contour, (c) density along the stagnation line, (d) velocity and temperature along the stagnation line, (e) pressure coefficient at the wall, (f) shear stress and heat flux coefficients at the wall.}
\end{figure}

\begin{figure}[H]
	\centering
	\subfigure[]{
			\includegraphics[width=0.22 \textwidth]{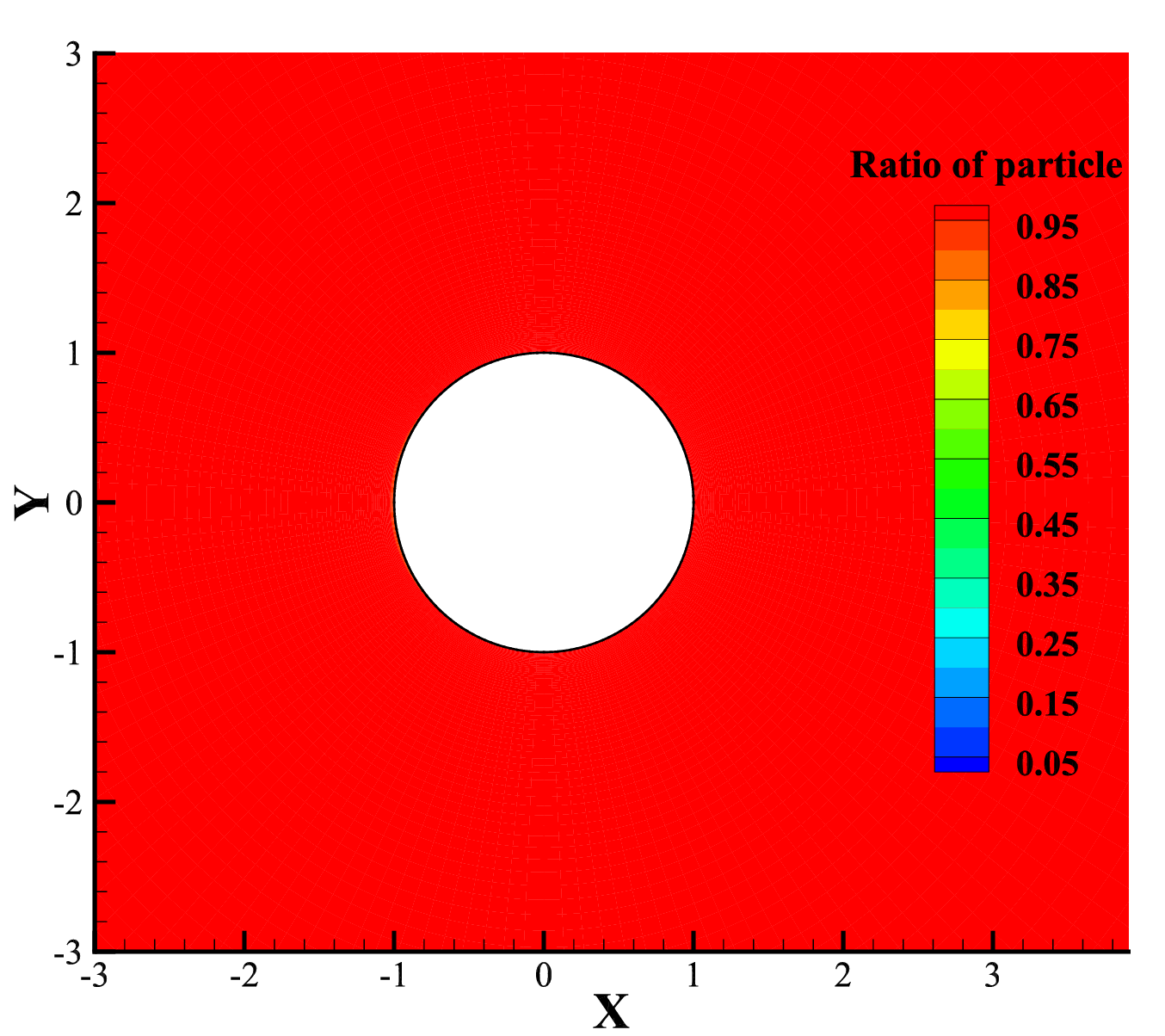}
		}
    \subfigure[]{
    		\includegraphics[width=0.22 \textwidth]{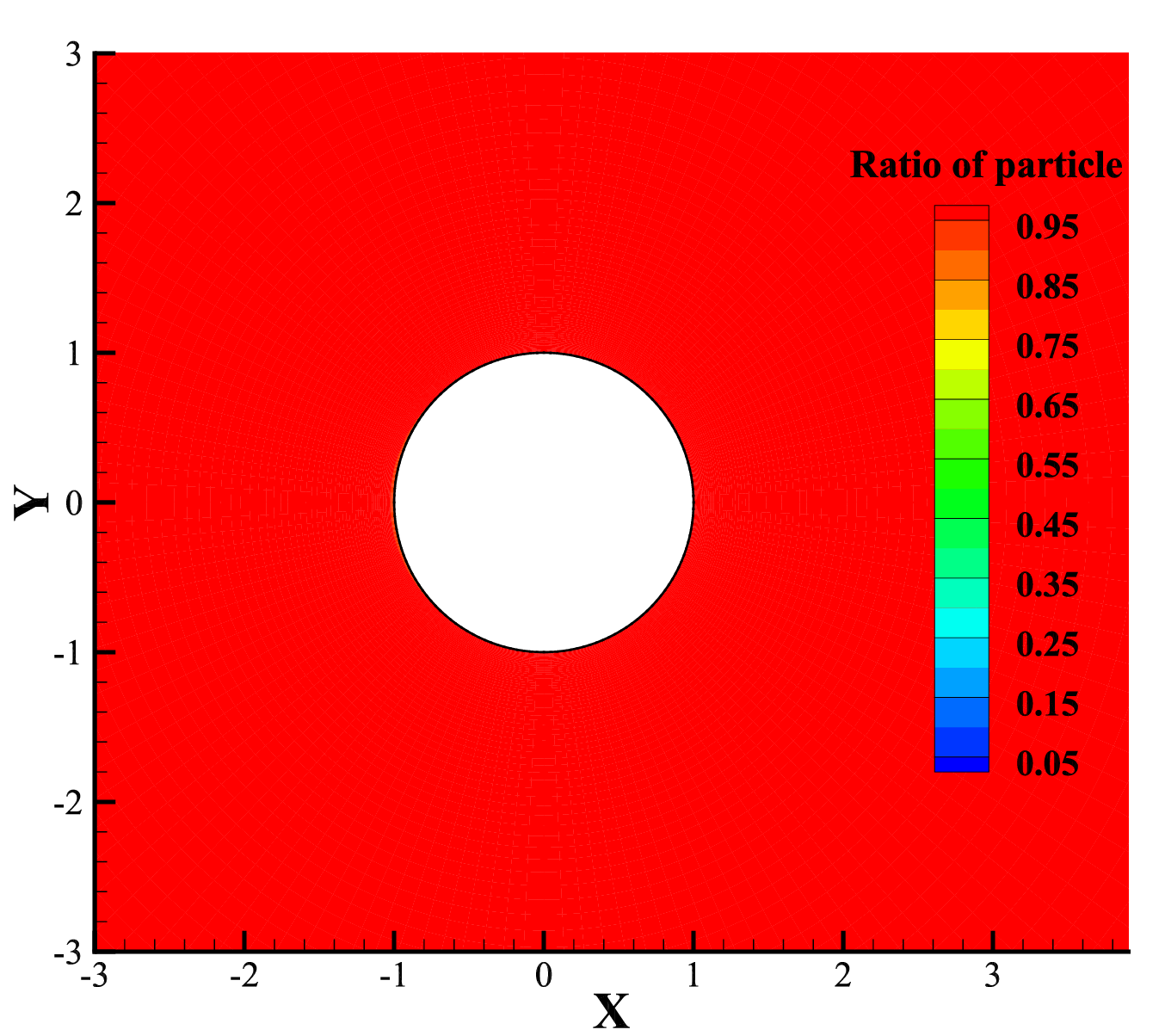}
    	}
    \subfigure[]{
    		\includegraphics[width=0.22 \textwidth]{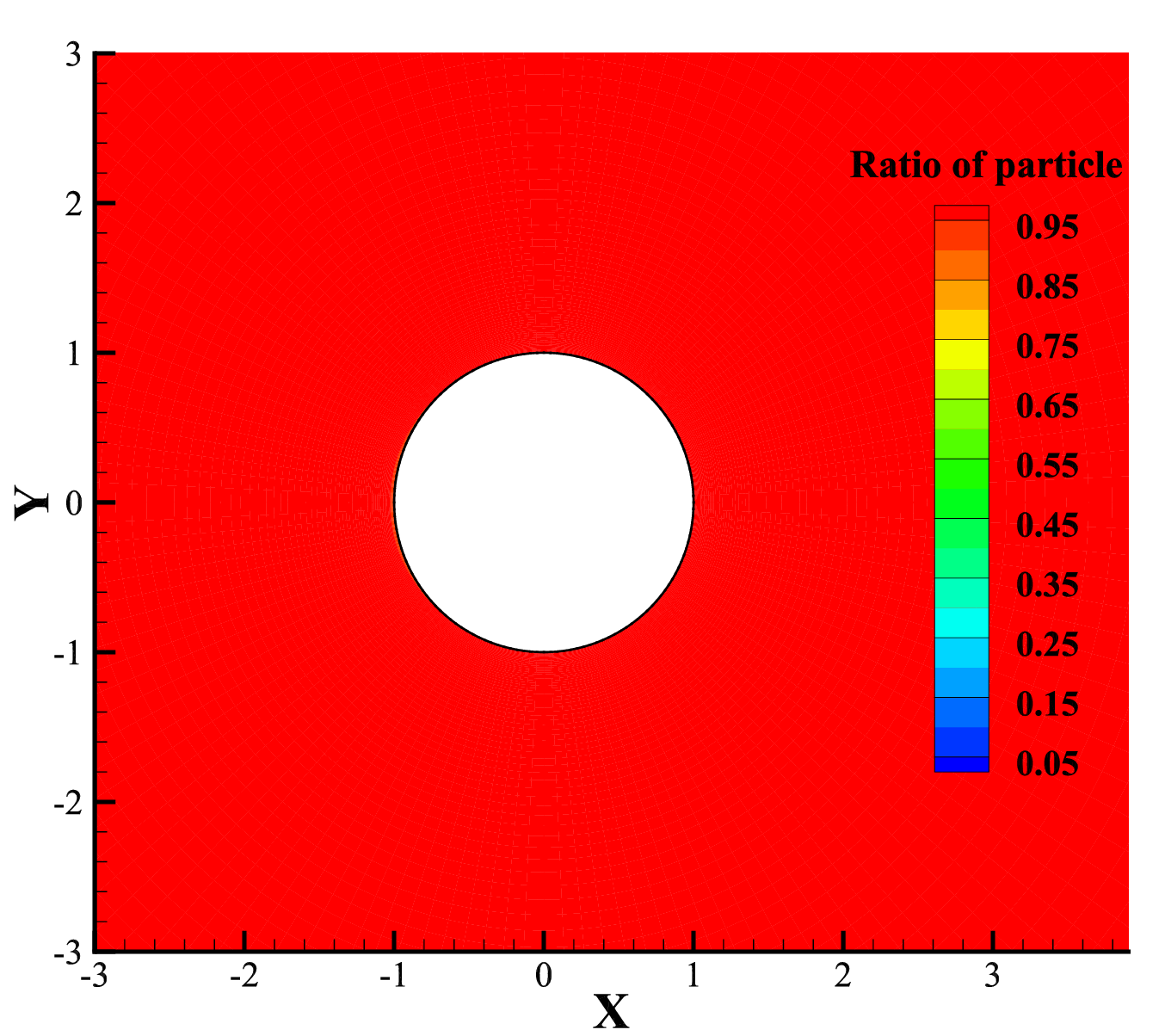}
    	}
    \subfigure[]{
    		\includegraphics[width=0.22 \textwidth]{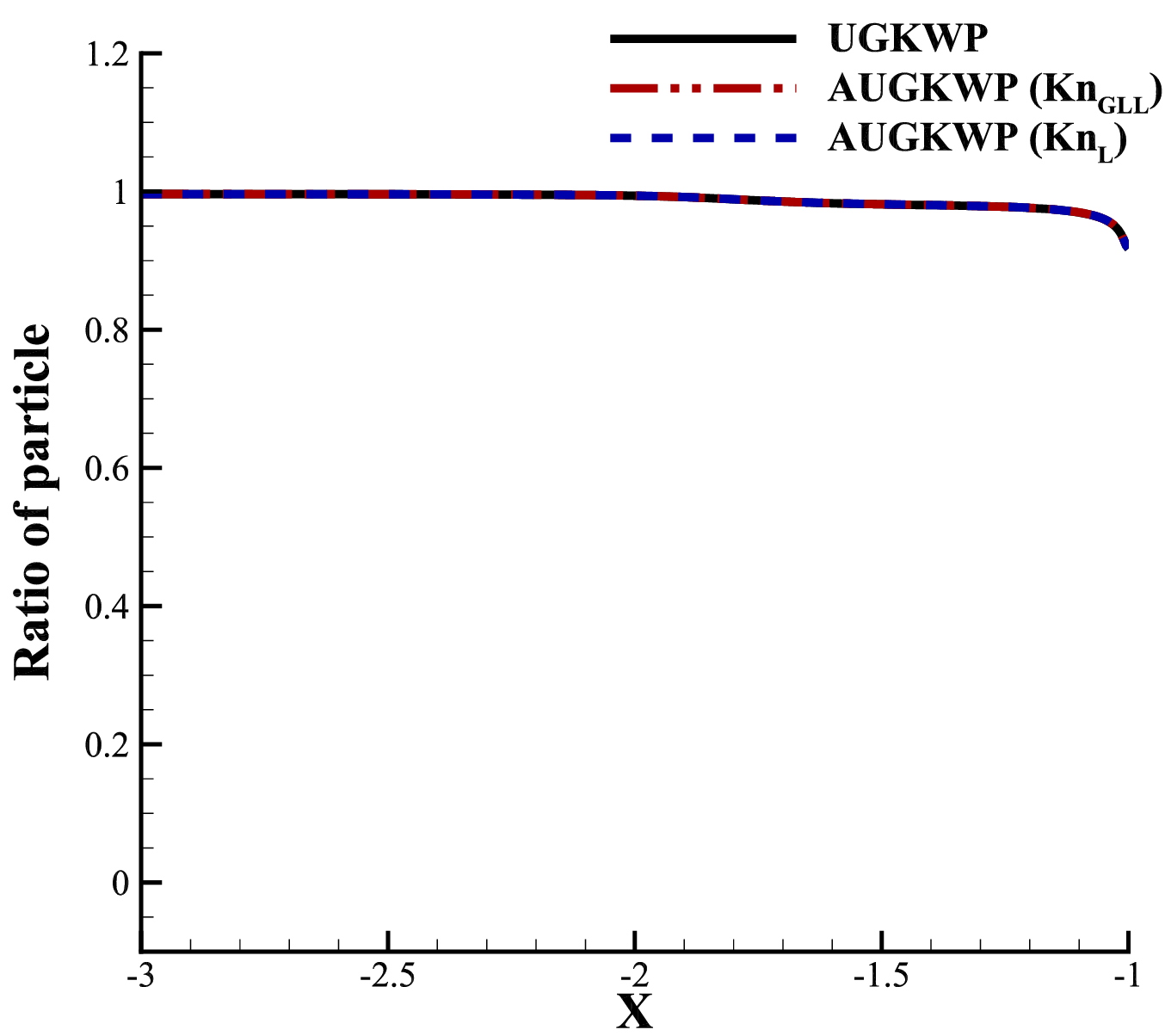}
    	}
	\caption{\label{cylinder_kn0.1_rop} Comparison of ratio of particles at ${\rm{Kn_{\infty}}}=0.1$: (a) Original UGKWP method, (b) AUGKWP method with ${\rm{Kn_{GLL}}}$, (c) AUGKWP method with ${\rm{Kn_{L}}}$, (d) details along the stagnation line of three approaches.}
\end{figure}

Then ${\rm{Kn_{\infty}}}$ is decreased to $0.001$ for a more continuum case. As shown in Fig.~\ref{cylinder_kn0.001_rop_c}, most regions are identified to be continuum by $\rm{Kn_L}$ except for a small part of the wake area. However, in the original UGKWP method, particles are still employed almost in the whole domain as Fig.~\ref{cylinder_kn0.001_rop_a}, because the global time step is so tiny. If only using ${\rm{Kn_{GLL}}}$ as Fig.~\ref{cylinder_kn0.001_rop_b}, many particles will be employed around the shock and boundary layer because of the huge density gradient. Since the mesh within the boundary layer is very precise, ${\rm{Kn_{L}}}$ saves many particles there, compared with ${\rm{Kn_{GLL}}}$. Meanwhile, even at the wake area, more particles are released by using ${\rm{Kn_{GLL}}}$ than by ${\rm{Kn_L}}$, mainly because the stochastic noise has a rather larger influence on ${\rm{Kn_{GLL}}}$ than on ${\rm{Kn_L}}$. Detailed exhibitions and comparisons are as Fig.~\ref{cylinder_kn0.001}, where results of AUGKWP method match well with the UGKS. Finally, ${\rm{Kn_{\infty}}}$ is increased to $0.1$ for a rarefied case. In this case, AUGKWP method brings about no improvement on efficiency any more. As Fig.~\ref{cylinder_kn0.1_rop}, particles are employed almost everywhere in all of three approaches. And the results of AUGKWP method are accurate compared with the UGKS.

Additionally, the iteration times are shown in Tab.~\ref{cylinder}. The total particle number is also given for an efficiency comparison of the three approaches. Since the efficiency loss is huge due to the distributed parallelism, the simulation is restarted by a single core, and the average CPU time per one step is shown in Tab.~\ref{cylinder} as well. We notice that at the low ${\rm{Kn_{\infty}}}$ in this case, the AUGKWP method brings about a $2$ to $3$ times increase in efficiency.

\begin{figure}[H]
	\centering
	\subfigure[]{
			\includegraphics[width=0.3 \textwidth]{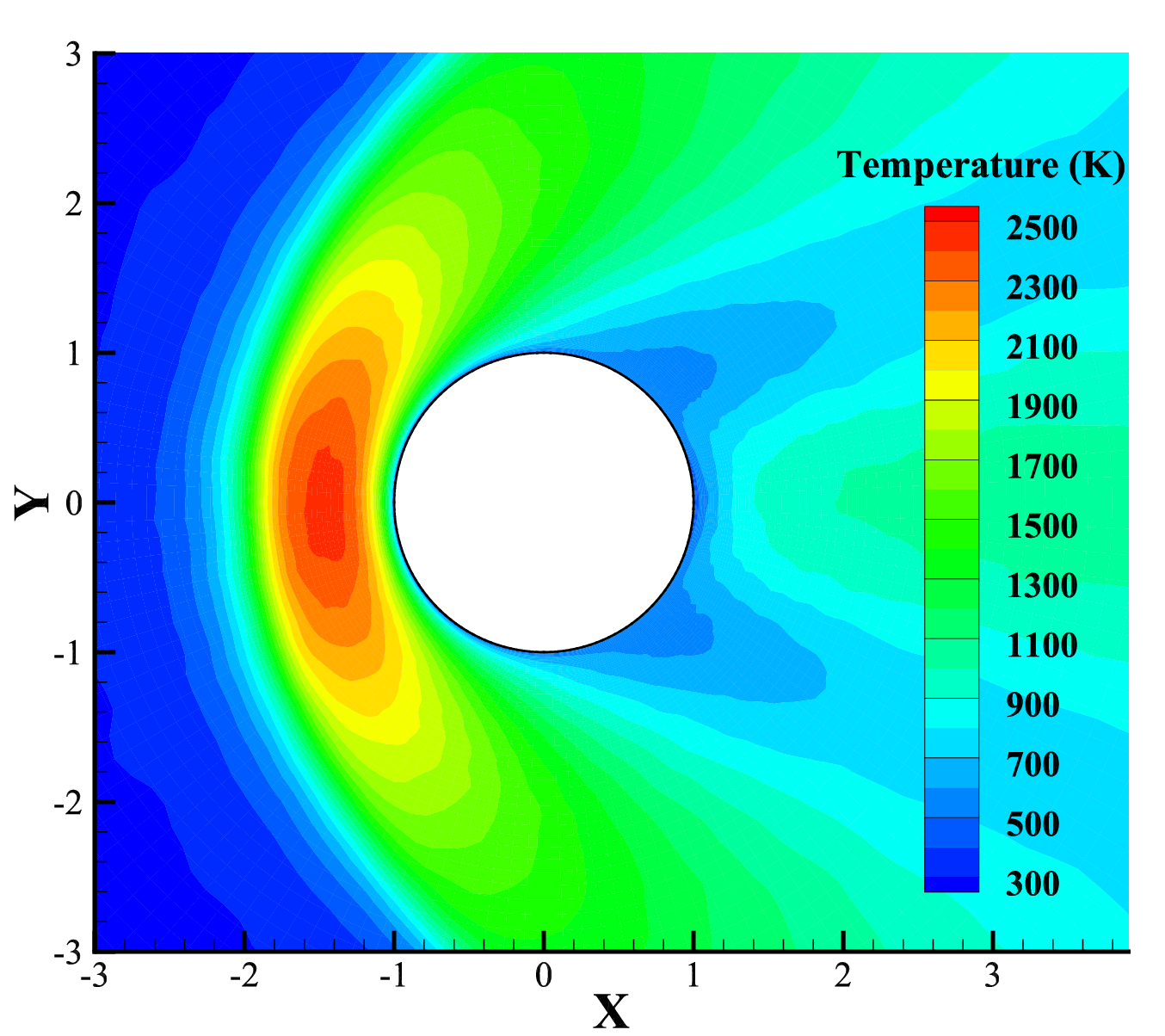}
		}
    \subfigure[]{
    		\includegraphics[width=0.3 \textwidth]{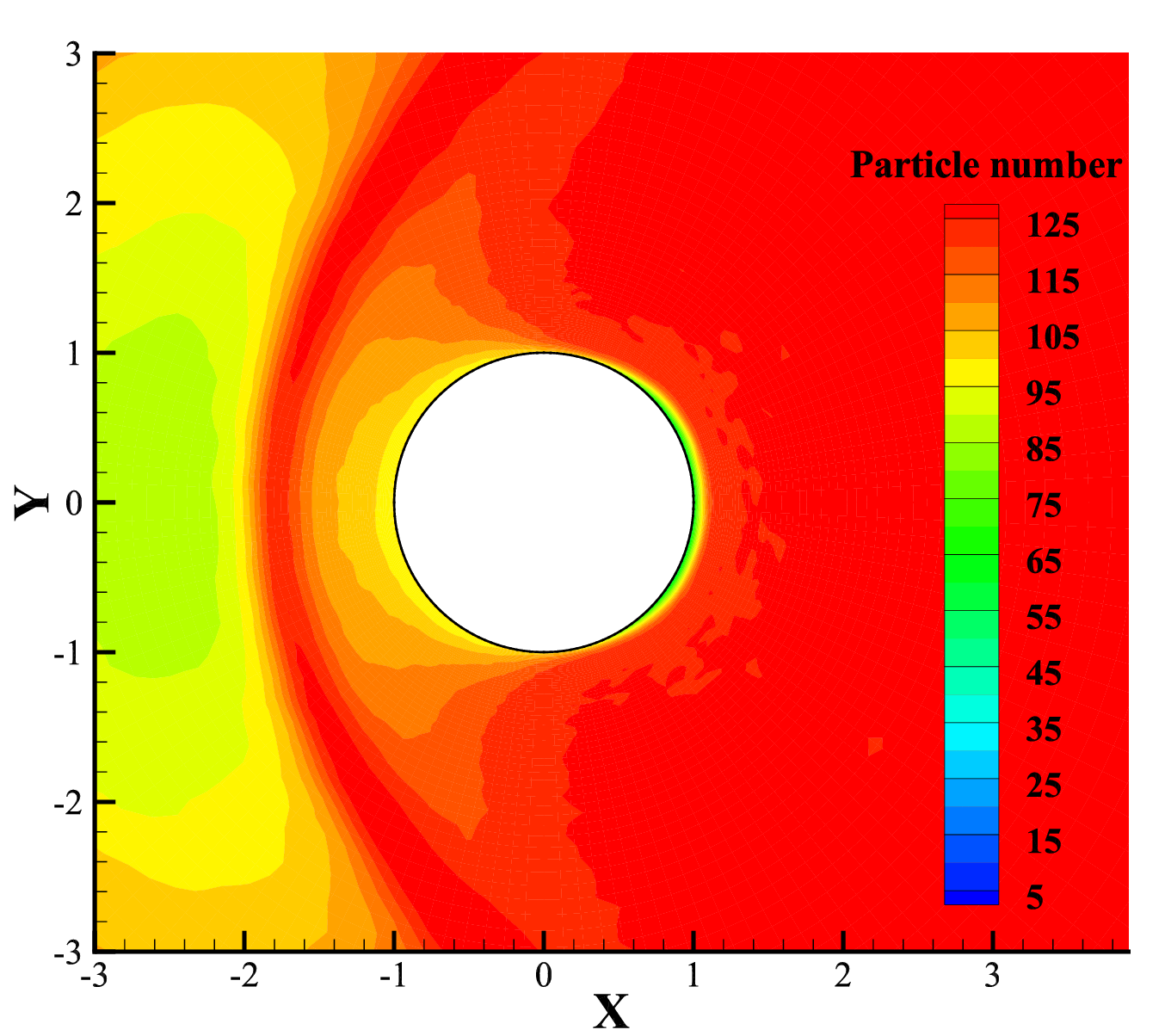}
    	}
    \subfigure[]{
    		\includegraphics[width=0.3 \textwidth]{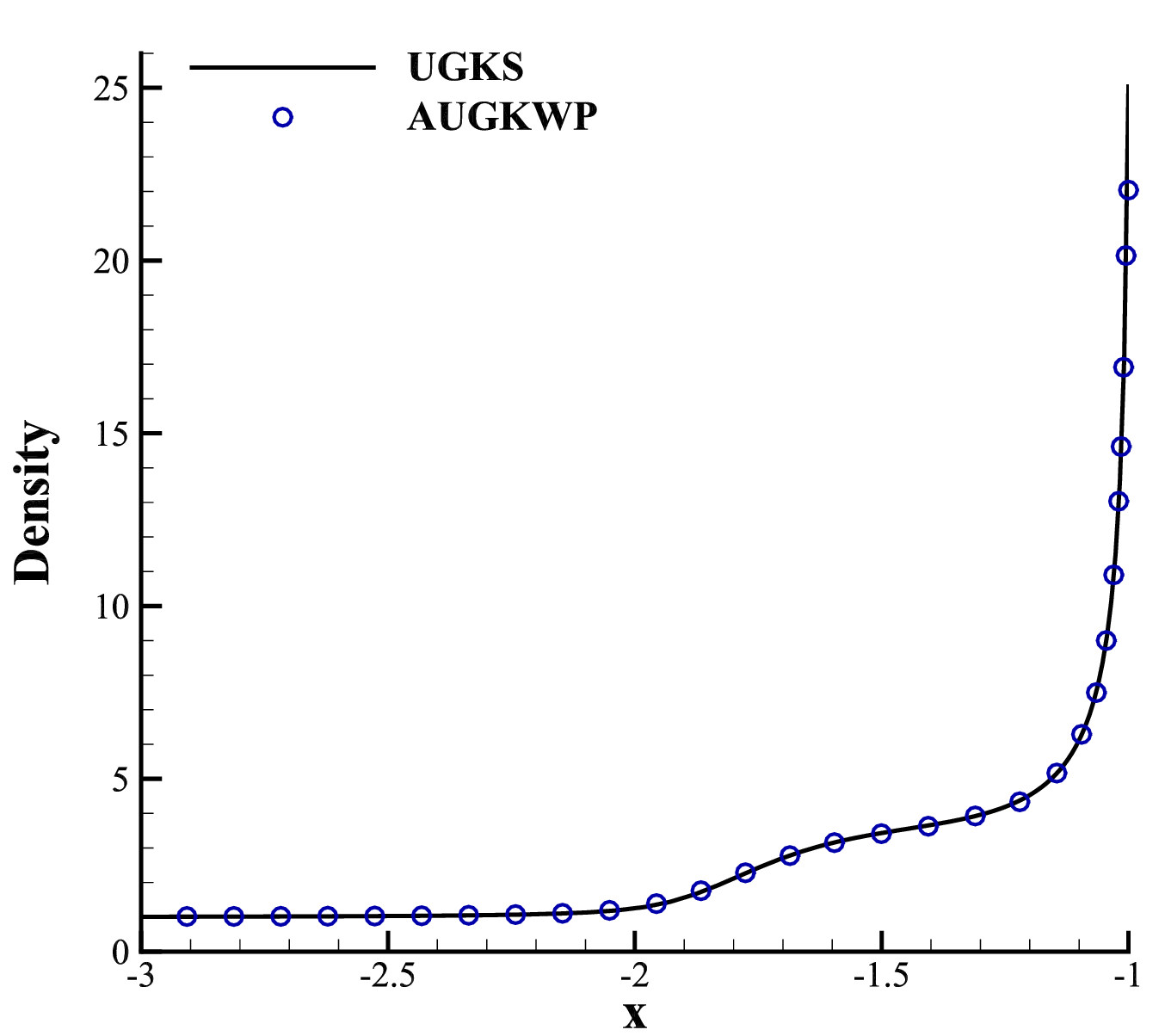}
    	}
    \subfigure[]{
			\includegraphics[width=0.3 \textwidth]{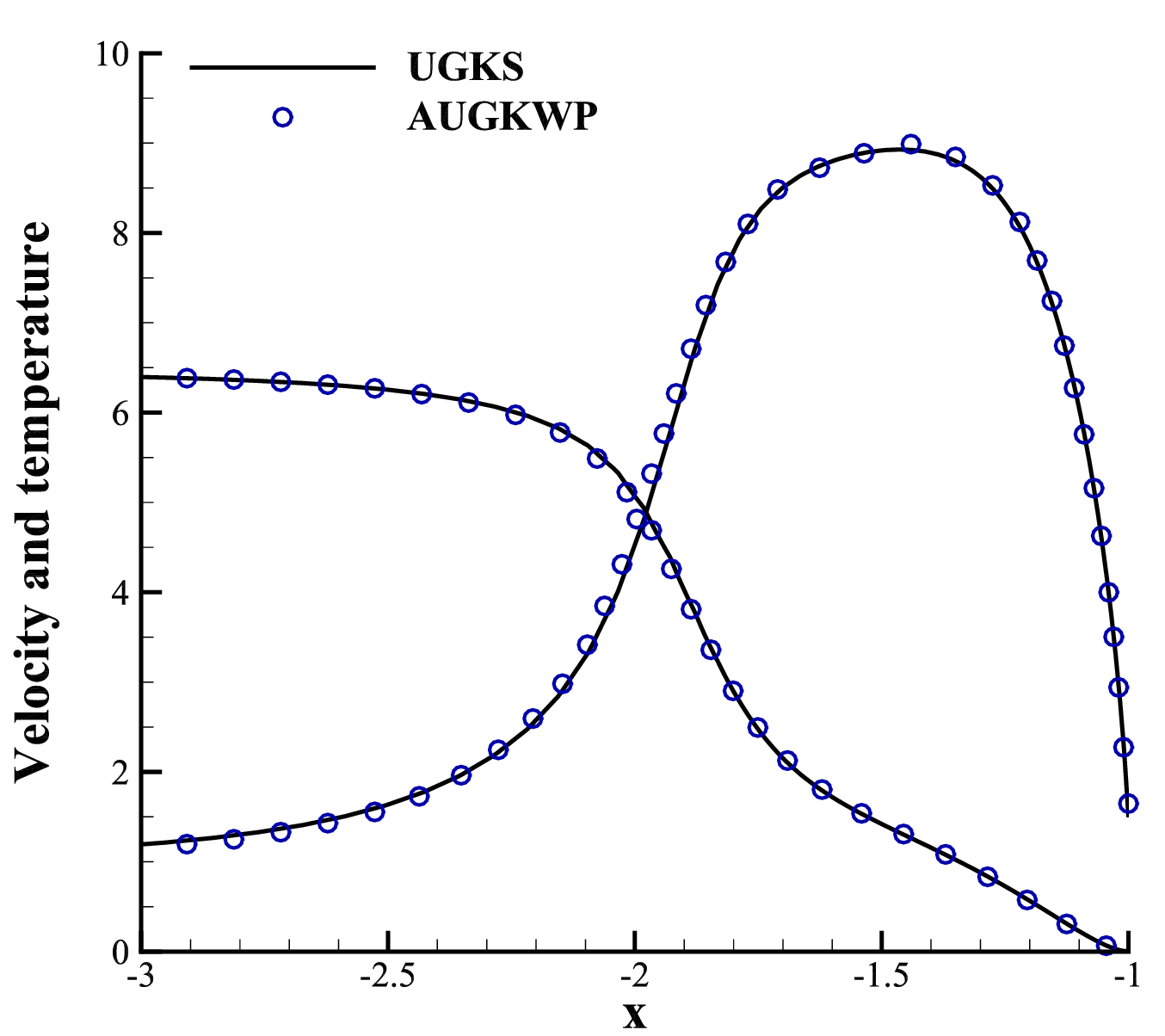}
		}
    \subfigure[]{
    		\includegraphics[width=0.3 \textwidth]{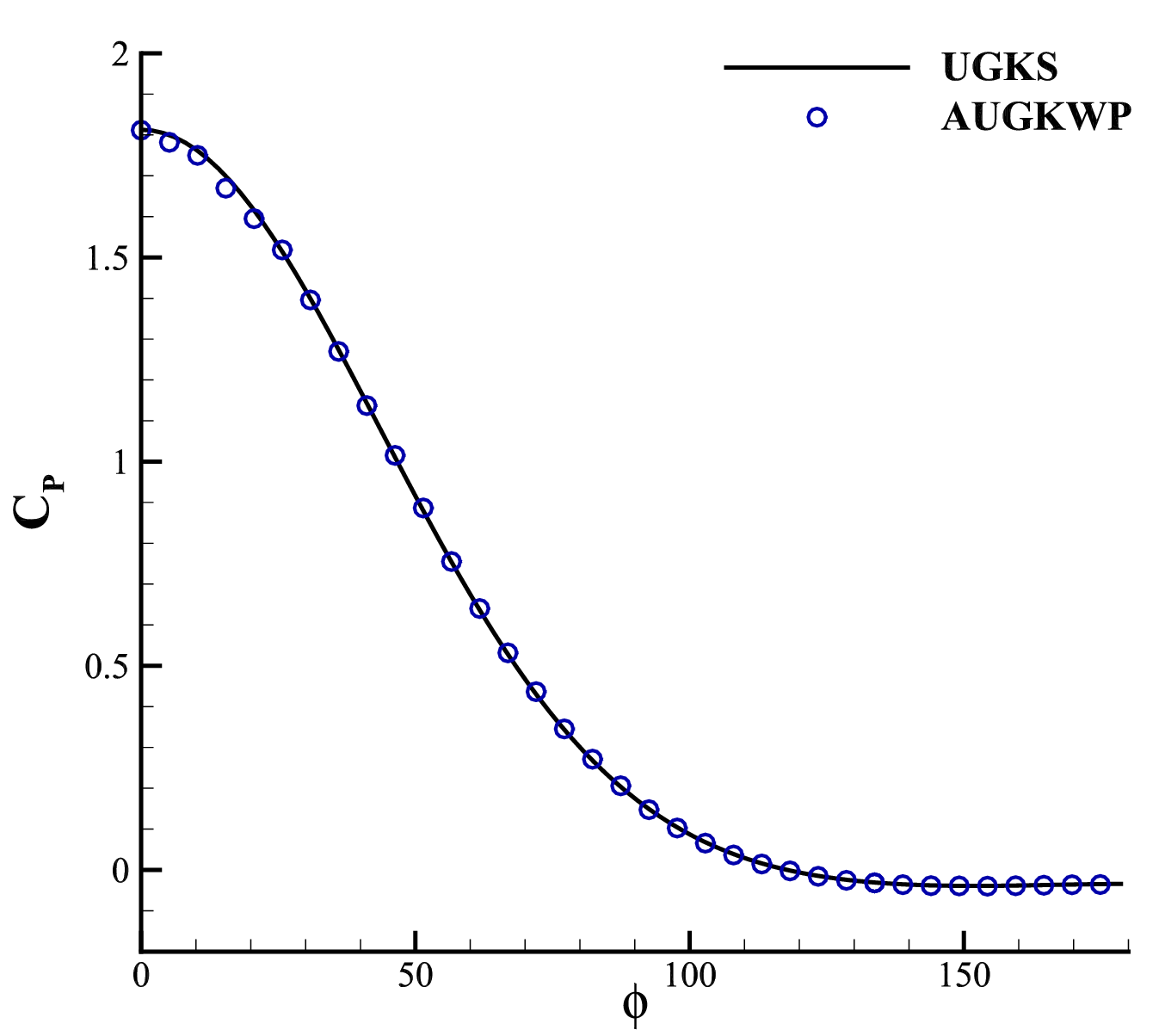}
    	}
    \subfigure[]{
    		\includegraphics[width=0.3 \textwidth]{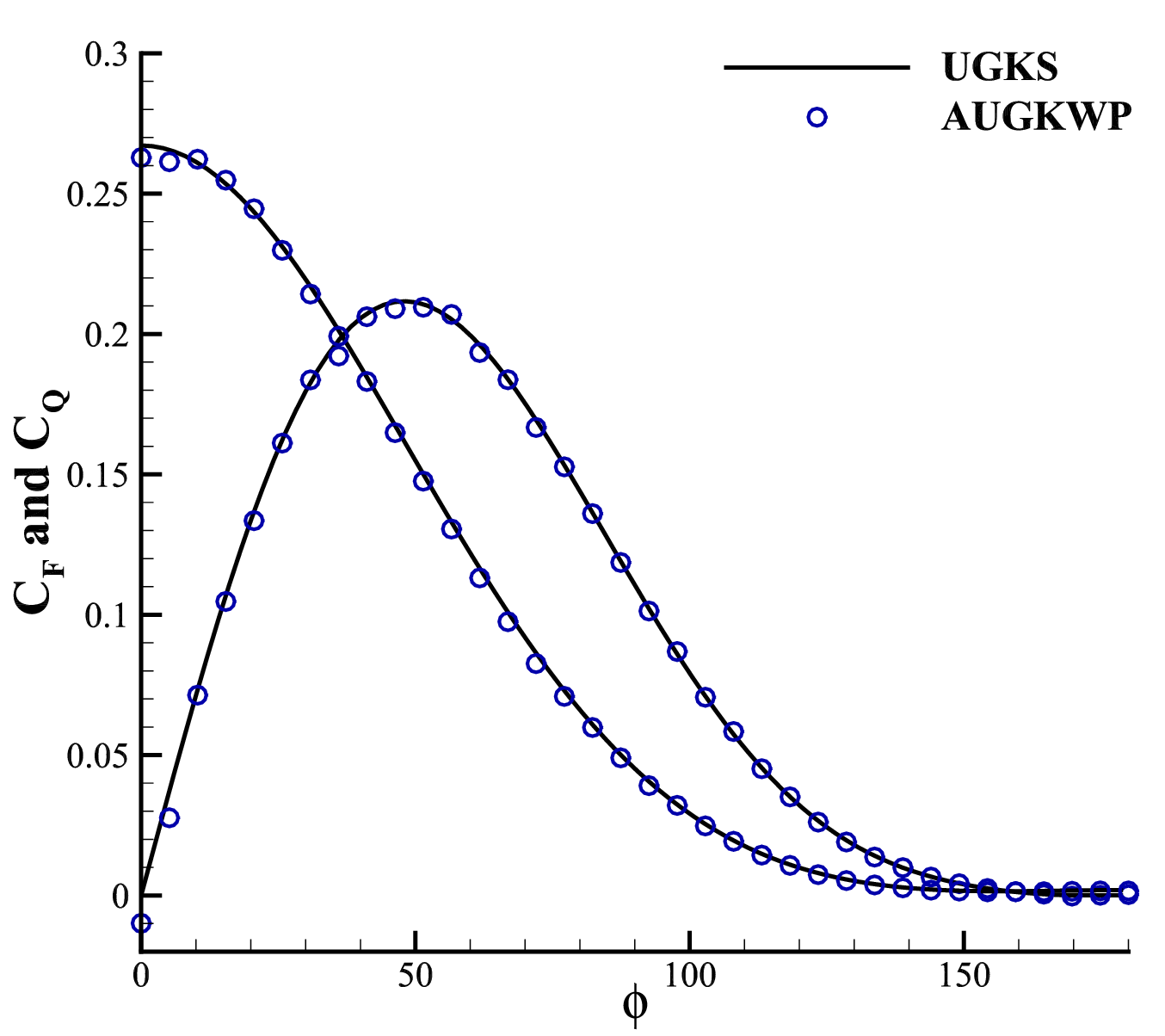}
    	}
	\caption{\label{cylinder_kn0.1} Results of AUGKWP method at ${\rm{Kn_{\infty}}}=0.1$: (a) Temperature contour, (b) particle number contour, (c) density along the stagnation line, (d) velocity and temperature along the stagnation line, (e) pressure coefficient at the wall, (f) shear stress and heat flux coefficients at the wall.}
\end{figure}

\begin{table}[H]
\centering
\caption{\label{cylinder} Efficiency comparison in the simulation of hypersonic flow around a cylinder}
\begin{tabular}{|c|c|c|c|c|}
\hline
    ${\rm{Kn_{\infty}}}$   &Steps (k)                       &Approach                  &Particle number (k) &CPU time per $1$ step$^{\rm a}$ (s) \\ \hline
	\multirow{3}*{$0.1$}   &\multirow{3}*{$30+10^{\rm b}$}  &UGKWP                     &$1246$              &$1.63$                              \\
    ~                      &~                               &AUGKWP(${\rm{Kn_{GLL}}}$) &$1113$              &$1.64$                              \\
    ~                      &~                               &AUGKWP(${\rm{Kn_{L}}}$)   &$1111$              &$1.64$                              \\ \hline
    \multirow{3}*{$0.01$}  &\multirow{3}*{$60+10^{\rm b}$}  &UGKWP                     &$1494$              &$1.84$                              \\
    ~                      &~                               &AUGKWP(${\rm{Kn_{GLL}}}$) &$1219$              &$1.64$                              \\
    ~                      &~                               &AUGKWP(${\rm{Kn_{L}}}$)   &$643$               &$1.10$                              \\ \hline
    \multirow{3}*{$0.001$} &\multirow{3}*{$320+10^{\rm b}$} &UGKWP                     &$1483$              &$2.24$                              \\
    ~                      &~                               &AUGKWP(${\rm{Kn_{GLL}}}$) &$741$               &$1.35$                              \\
    ~                      &~                               &AUGKWP(${\rm{Kn_{L}}}$)   &$313$               &$0.83$                              \\
\hline
\end{tabular}
\begin{tablenotes}
\item{$^{\rm a}$ Tested with only one core.}\\
\item{$^{\rm b}$ For time average.}
\end{tablenotes}
\end{table}

\subsection{Hypersonic flow over a slender cavity}\label{sec:slender}
The hypersonic flow over a slender cavity is an important case in hypersonic vehicles. As shown in Ref.~\cite{palharini2018}, a large number of individual thermal protections system (TPS) tiles are used on the vehicle surface, which create numerous panel-to-panel joints, such as cavities, gaps and steps. An accurate prediction of thermal load can help to design the TPS. Moreover, as Ref.~\cite{slenderexp}, in response to the accident of the Space Shuttle Columbia in $2003$, a TPS damage assessment process was developed. The heating augmentation bump factors need to be calculated for cavity damage sites by ice, detached pieces of insulation, etc. Since the size of slender cavity is tiny, the local flow is rarefied even at a low attitude. In this work, the condition at $50~{\rm{km}}$ attitude is taken for simulation, as Ref.~\cite{guo2024ast}, $\rho_{\infty}=1.027\times10^{-3}~{\rm{kg/m^3}}$, $T_{\infty}=270.65~{\rm{K}}$, $p_{\infty}=79.779~{\rm{Pa}}$, $\lambda_{\infty}=7.913\times10^{-5}~{\rm{m}}$. Inflow ${\rm{Ma_{\infty}}}$ is set to be $20$. The angle of attack of freestream is $0$ degree. The temperature at the wall is set to be $300~{\rm{K}}$. The gas is set to be diatomic, with $\gamma=1.4$. The Shakhov model is used as the kinetic model with $Pr=0.72$. The dynamic viscosity coefficient is calculated as $\mu\sim T^{\omega}$ and $\omega$ is set to be $0.77$ according to the parameter of air~\cite{dsmcre}. The depth-to-width ratio of the slender cavity is set to be $3$, the width is $2mm$ and the flat length before the slender cavity is $40mm$. Though the attitude is $50~{\rm{km}}$, ${\rm{Kn_{\infty}}}$ is about $0.04$ once taking the width of slender cavity as the reference length.

\begin{figure}[H]
	\centering
	\subfigure[]{
			\includegraphics[width=0.3 \textwidth]{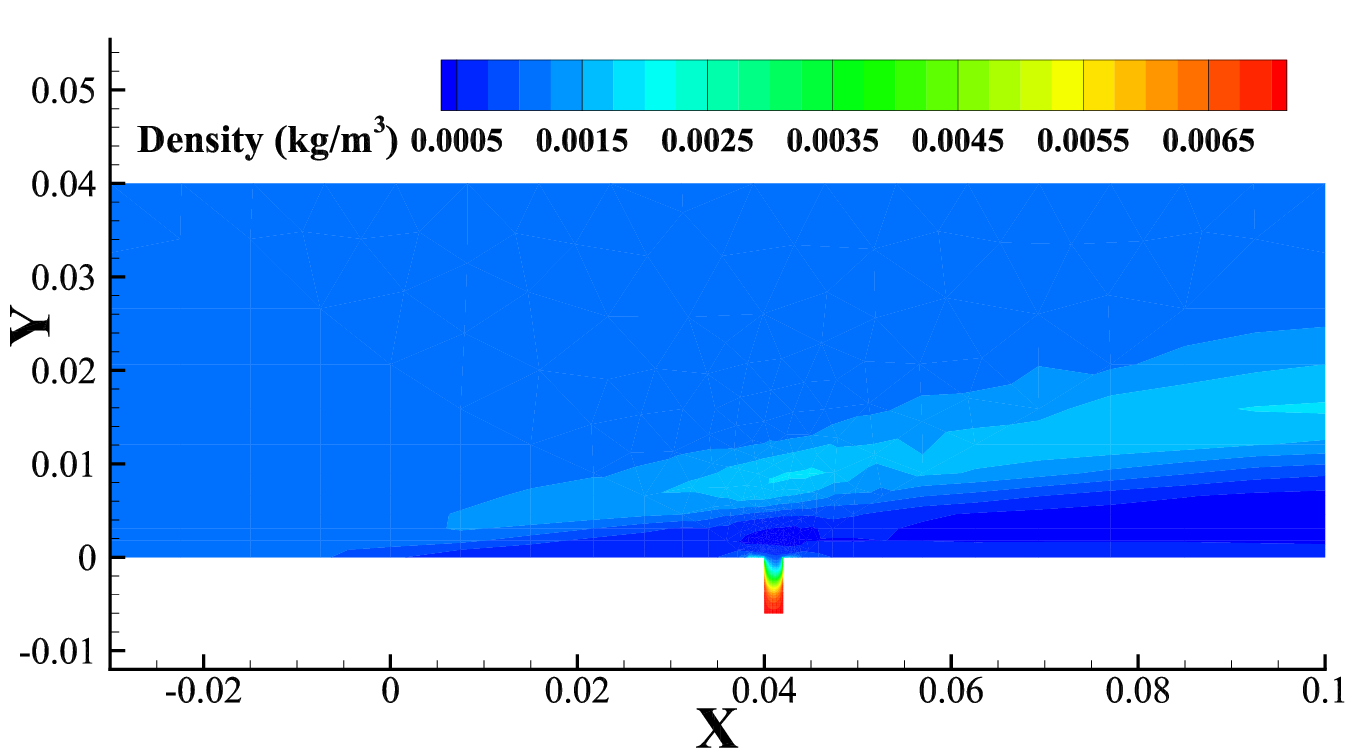}
		}
    \subfigure[]{
    		\includegraphics[width=0.3 \textwidth]{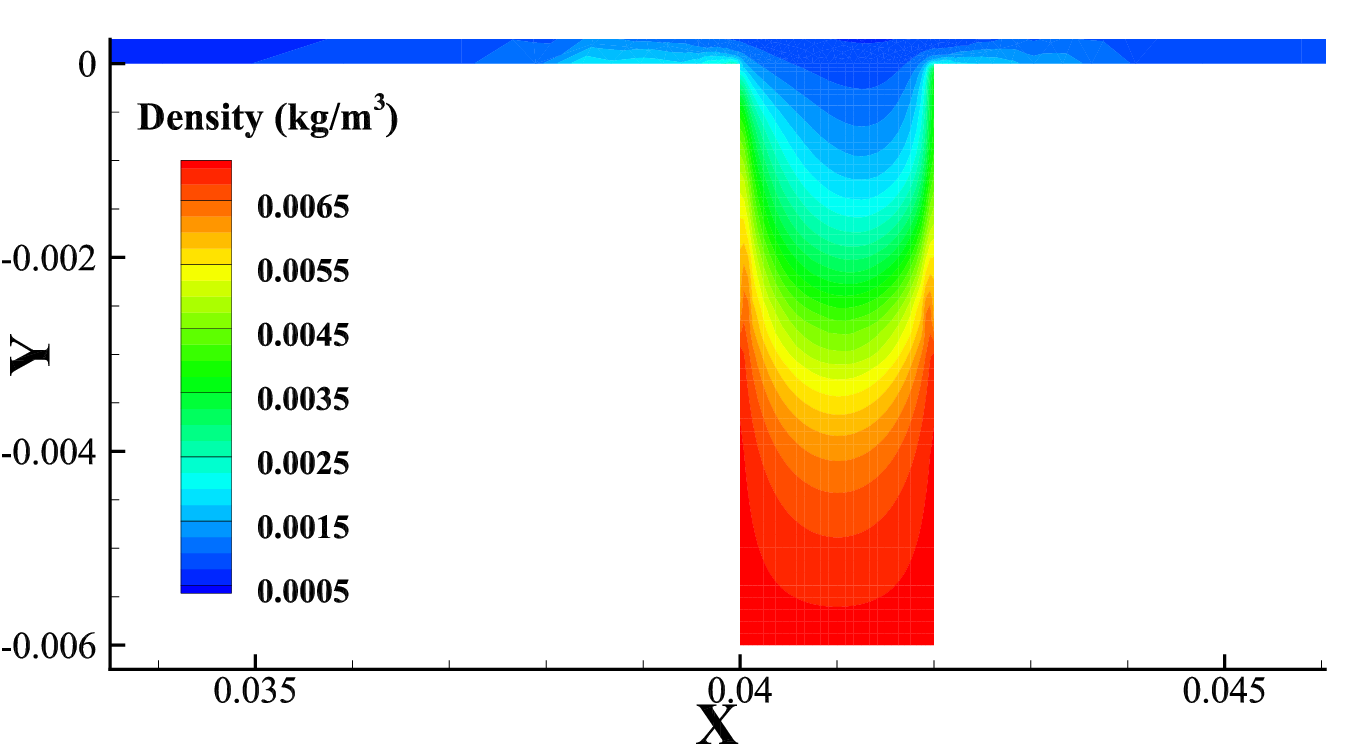}
    	}
    \subfigure[]{
    		\includegraphics[width=0.3 \textwidth]{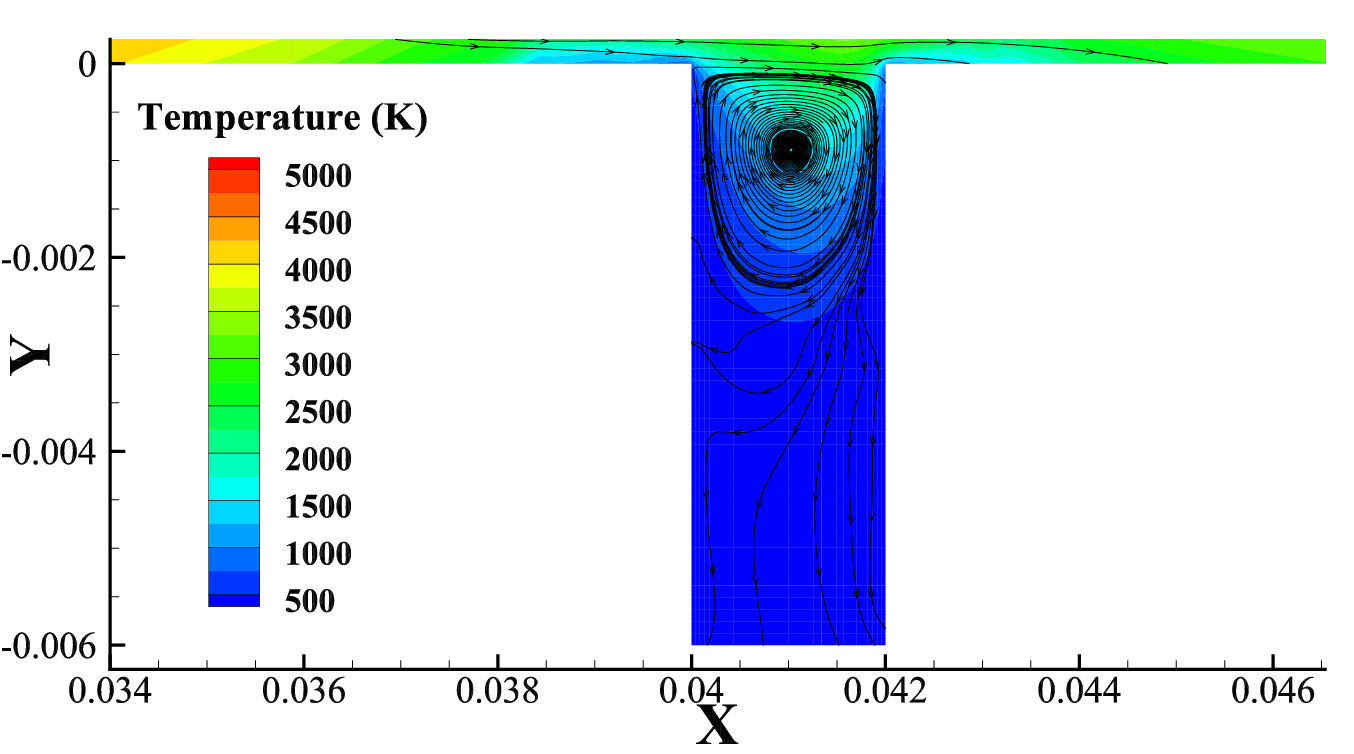}
    	}
	\caption{\label{slender_contour} Contours simulated by AUGKWP method (unit: m): (a) Density (global), (b) density (local), (c) temperature (local and streamline).}
\end{figure}

\begin{figure}[H]
	\centering
	\subfigure[]{
			\includegraphics[width=0.3 \textwidth]{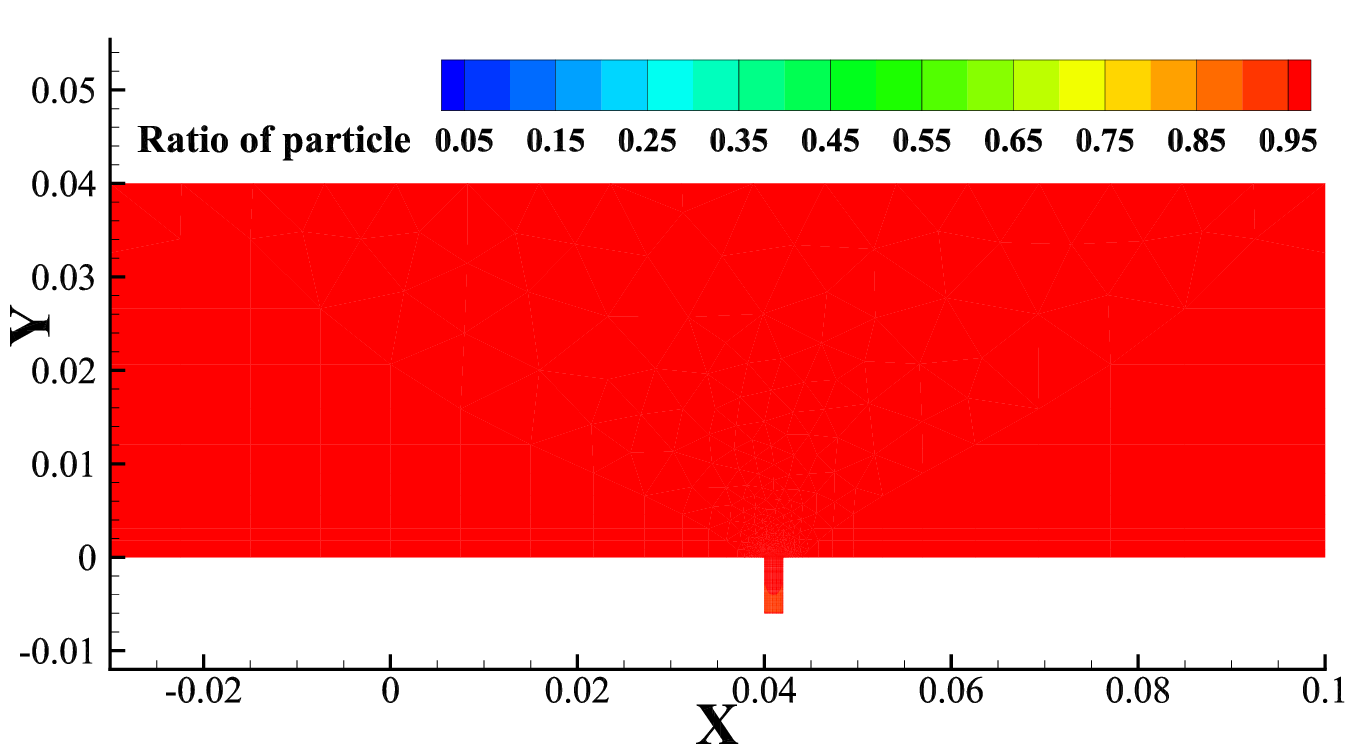}
		}
    \subfigure[]{
    		\includegraphics[width=0.3 \textwidth]{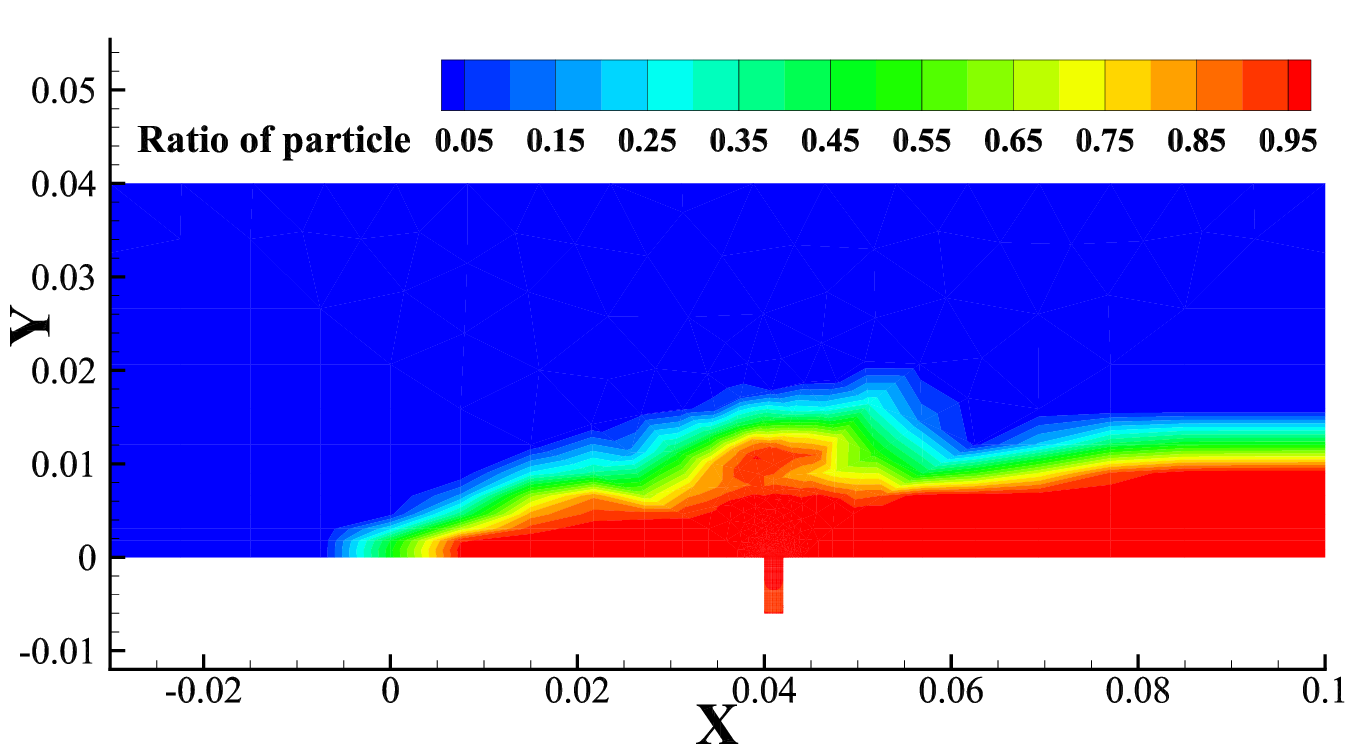}
    	}
    \subfigure[]{
    		\includegraphics[width=0.3 \textwidth]{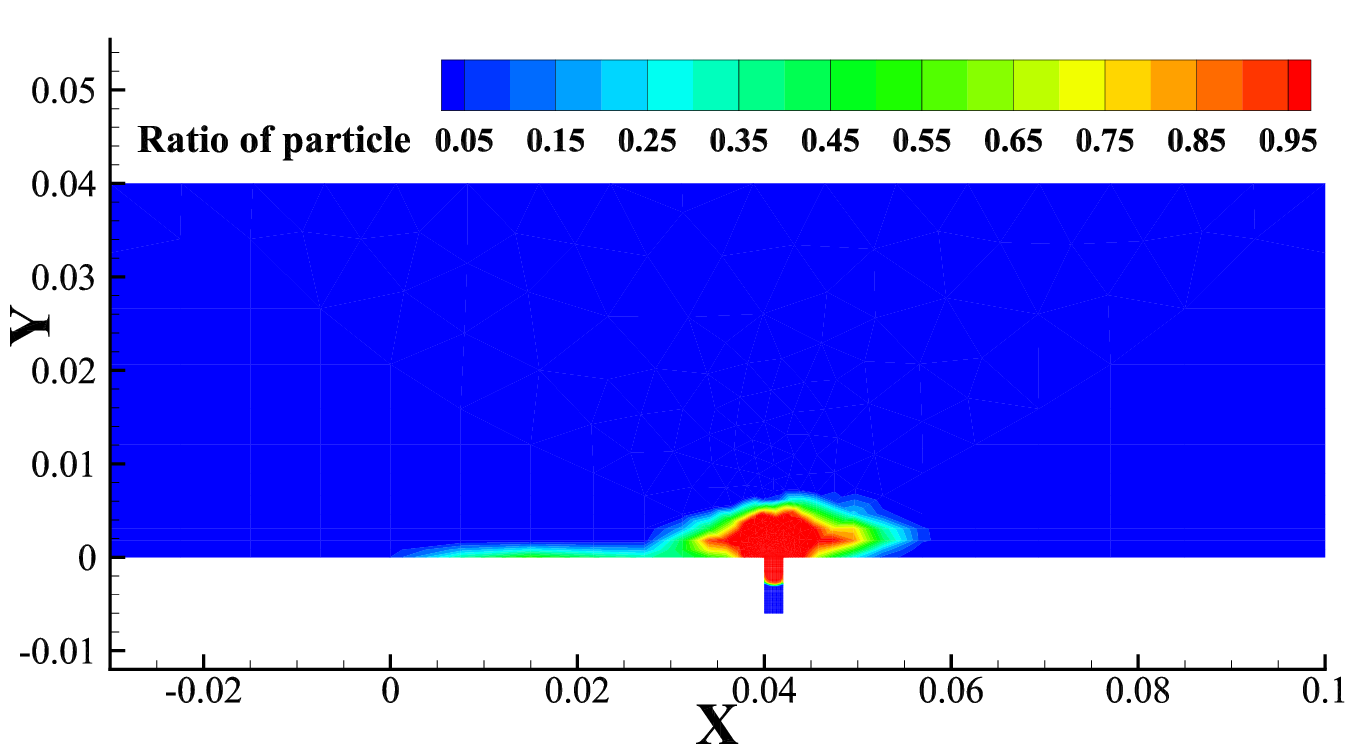}
    	}
	\caption{\label{slender_rop} Comparison of ratio of particles (unit: m): (a) Original UGKWP method, (b) AUGKWP method with ${\rm{Kn_{GLL}}}$, (c) AUGKWP method with ${\rm{Kn_{L}}}$.}
\end{figure}

\begin{figure}[H]
	\centering
	\subfigure[]{
			\includegraphics[width=0.45 \textwidth]{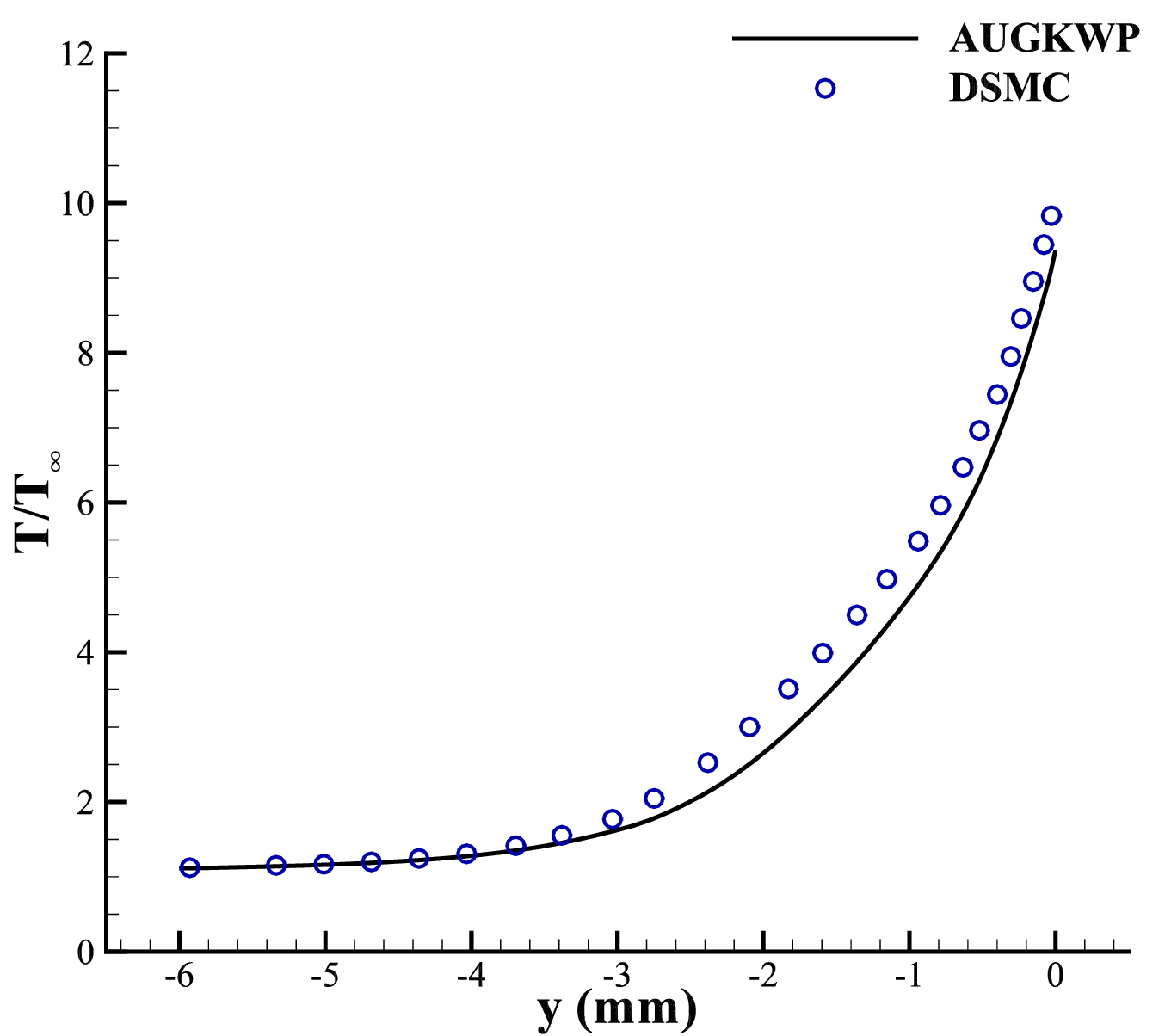}
		}
    \subfigure[]{
    		\includegraphics[width=0.45 \textwidth]{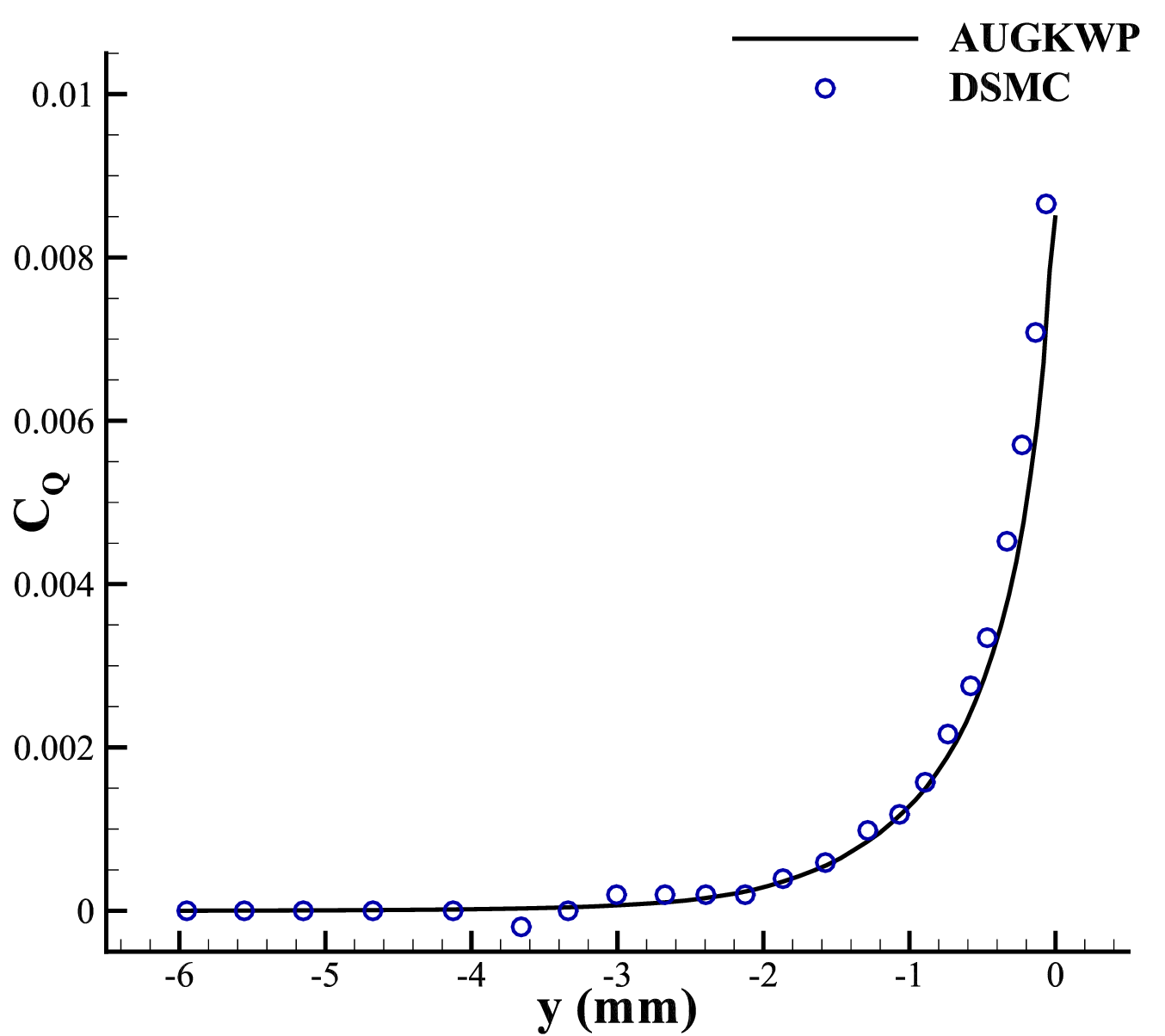}
    	}
	\caption{\label{slender_line} Results simulated by AUGKWP method: (a) Temperature at $x=0.041~{\rm{m}}$, (b) heat flux at wall at $x=0.042~{\rm{m}}$.}
\end{figure}

The simulation consists of $0.5$ million iteration steps, followed by another $0.1$ million steps for time-averaging. The density contour and temperature contour are shown in Fig.~\ref{slender_contour}. For a more obvious distinction of scale changing, the domain is set to be larger than Ref.~\cite{guo2024ast}. The particle ratio contours of three different approaches are shown in Fig.~\ref{slender_rop}. When using the original UGKWP method, almost the whole domain is identified to be rarefied since the smallest mesh is restricted by the size of the tiny slender cavity and the global $\Delta t$ is small as well. When using AUGKWP method with ${\rm{Kn_{L}}}$, particle are released in the vicinity of the cavity, mainly according to the mesh size there. Because the mass is accumulated at the bottom of the cavity, that regime is judged to be continuum by ${\rm{Kn_{L}}}$. The particle ratio contour calculated by AUGKWP method with ${\rm{Kn_{GLL}}}$ is also shown for comparison. Though particles are still saved in the farfield, they are released around the boundary layer and cavity because of the large density gradient there. Results of AUGKWP method are compared with DSMC~\cite{guo2024ast} as shown in Fig.~\ref{slender_line}, and $C_H$ is nondimensionalized by $0.5\rho_{\infty}\left|U\right|_{\infty}^3$. It is noticed that though multiple components (nitrogen and oxygen) are employed in the DSMC simulation, AUGKWP results match well with the DSMC. Additionally, total particle number is shown in Tab.~\ref{slender}, together with the average CPU time per one thousand steps. Compared with original UGKWP method, in this case, the AUGKWP method brings about a $1.64$ times increase in efficiency.

\begin{table}[H]
\centering
\caption{\label{slender} Efficiency comparison in the simulation of hypersonic flow over a slender cavity}
\begin{tabular}{|c|c|c|}
\hline
    Approach                  &Particle number (k) &CPU time per $1000$ steps$^{\rm a}$ (min) \\ \hline
	UGKWP                     &$315$               &$8.36$                                    \\
    AUGKWP(${\rm{Kn_{GLL}}}$) &$290$               &$7.99$                                    \\
    AUGKWP(${\rm{Kn_{L}}}$)   &$194$               &$5.10$                                    \\
\hline
\end{tabular}
\begin{tablenotes}
\item{$^{\rm a}$ Tested with only one core.}
\end{tablenotes}
\end{table}

\subsection{Side-jet impingement on hypersonic flow}\label{sec:sidejet}
Side jet is widely used in the reaction control system to improve the controllability of high-${\rm{Ma_{\infty}}}$ vehicles. Similar to case in Sec.~\ref{sec:gkt}, the observation scale is locally decreased by minor perturbation. But since the density of jet flow is rather higher than the inflow, the local rarefied region is much smaller. In this case, the side jet is placed in the middle of a wedge, vertically with the wall. The half angle of the wedge is $7$ degree. The height of the wedge is $r_b=0.1~{\rm{m}}$ and the length $L_{wedge}$ is $0.4~{\rm{m}}$. The width of jet $L_{jet}$ is $1~{\rm{mm}}$. Parameters of freestream and jet flow are set as Ref.~\cite{jetdsmc}, shown in Tab.~\ref{sidejet1}. The temperature at the wall is set to be $300~{\rm{K}}$. The angle of attack of freestream is $0$ degree. The gas used for simulation is nitrogen, with $\gamma=1.4$. The molecular diameter is set to be $4.467{\AA}$ at $273~{\rm{K}}$. The Shakhov model is used as the kinetic model with $Pr=0.72$. The dynamic viscosity coefficient is calculated by $\mu\sim T^{\omega}$ and $\omega$ is set to be $0.76$.

\begin{figure}[H]
	\centering
	\subfigure[]{
			\includegraphics[width=0.45 \textwidth]{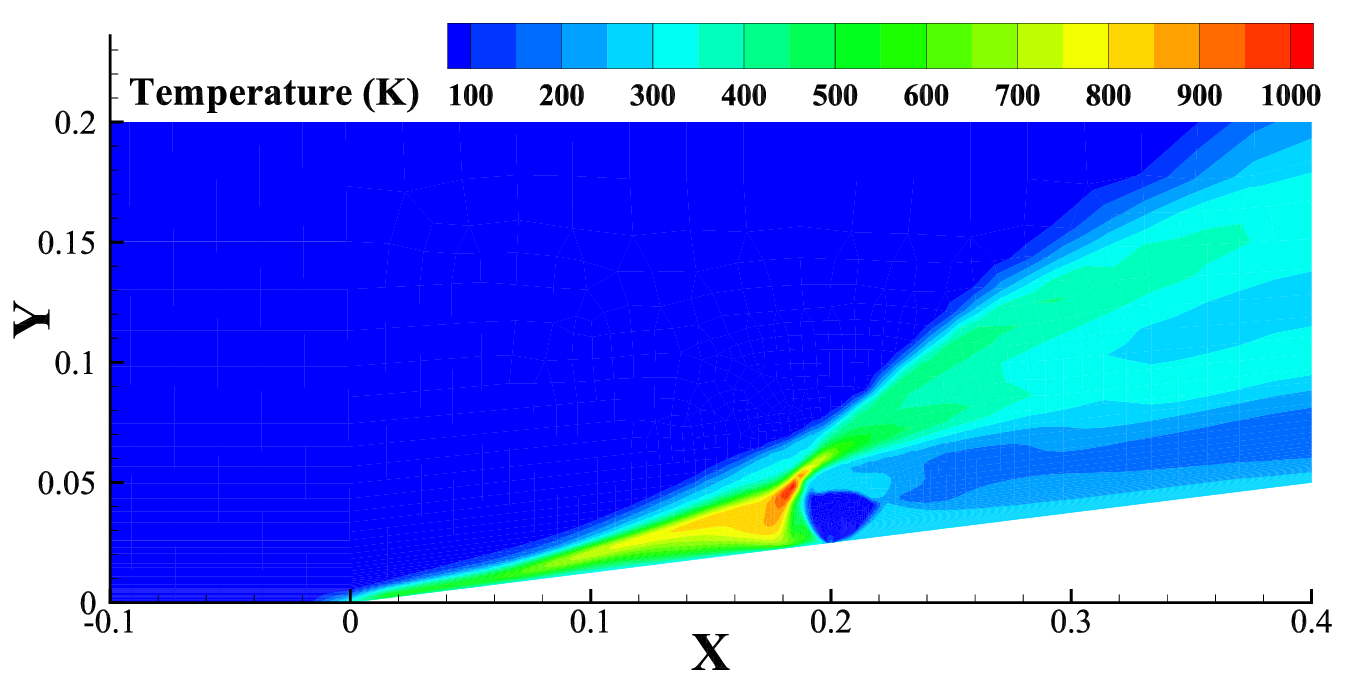}
		}
    \subfigure[]{
			\includegraphics[width=0.45 \textwidth]{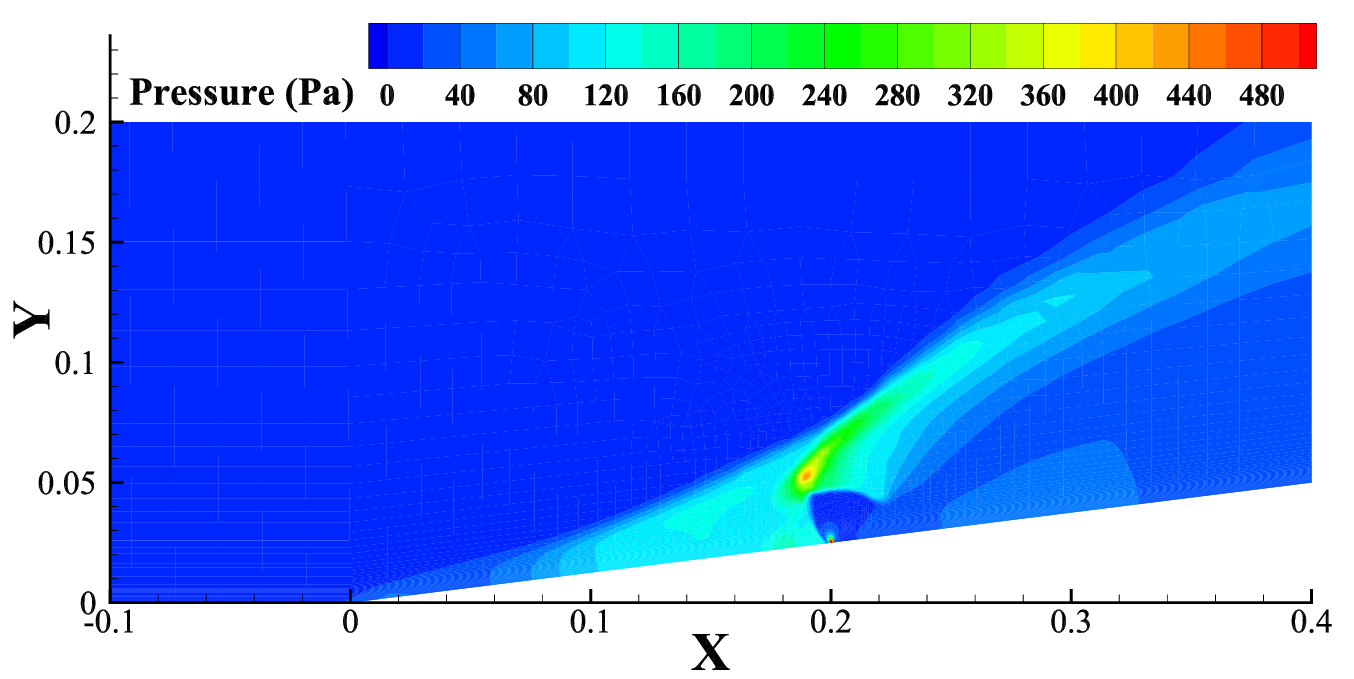}
		}
    \subfigure[]{
    		\includegraphics[width=0.45 \textwidth]{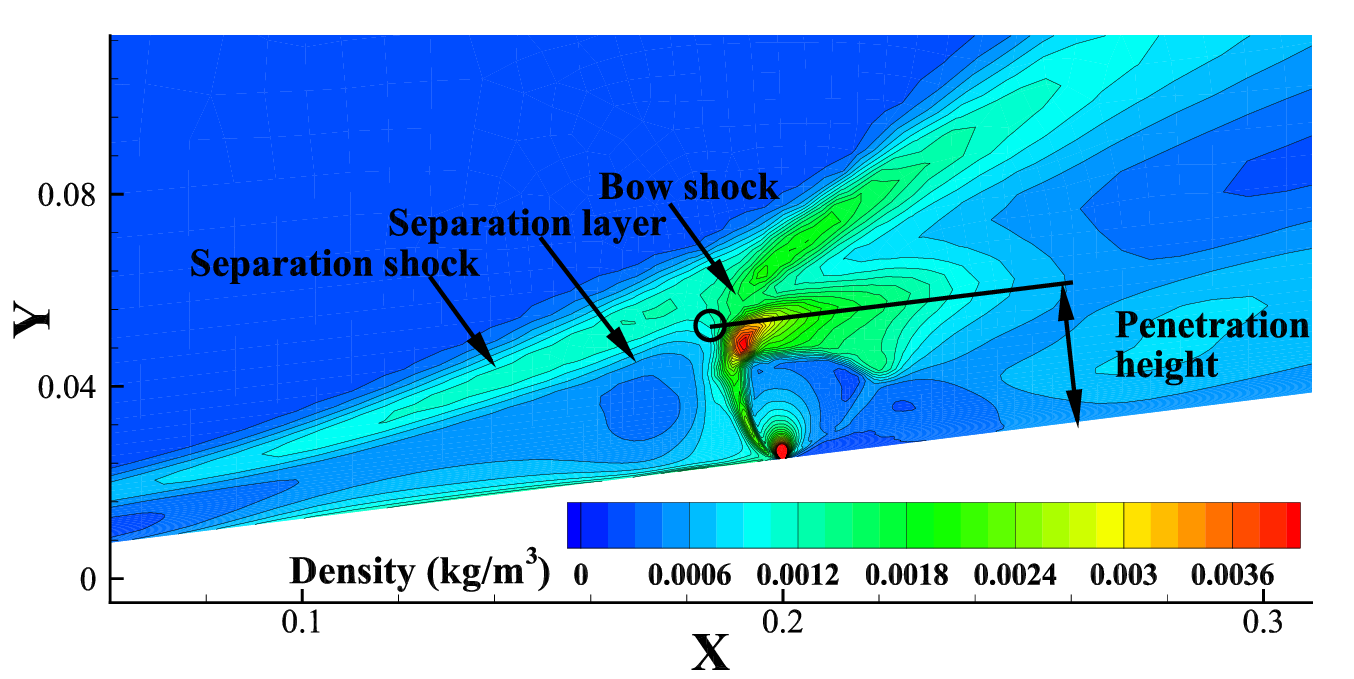}
    	}
    \subfigure[]{
    		\includegraphics[width=0.45 \textwidth]{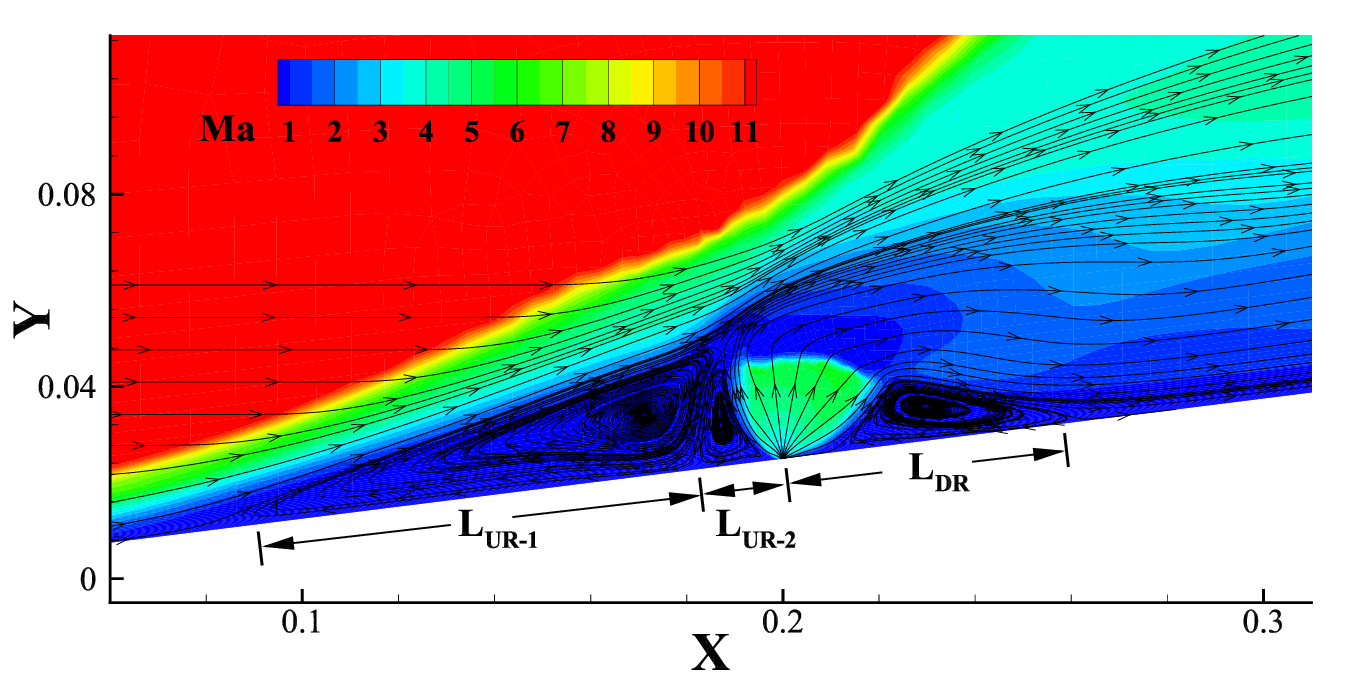}
    	}
	\caption{\label{sidejet_contour} Contours simulated by AUGKWP method (unit: m): (a) Temperature (global), (b) pressure (global), (c) density (local), (d) {\rm{Ma}} (local) and streamlines.}
\end{figure}

The simulation consists of $150$ thousand iteration steps, followed by another $20$ thousand steps for time-averaging. The temperature, pressure, density and ${\rm{Ma}}$ contours are shown as Fig.~\ref{sidejet_contour}. Due to its high density, the jet flow expands and speeds up quickly. Since the density and velocity change intensely around the jet, the shocks surrounding this region is partially curving. Recirculation zones are produced upstream and downstream, separated by the two shocks. Upon the upstream recirculation zone, a high-speed flow region is surrounded by the separation shock, separation layer and bow shock. The flow structure between the bow shock and the normal shock of the expanded jet flow is complicated. More descriptions can be referred to Ref.~\cite{jetdsmc}, where the penetration height is defined as the normal distance from the surface to the location where there is a black ``streak'' between the bow shock and the normal shock. In addition to the penetration height $h$, lengths of recirculation zones are measured and shown in Tab.~\ref{sidejet2}, compared with results of DSMC~\cite{jetdsmc} and UGKWP method with BGK model~\cite{wpwp}. Differences of results are limited within $20\%$.

\begin{figure}[H]
	\centering
	\subfigure[]{
			\includegraphics[width=0.3 \textwidth]{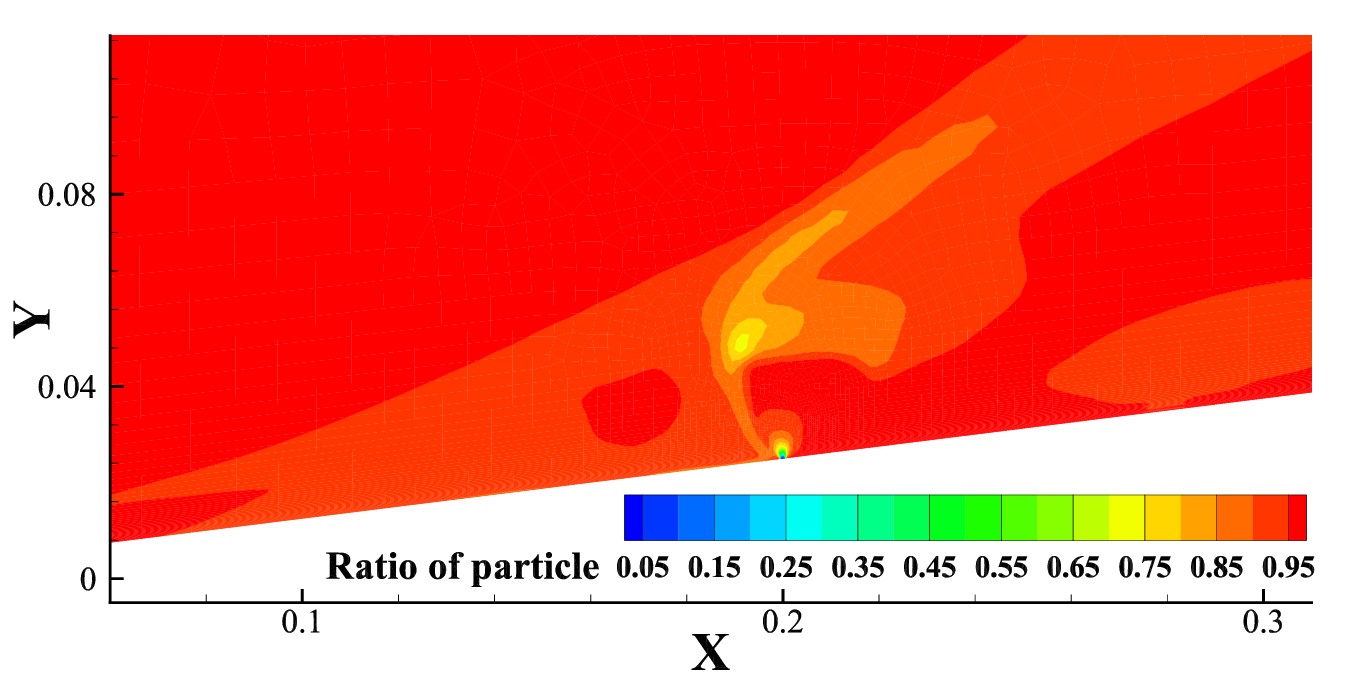}
		}
    \subfigure[]{
			\includegraphics[width=0.3 \textwidth]{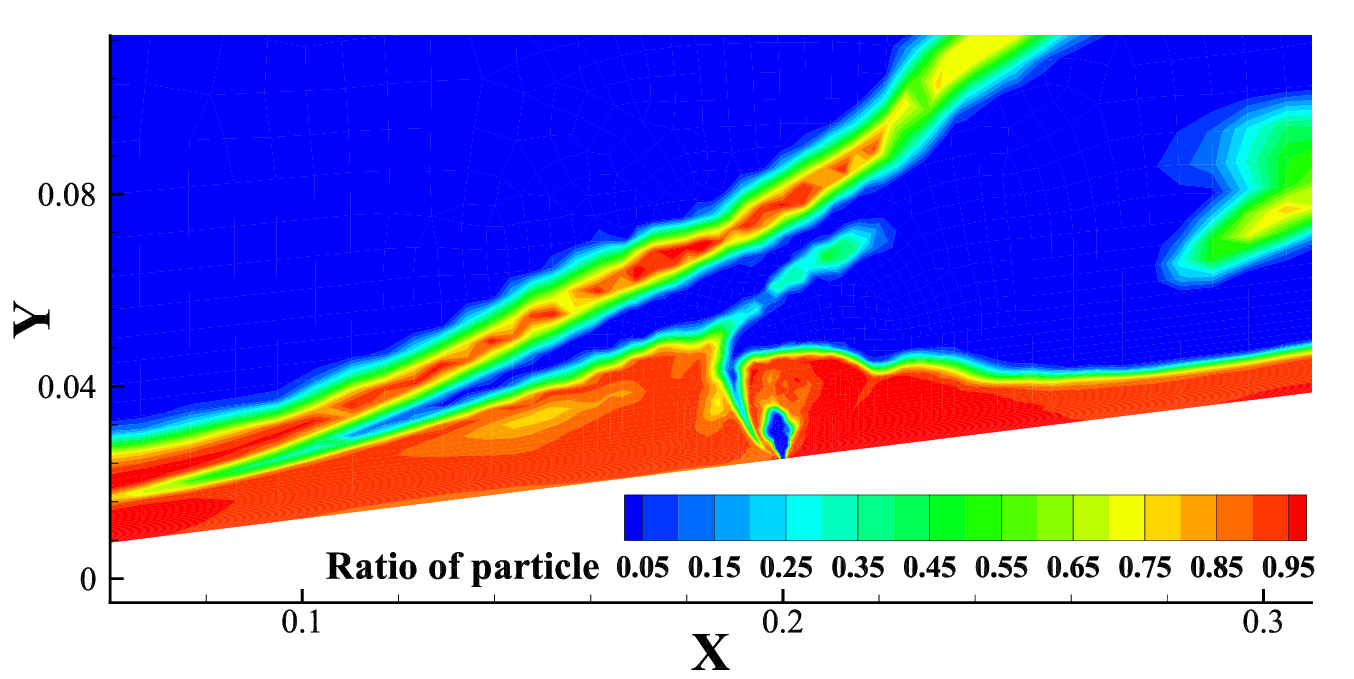}
		}
    \subfigure[]{
    		\includegraphics[width=0.3 \textwidth]{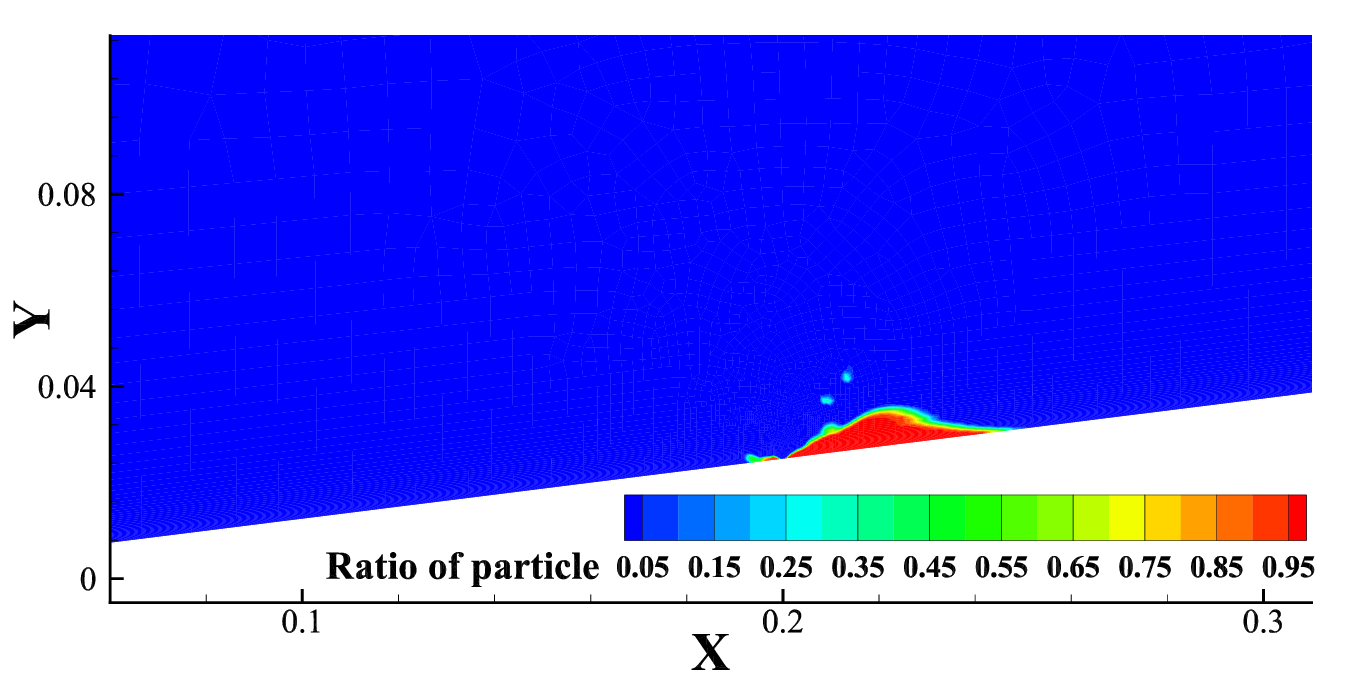}
    	}
	\caption{\label{sidejet_contour_rop} Comparison of ratio of particles (unit: m): (a) Original UGKWP method, (b) AUGKWP method with ${\rm{Kn_{GLL}}}$, (c) AUGKWP method with ${\rm{Kn_{L}}}$.}
\end{figure}

The contours of particle ratio are shown in Fig.~\ref{sidejet_contour_rop}. Due to the high-density jet flow, the influence of tiny mesh size is reduced. By the AUGKWP method with ${\rm{Kn_L}}$, most of particles are released in the downstream reattachment region, because the density is lower there. However, it is noticed that with only ${\rm{Kn_{GLL}}}$, both AUGKWP method and DSMC method~\cite{jetdsmc} identify many regions to be rarefied, even though the shock is definitely under-resolved by only $2$ to $3$ mesh cells. Finally, the efficiency of different approaches are shown in Tab.~\ref{sidejet3}, where particle number and average CPU time per one thousand steps are taken for comparison. Since this case is much more continuum, the AUGKWP method brings about a $3.7$ times improvement in efficiency.

\begin{table}[H]
\centering
\caption{\label{sidejet1} Parameters of freestream and jet flow}
\begin{tabular}{|c|c|c|}
\hline
    ~                         &Freestream             &Side jet              \\ \hline
	Density (${\rm{kg/m^3}}$) &$2.325\times10^{-4}$   &$38.363\times10^{-3}$ \\
    Temperature (${\rm{K}}$)  &$50$                   &$250$                 \\
    Pressure (${\rm{Pa}}$)    &$3.45$                 &$2847.4$              \\
    {\rm{Ma}}                 &$12$                   &$1$                   \\
    {\rm{Kn}} ($L_{wedge}$)   &$3.8\times10^{-4}$     &$3.3\times10^{-6}$    \\
    {\rm{Kn}} ($L_{jet}$)     &$0.15$                 &$0.0013$              \\
\hline
\end{tabular}
\end{table}

\begin{table}[H]
\centering
\caption{\label{sidejet2} Comparison of simulation results of side-jet impingement on hypersonic flow}
\begin{tabular}{|c|c|c|c|c|c|c|c|}
\hline
    Numerical method and literature        &$h/r_b$ &$L_{UR-1}/h$ &$L_{UR-2}/h$ &$L_{DR}/h$ \\ \hline
	DSMC~\cite{jetdsmc}                    &$0.5$   &$3.32$       &$0.55$       &$1.81$     \\
    UGKWP (BGK model)~\cite{wpwp}          &$0.564$ &$3.26$       &$0.66$       &$1.71$     \\
    AUGKWP (Shakhov model)                 &$0.56$  &$3.30$       &$0.64$       &$2.09$     \\
\hline
\end{tabular}
\end{table}

\begin{table}[H]
\centering
\caption{\label{sidejet3} Efficiency comparison in the simulation of side-jet impingement on hypersonic flow}
\begin{tabular}{|c|c|c|}
\hline
    Approach                  &Particle number (k) &CPU time per $1000$ steps$^{\rm a}$ (min) \\ \hline
	UGKWP                     &$724$               &$21.1$                                    \\
    AUGKWP(${\rm{Kn_{GLL}}}$) &$490$               &$16.0$                                    \\
    AUGKWP(${\rm{Kn_{L}}}$)   &$97$                &$5.7$                                     \\
\hline
\end{tabular}
\begin{tablenotes}
\item{$^{\rm a}$ Tested with only one core.}
\end{tablenotes}
\end{table}

\subsection{Hypersonic flow over a $70^{\circ}$ blunted cone with a cylindrical sting}\label{sec:cone}
\begin{figure}[H]
	\centering
	\subfigure{
			\includegraphics[width=0.5 \textwidth]{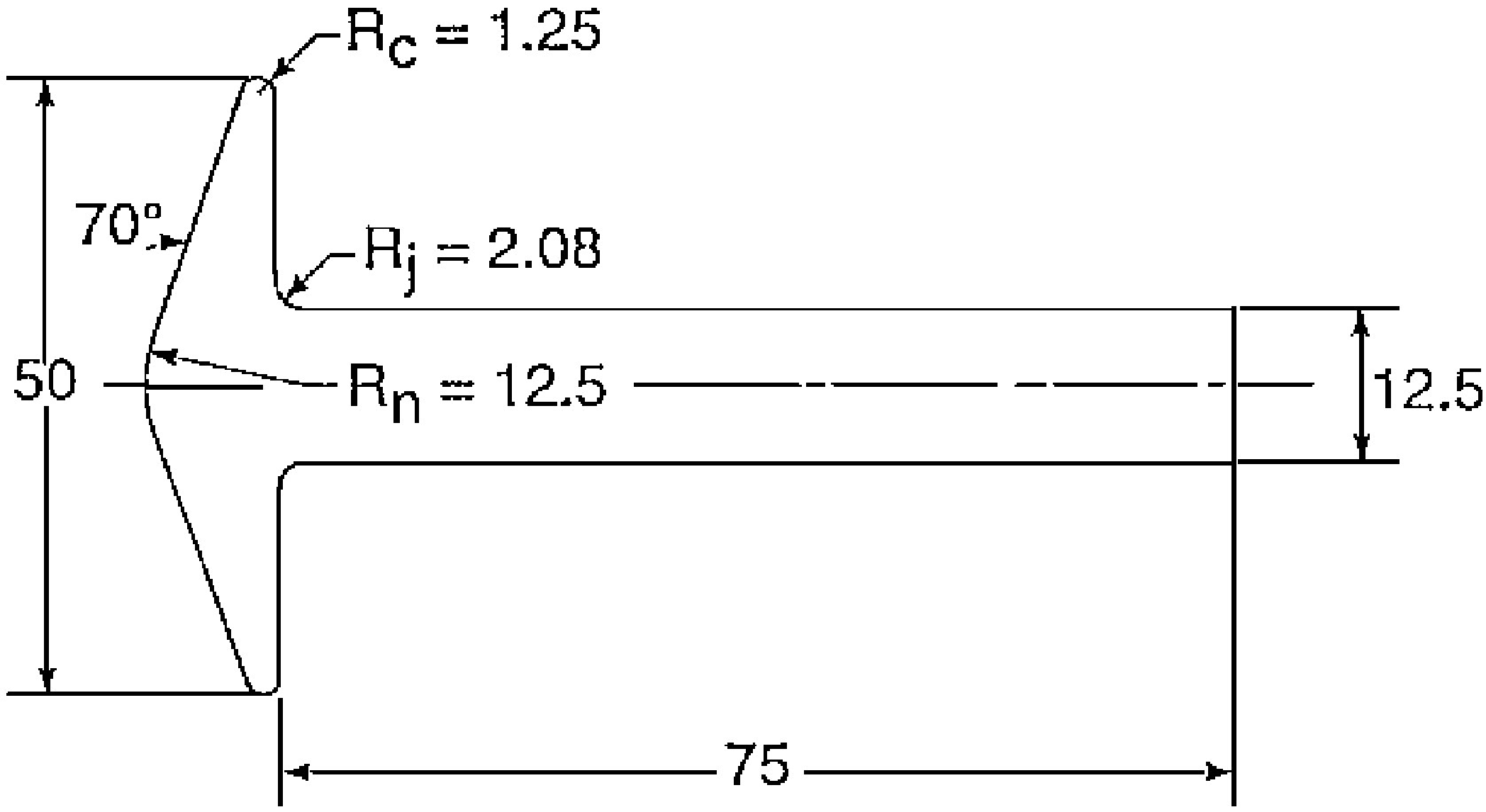}
		}
	\caption{\label{cone_shape} Geometric profile at symmetry plane of $70^{\circ}$ blunted cone~\cite{bcexp2} (All dimensions are in millimeters.)}
\end{figure}

\begin{figure}[H]
	\centering
	\subfigure[]{
			\includegraphics[width=0.22 \textwidth]{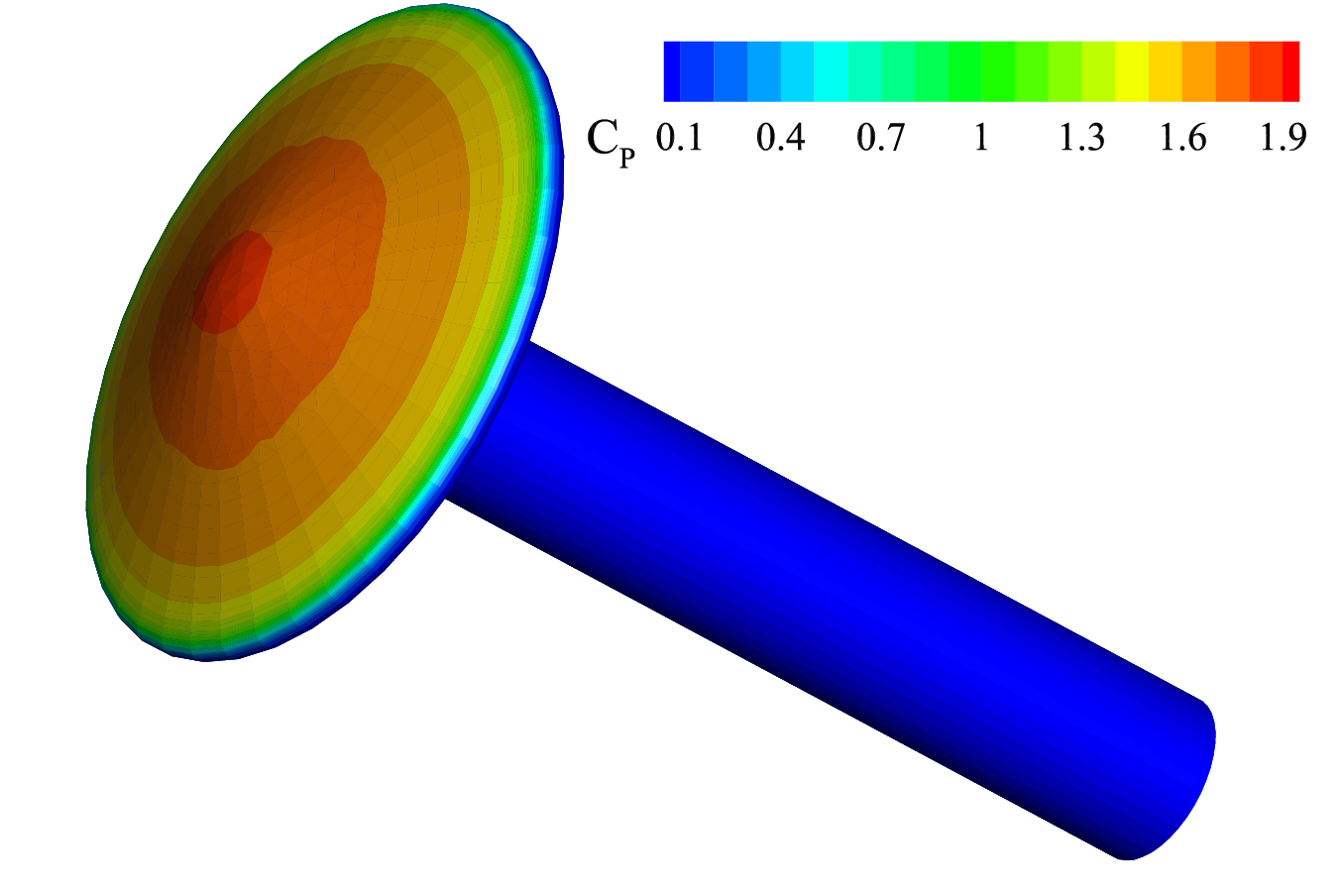}
		}
    \subfigure[]{
    		\includegraphics[width=0.22 \textwidth]{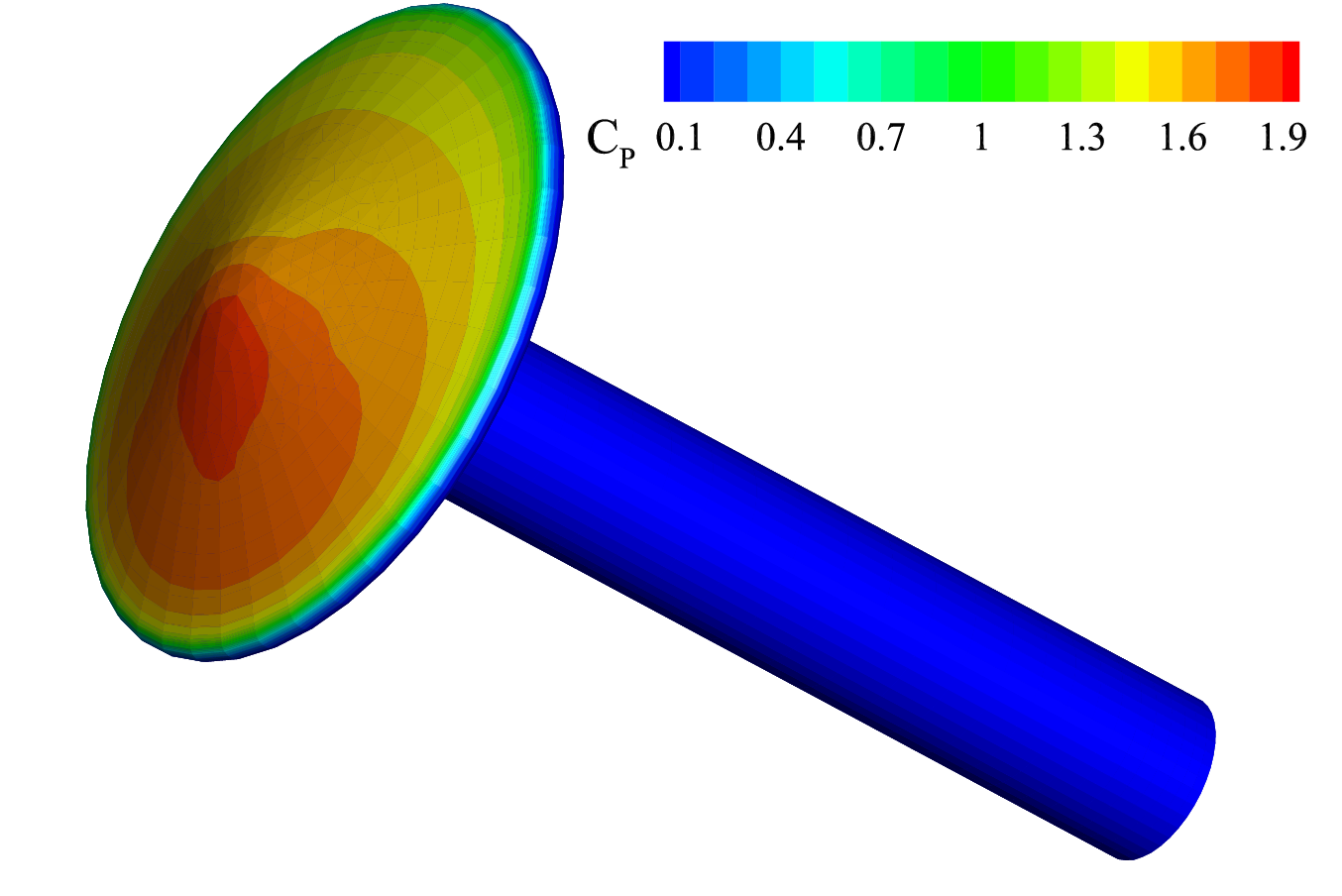}
    	}
    \subfigure[]{
    		\includegraphics[width=0.22 \textwidth]{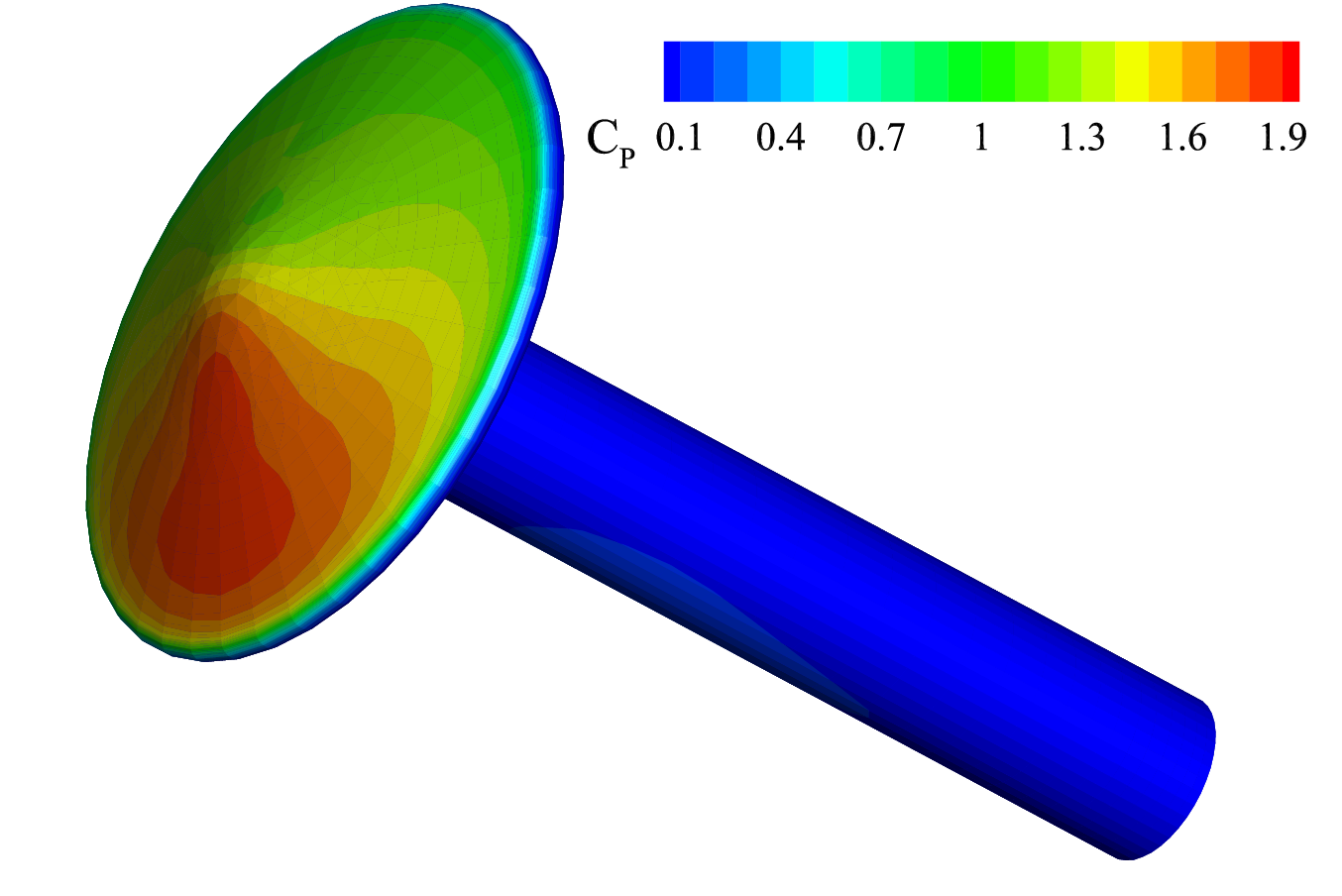}
    	}
    \subfigure[]{
    		\includegraphics[width=0.22 \textwidth]{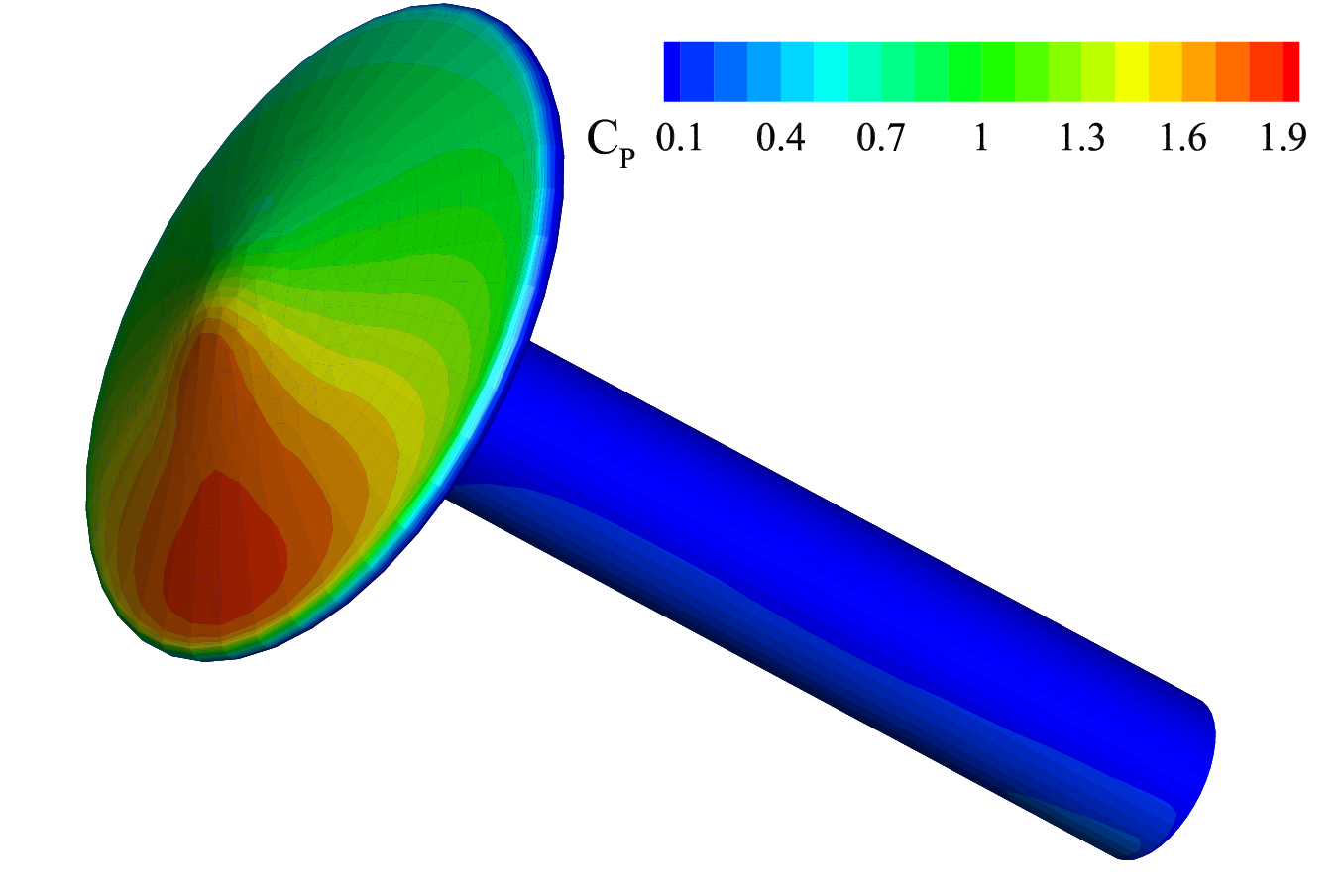}
    	}
    \subfigure[]{
			\includegraphics[width=0.22 \textwidth]{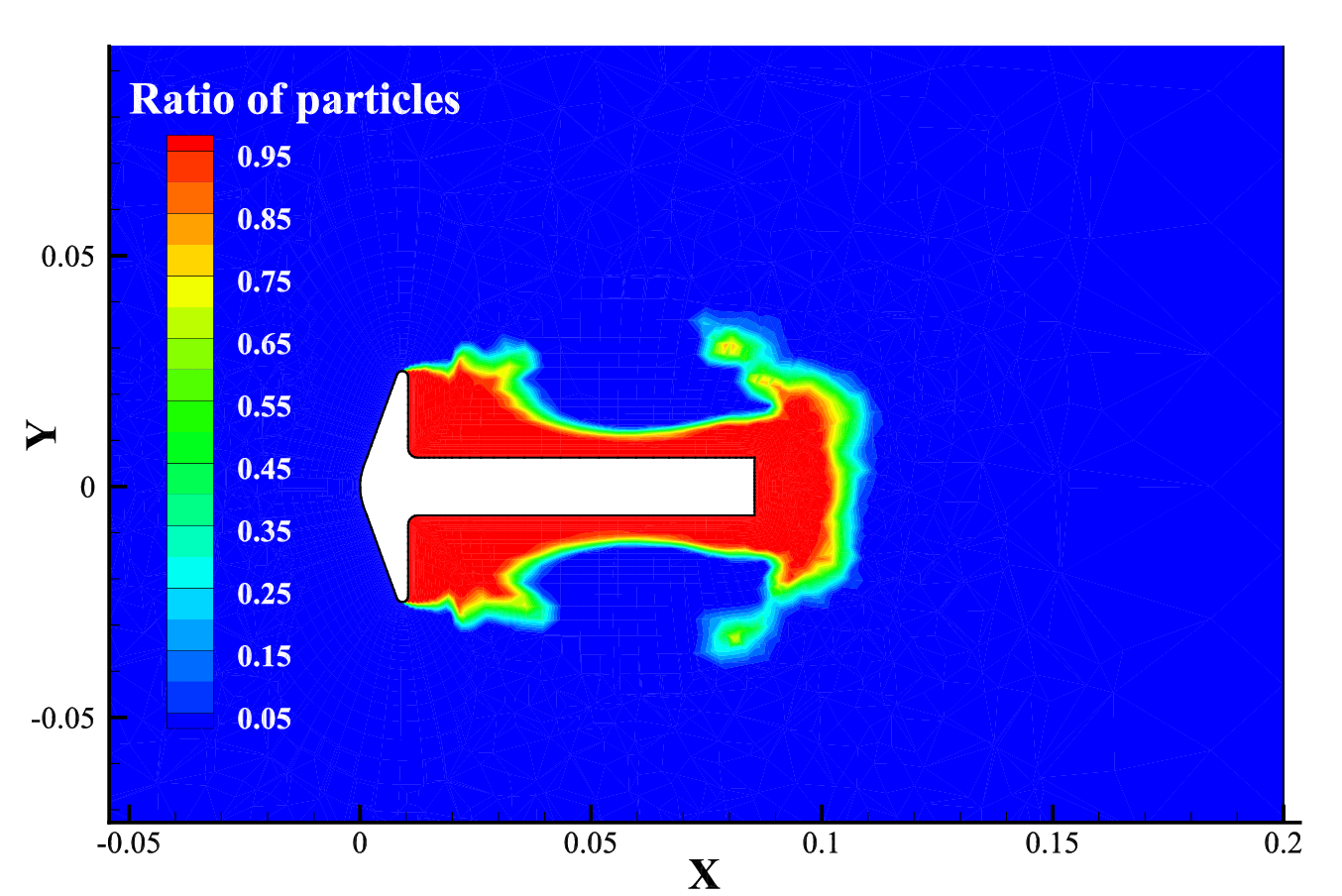}
		}
    \subfigure[]{
    		\includegraphics[width=0.22 \textwidth]{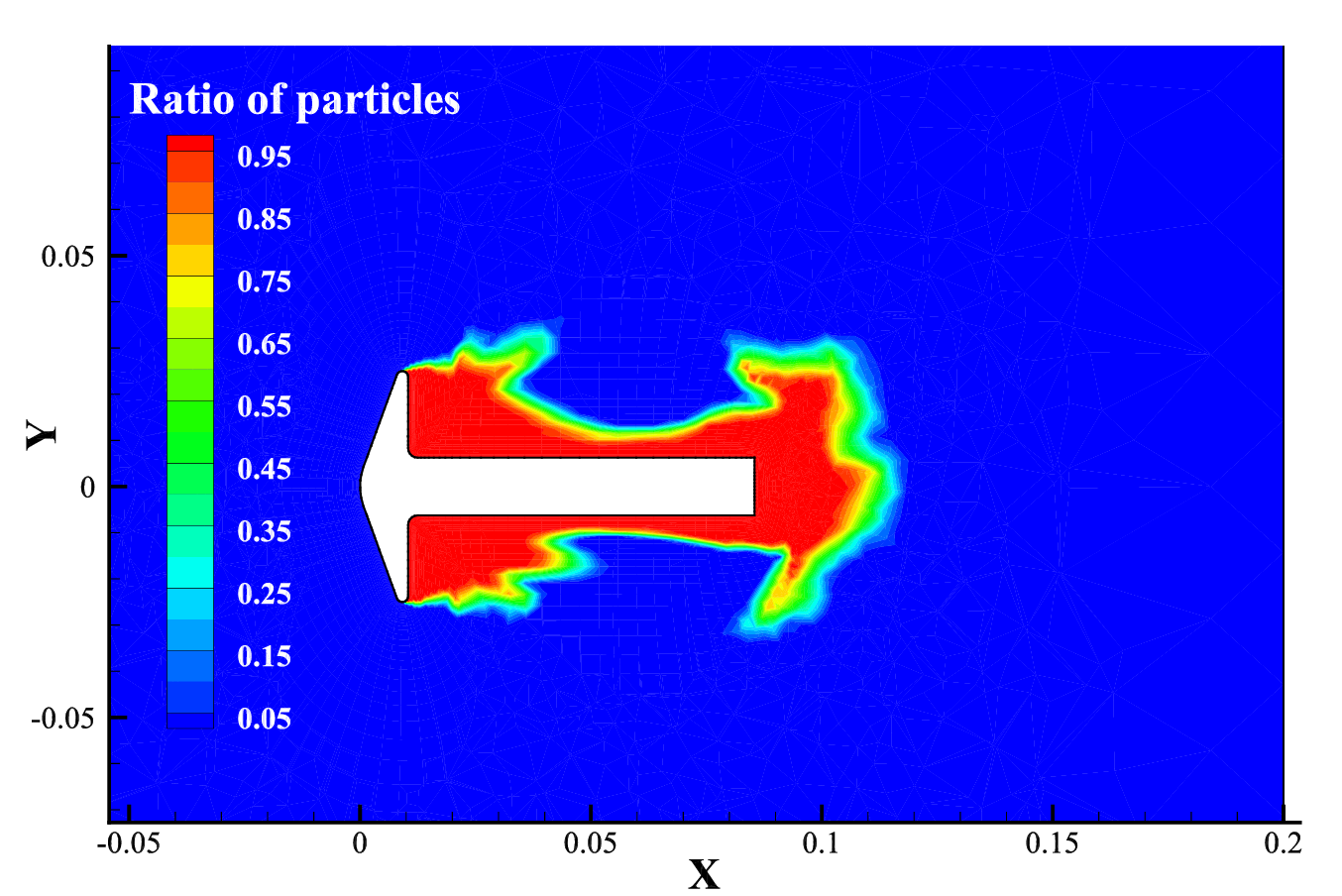}
    	}
    \subfigure[]{
    		\includegraphics[width=0.22 \textwidth]{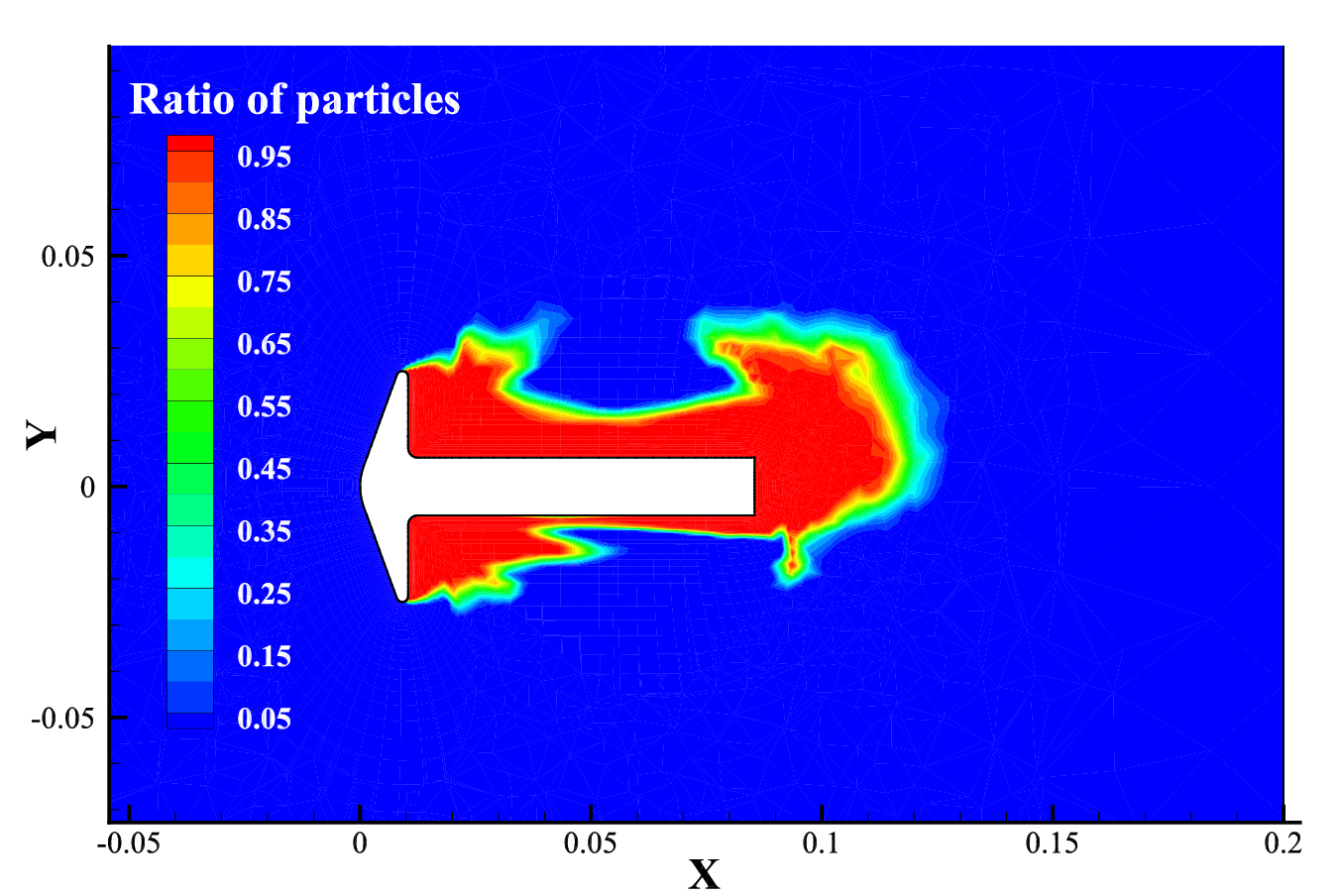}
    	}
    \subfigure[]{
    		\includegraphics[width=0.22 \textwidth]{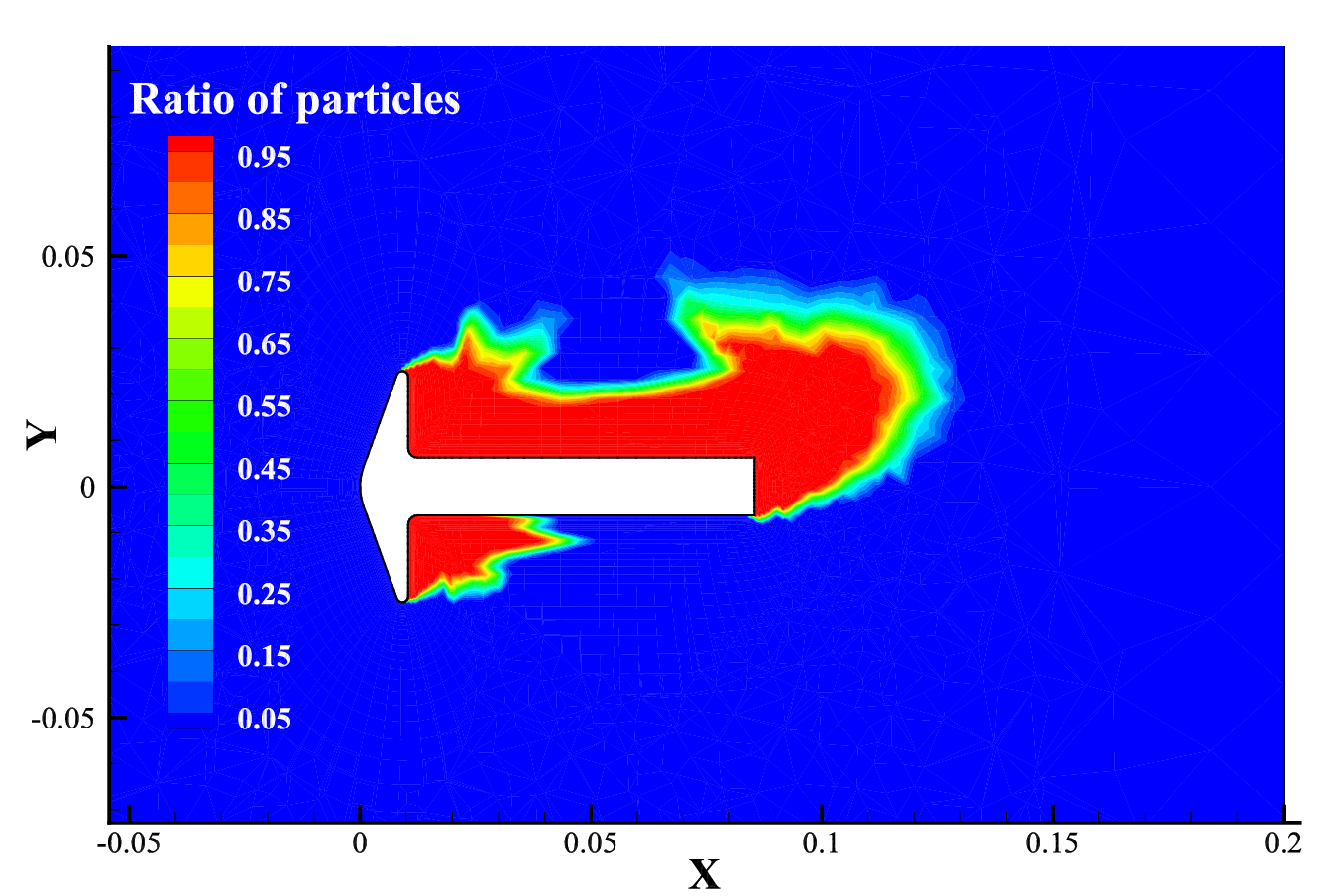}
    	}
	\caption{\label{cone1} Pressure coefficient contours at the surface and particle ratio contours at the symmetry plane (unit: m): (a) $C_P$ ($0^{\circ}$), (b) $C_P$ ($10^{\circ}$), (c) $C_P$ ($20^{\circ}$), (d) $C_P$ ($30^{\circ}$), (e) particle ratio contour ($0^{\circ}$), (f) particle ratio contour ($10^{\circ}$), (g) particle ratio contour ($20^{\circ}$), (h) particle ratio contour ($30^{\circ}$).}
\end{figure}

Finally, a three-dimensional case is simulated. The shape is taken from the experiment~\cite{bcexp2,bcexp3} as Fig.~\ref{cone_shape}, which is identical to the forebody configuration of the Mars Pathfinder probe. This test case is widely used for method validation~\cite{sparta,nccr1,nccr2}. The third case of the experiment is taken as the flow condition, such as $\rho_{\infty}=46.62\times10^{-5}~{\rm{kg/m^3}}$, $U_{\infty}=1634~{\rm{m/s}}$, $T_{\infty}=15.3~{\rm{K}}$ and ${\rm{Ma_{\infty}}}=20.5$. The temperature at the wall is set to be $300~{\rm{K}}$ according to Ref.~\cite{bcdsmc}. For a detailed study on aerodynamic force, the attack angles are set to $0^{\circ}$, $10^{\circ}$, $20^{\circ}$ and $30^{\circ}$ for simulation. The gas is set to be nitrogen and parameters are taken from Ref.~\cite{dsmcre}. At the temperature $273~{\rm{K}}$, the dynamic viscosity coefficient is set to be $1.6579\times10^{-5}~{\rm{Ns/m^2}}$, and $\omega$ is set to be $0.74$ to calculate as $\mu\sim T^{\omega}$. The Shakhov model is used as the kinetic model with $Pr=0.72$. In the experiment, the dynamic viscosity coefficient is calculated by a combination of Sutherland viscosity law and a linear viscosity law. In this way, the inflow Reynolds number is calculated to be $36265$, with the reference length to be the base diameter, $50~{\rm{mm}}$. Though the inflow is in the continuum flow regime, the flow at the wake area is rather rarefied, because the radius $R_c$ at the transition region is so tiny, bringing about a strong expansion. In Ref.~\cite{bcexp3,bcdsmc}, results by DSMC are rather different from N-S solver with slip boundary condition when predicting heat flux at the wake area. As shown in Fig.~\ref{cone1}, in the simulation by the AUGKWP method, particles are mainly employed at the wake area as well.

\begin{figure}[H]
	\centering
	\subfigure[]{
			\includegraphics[width=0.45 \textwidth]{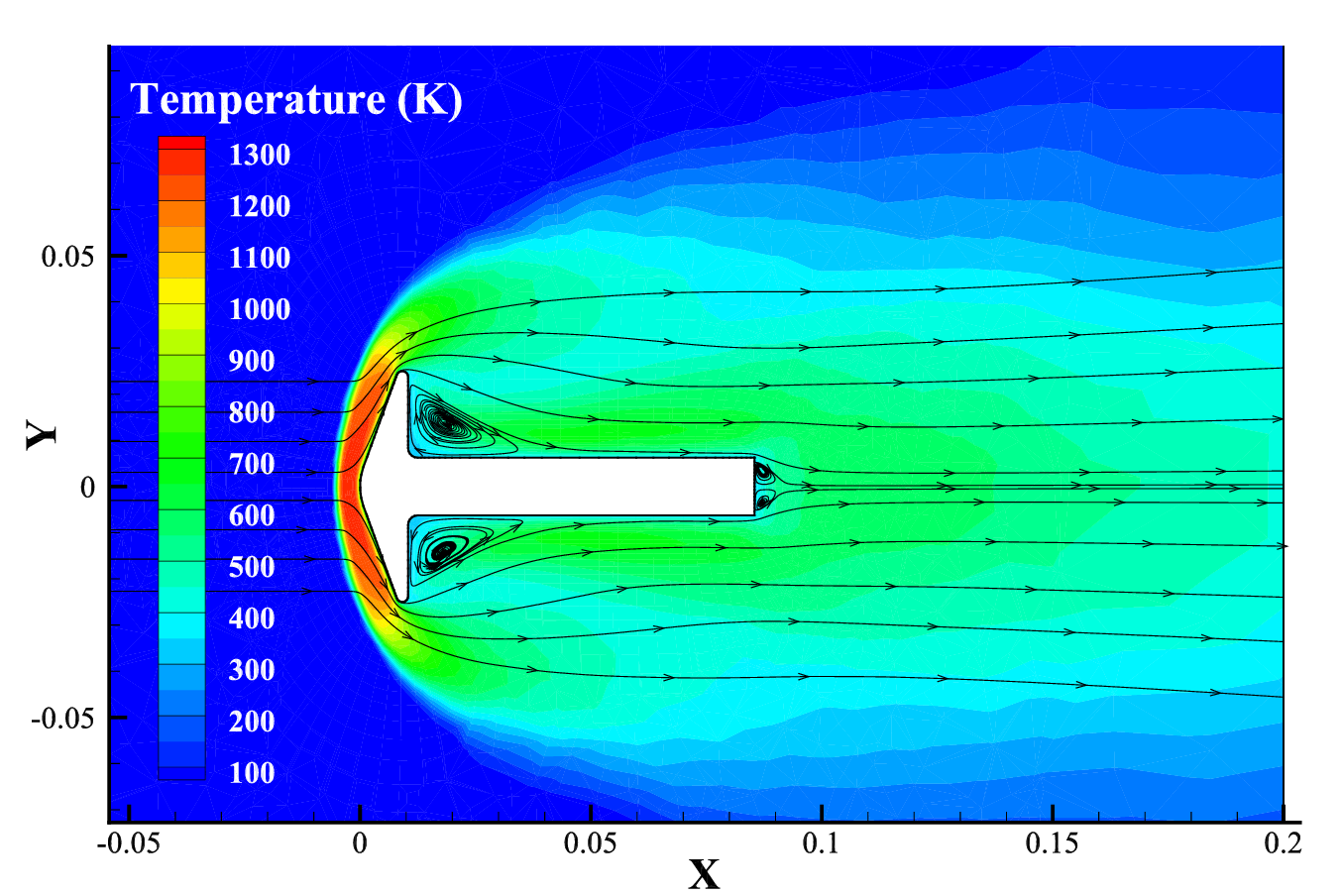}
		}
    \subfigure[]{
			\includegraphics[width=0.45 \textwidth]{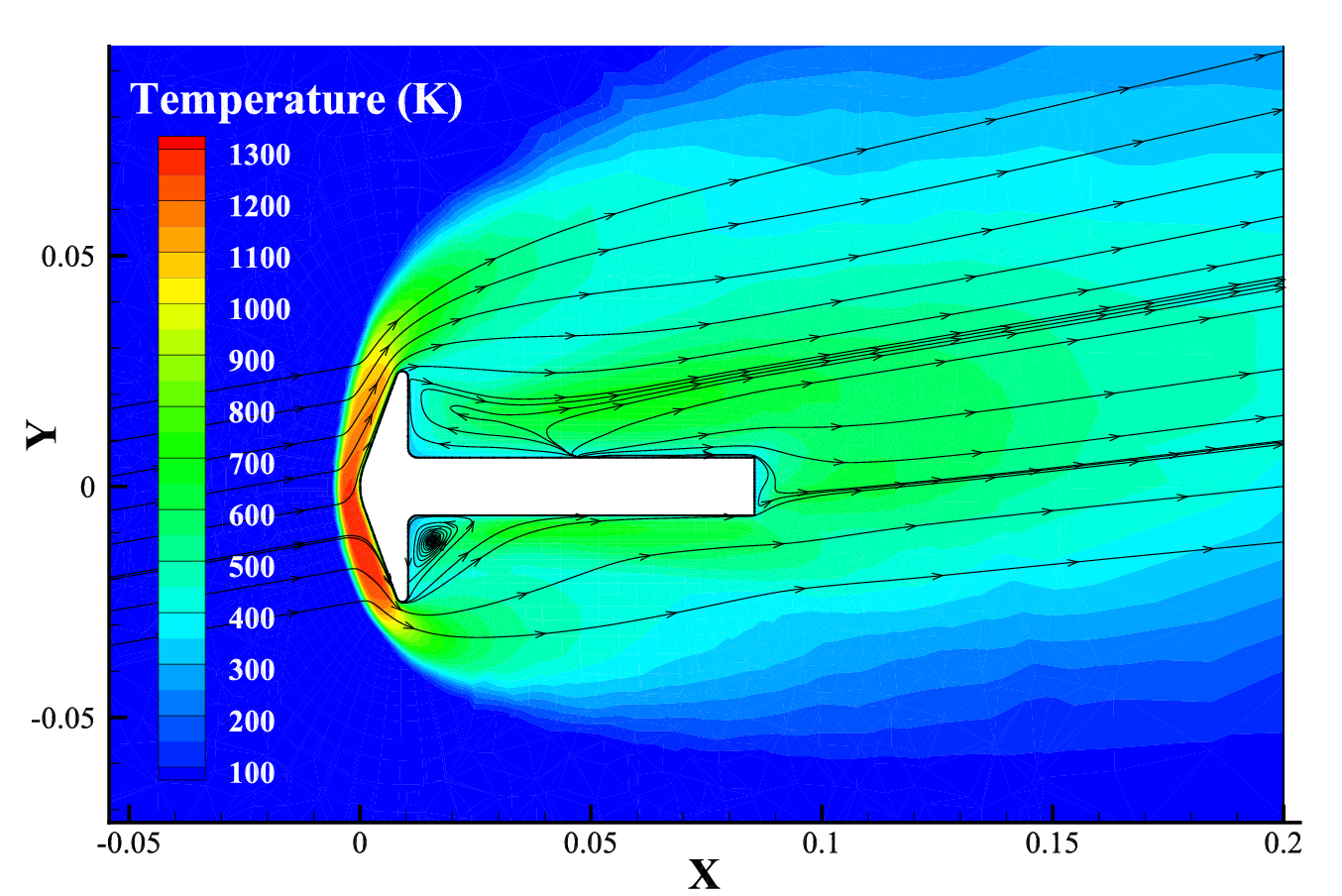}
		}
    \subfigure[]{
    		\includegraphics[width=0.45 \textwidth]{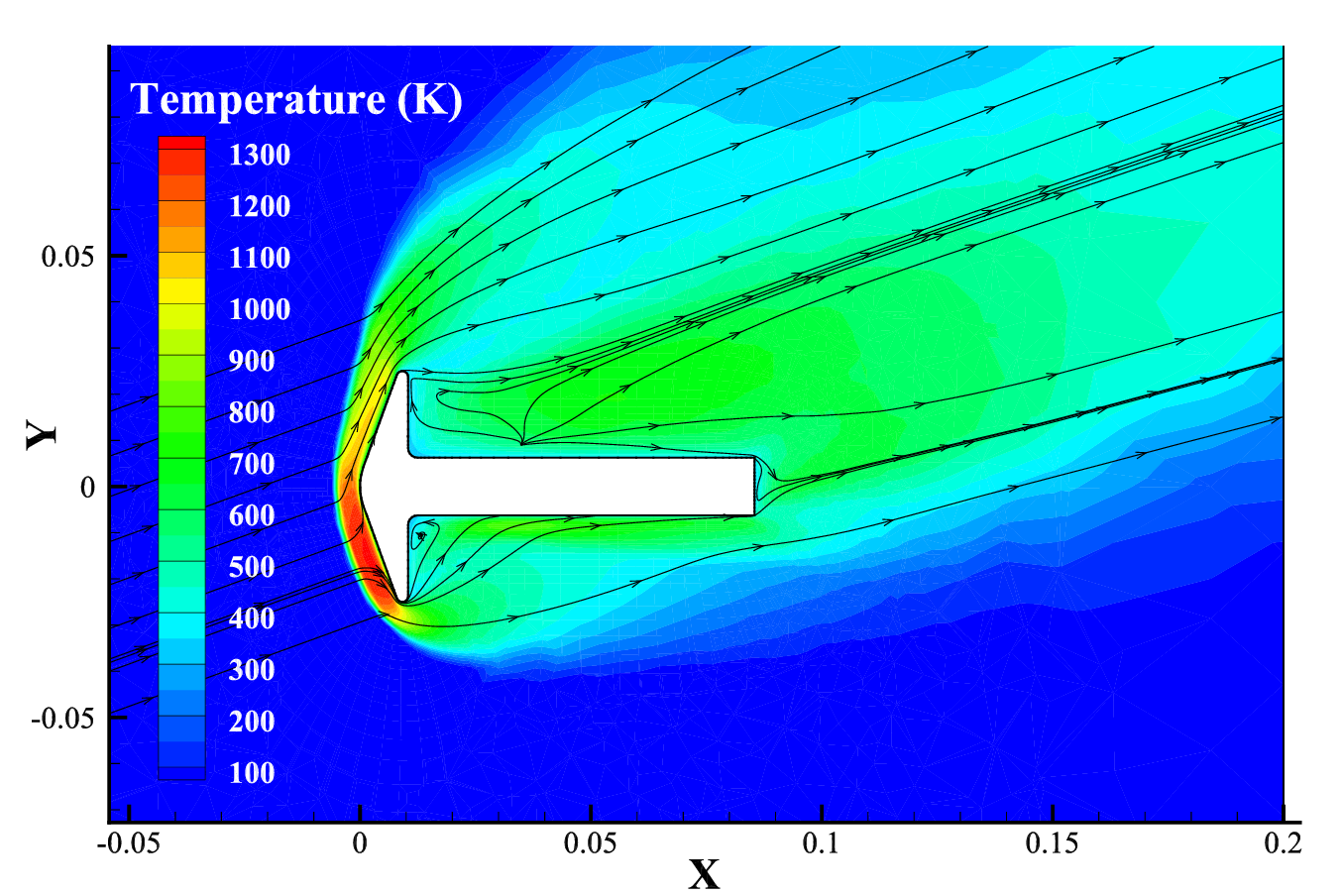}
    	}
    \subfigure[]{
    		\includegraphics[width=0.45 \textwidth]{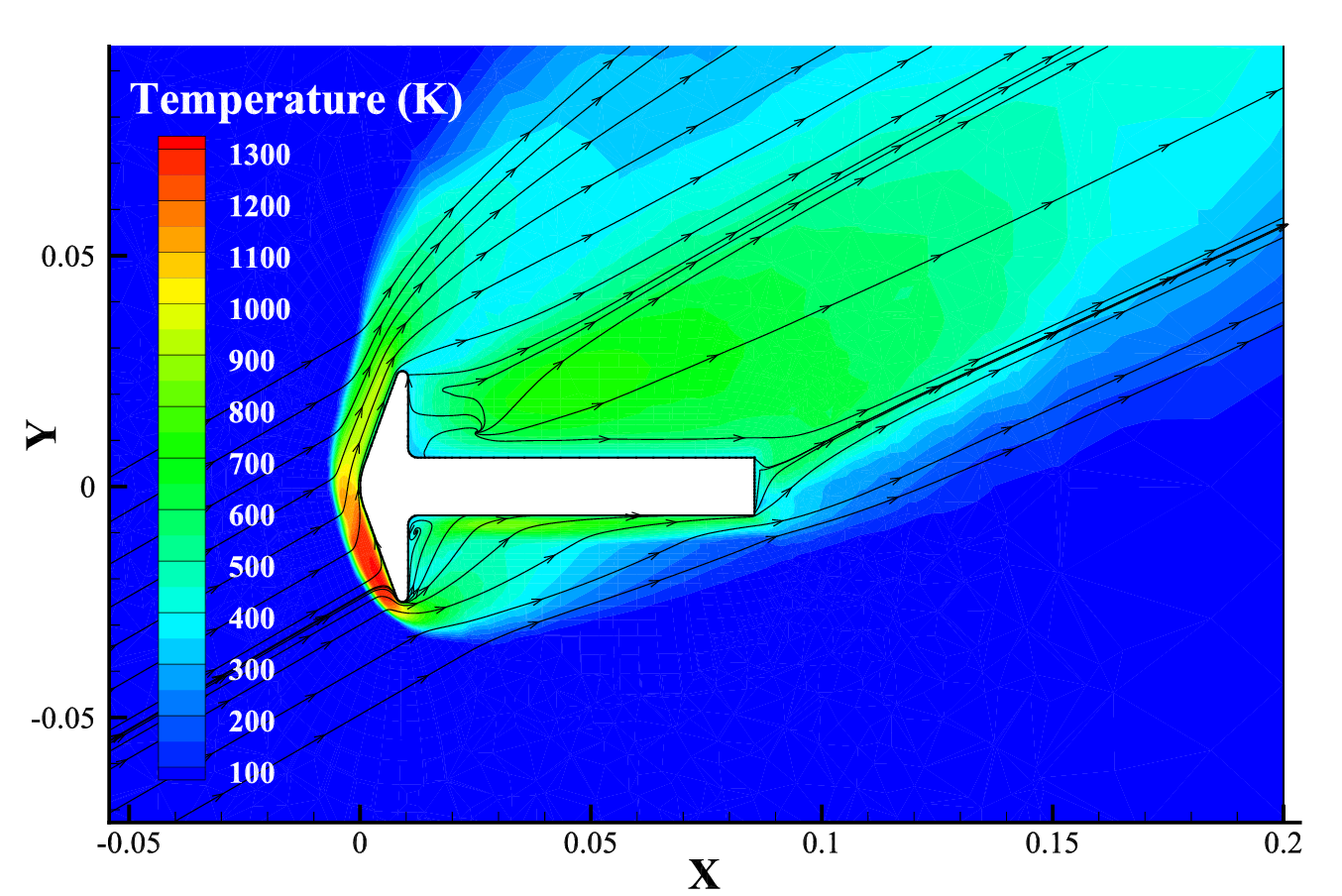}
    	}
	\caption{\label{cone2} Temperature contours and streamlines at the symmetry plane (unit: m): (a) $0^{\circ}$, (b) $10^{\circ}$, (c) $20^{\circ}$, (d) $30^{\circ}$.}
\end{figure}

\begin{figure}[H]
	\centering
	\subfigure{
			\includegraphics[width=0.8 \textwidth]{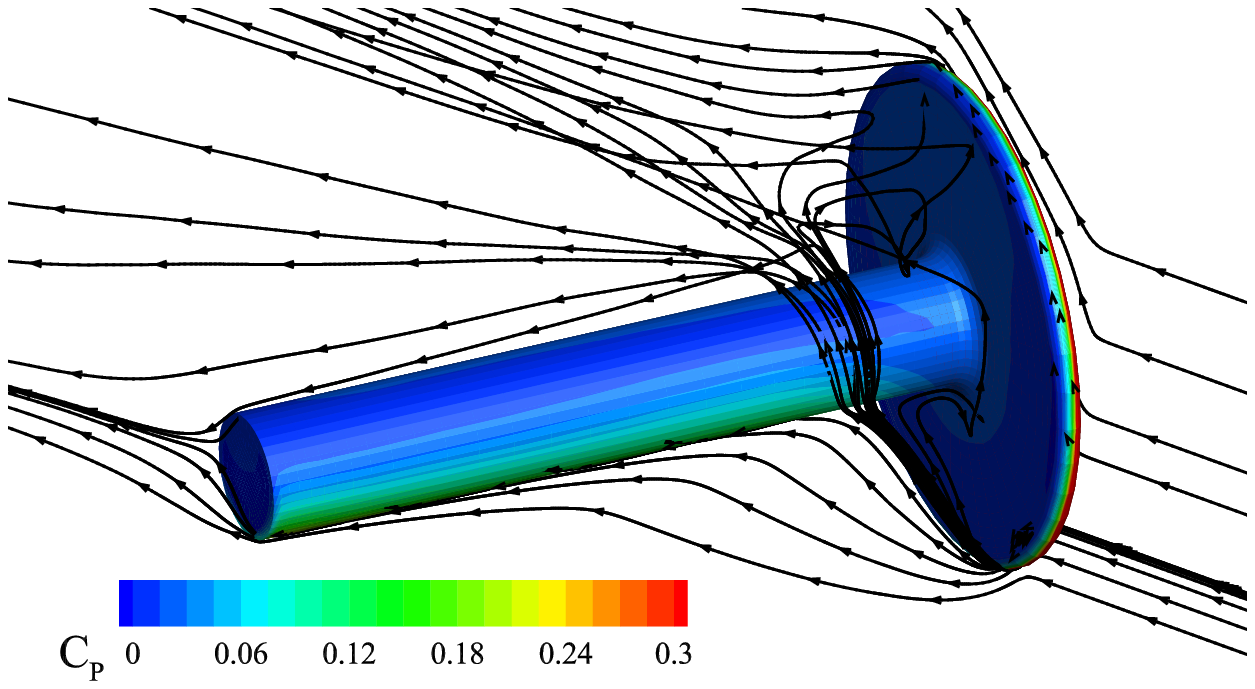}
		}
	\caption{\label{cone3} Three-dimensional streamlines at $30^{\circ}$ angle attack.}
\end{figure}

\begin{figure}[H]
	\centering
	\subfigure[]{
			\includegraphics[width=0.45 \textwidth]{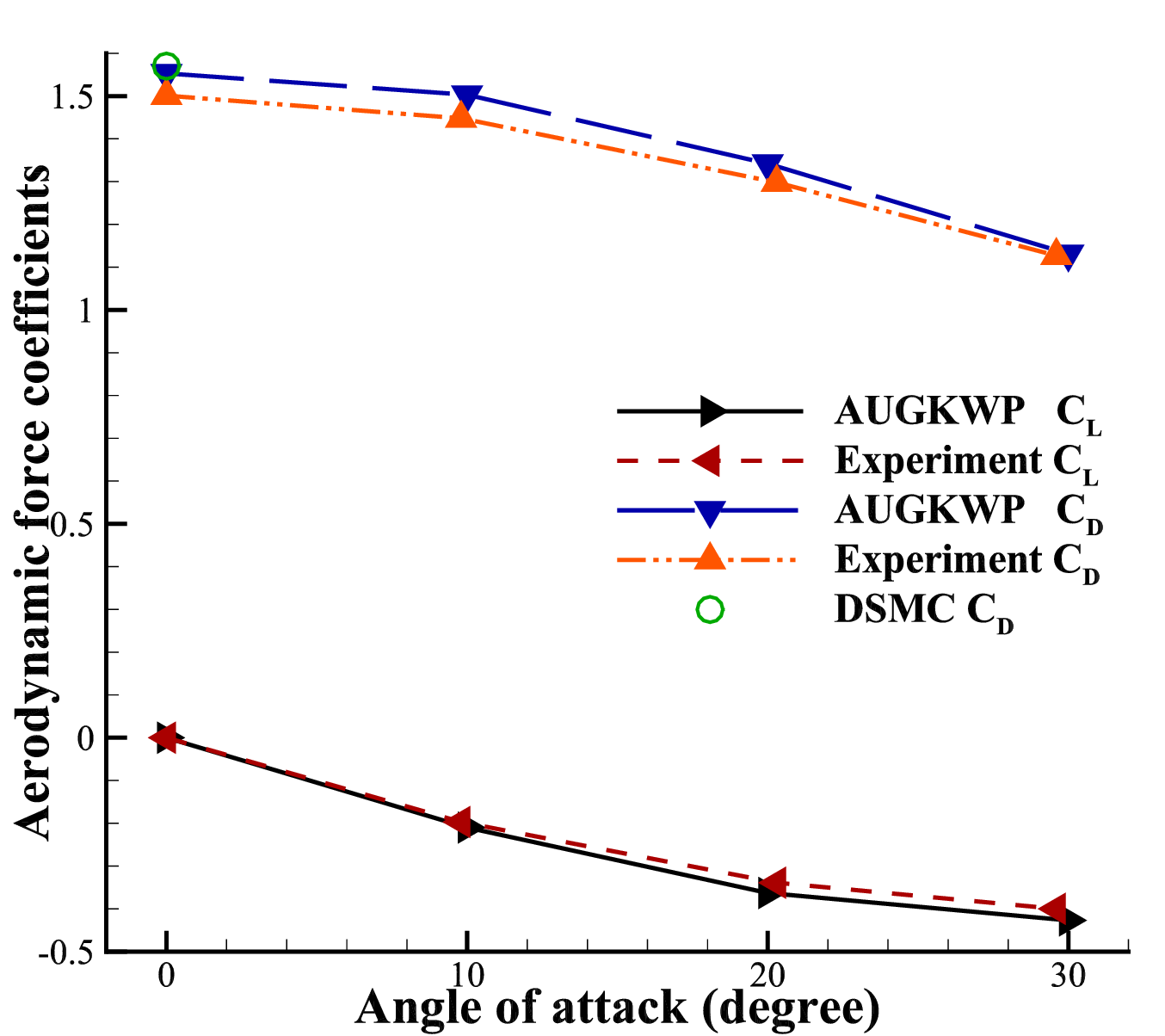}
		}
    \subfigure[]{
			\includegraphics[width=0.45 \textwidth]{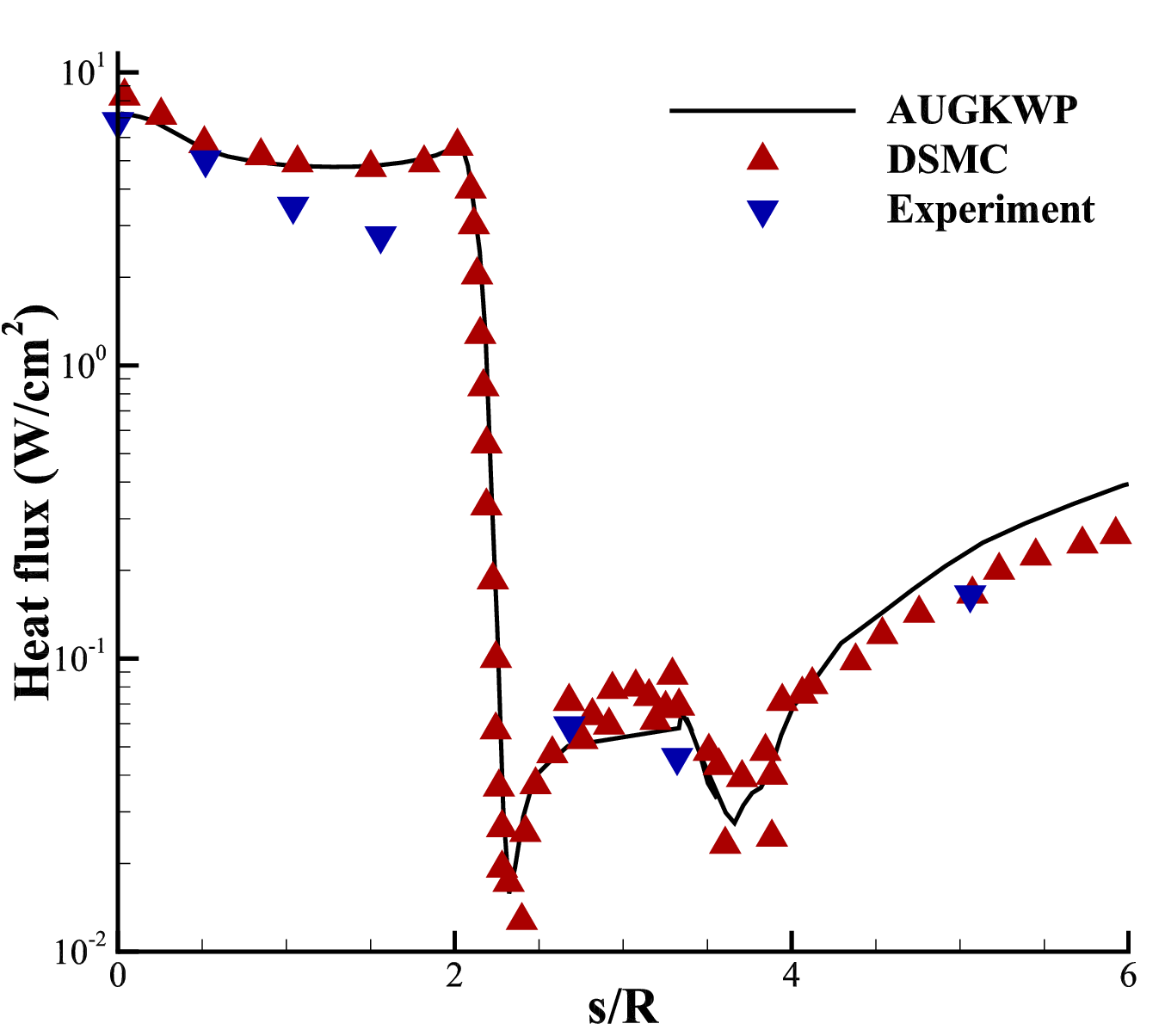}
		}
	\caption{\label{cone4} Quantitative comparison between experiment, DSMC and AUGKWP method: (a) Aerodynamic force coefficients, (b) heat flux at the central axis at $0^{\circ}$ attack angle ($s$ is the curvilinear abscissa).}
\end{figure}

The simulation consists of $240$ thousand iteration steps, followed by another $20$ thousand steps for time-averaging. After the variables at the symmetry plane are extracted, temperature contours and streamlines are drawn as Fig.~\ref{cone2}. It is noticed that at the wake area, there is a point source. Streamlines all go out of it, but never into it. To illustrate this, the three-dimensional streamline at $30^{\circ}$ angle attack is drawn as Fig.~\ref{cone3}. It is found that the streamline comes from the third dimension, and spreads out in all direction at the symmetry plane. Moveover, the aerodynamic force and heat flux are shown as Fig.~\ref{cone4}. Though there is deviation between AUGKWP method and experiment, results of AUGKWP method match well with DSMC~\cite{bcdsmc}. As to the efficiency, at $20^{\circ}$ angle attack, $22.78$ millions particles are employed in the original UGKWP method, while only $7.88$ millions particles are needed in the AUGKWP method, using ${\rm{Kn_L}}$ as the adaptive criterion.

\section{Conclusions}\label{sec:conclusion}
This study aims to further improve the AUGKWP method for high-speed multiscale flow simulation. From all perspectives of the time, space and gradient, the adaptive criteria for wave-particle decomposition are developed and tested in various cases. Since the local mesh size is taken into account in the formula of ${\rm{Kn_L}}$, the numerical multiscale modeling is directly related to the mesh generation by the present method. In typical cases, such as the hypersonic flow around a cylinder, the present criteria show improvement to $\rm{Kn_{GLL}}$ by correctly identifying the post-shock and boundary layer as continuum regions. In addition, the current method also has good performance capturing local non-equilibrium phenomenon induced by localized small scale flow structures in a global continuum flow regime. By reducing the computation consumption of particles in the near-continuum regime, the efficiency of the present AUGKWP method improves by a factor of about $1.5$ to $4$. Moreover, in regions with significant scale variation, the equilibrium flux is improved to be more consistent with the particle evolution, according to the integration solution of gas kinetic model. The accuracy of the modified method is verified by a series of test cases.

\section*{Acknowledgements}
The current research is supported by National Key R$\&$D Program of China (Grant Nos. 2022YFA1004500), National Science Foundation of China (12172316, 92371107), and Hong Kong research grant council (16301222, 16208324).

\bibliography{maugkwpref}

\end{document}